\documentclass[12pt]{article}
\pdfoutput=1
\usepackage{yfonts}
\usepackage{color}
\usepackage{mhchem}
\usepackage{xcolor}
\usepackage{cite}
\usepackage{hyperref}
\hypersetup{colorlinks=true,linkcolor=red,anchorcolor=black,citecolor=green}
\usepackage[toc,page]{appendix}
\usepackage{amsfonts}
\usepackage{bbold}
\usepackage{textcomp}
\usepackage[DIV13]{typearea}
\usepackage{amsmath, amsthm, amssymb, mathtools,empheq,latexsym,dsfont}
\usepackage{bbm}
\usepackage{slashed, simplewick}
\usepackage[utf8]{inputenc}
\usepackage{graphicx,placeins}
\usepackage{makeidx}
\usepackage[font=small,labelfont=bf]{caption}
\usepackage{nicefrac}
\usepackage{subfigure}
\usepackage{array, bigdelim,multirow,multicol}
\usepackage[integrals]{wasysym}
\usepackage{fancybox}
\usepackage{bm}
\usepackage{float}
\usepackage{rotating}
\usepackage{colortbl}
\usepackage{booktabs}
\usepackage[top=2cm,textwidth=16.6cm,textheight=22.75cm]{geometry}
\usepackage{doi}
\usepackage{fancybox}
\usepackage[compat=1.0.0]{tikz-feynman}
\usepackage{tikz}
\usepackage{empheq}
\allowdisplaybreaks
\newcommand{\ignore}[1]{}
\graphicspath{{immagini/}}
\definecolor{Gray}{gray}{0.92}
\DeclareMathAlphabet{\mathscr}{OT1}{pzc}{m}{it}

\newcommand{\taubar}{\overline{\tau}}

\newcommand{\partialtaubar}[1]{\dfrac{\partial #1}{\partial \overline{\tau}}}

\renewcommand{\theequation}{\thesection.\arabic{equation}}

\makeatletter
\@addtoreset{equation}{section}
\makeatother

\begin{document}
%%%%%%%%%%%%%%%%%%%%%%%%%%%%%%%%%%%%%%%%%%%%%%%%%%%%%%
\unitlength = 1mm
\setlength{\extrarowheight}{0.2 cm}
\thispagestyle{empty}
\bigskip
\vskip 1cm

\title{\Large \bf Non-holomorphic modular flavor symmetry \\[2mm] }

\date{}

\author{
Bu-Yao Qu\footnote{E-mail: {\tt
qubuyao@mail.ustc.edu.cn}},  \
Gui-Jun Ding\footnote{E-mail: {\tt
dinggj@ustc.edu.cn}}
\\*[20pt]
\centerline{
\begin{minipage}{\linewidth}
\begin{center}
{\it \small
Department of Modern Physics, University of Science and Technology of China,\\
Hefei, Anhui 230026, China}
\end{center}
\end{minipage}}
\\[10mm]}
\maketitle
\thispagestyle{empty}

\centerline{\large\bf Abstract}

\begin{quote}
\indent The formalism of non-holomorphic modular flavor symmetry is developed, and the Yukawa couplings are level $N$ polyharmonic Maa{\ss} forms satisfying the Laplacian condition. We find that the integer (even) weight polyharmonic Maa{\ss} forms of level $N$ can be decomposed into multiplets of the finite modular group $\Gamma'_N$ ($\Gamma_N$).
The original modular invariance approach is extended by the presence of negative weight polyharmonic Maa{\ss} forms. The non-holomorphic modular flavor symmetry can be consistently combined with the generalized CP symmetry. We present three example models for lepton sector based on the $\Gamma_3\cong A_4$ modular symmetry, the charged lepton masses and the neutrino oscillation data can be accommodated very well, and the predictions for the leptonic CP violation phases and the effective Majorana neutrino mass are studied.

\end{quote}

\clearpage

%%%%%%%%%%%%%%%%%%%%%%%%%%%%%%%%%%%%%%%%%%%%%%%%%%%%%%%%%%%%%%%%%%%%%%%%%%%%%
\section{Introduction and motivation }
%%%%%%%%%%%%%%%%%%%%%%%%%%%%%%%%%%%%%%%%%%%%%%%%%%%%%%%%%%%%%%%%%%%%%%%%%%%%%

It is well-established that the standard model (SM) of particle physics has successfully described the electromagnetic, strong and weak forces in Nature.
It can explain almost all experimental results and it has passed through precision tests of many experiments. Now we have a very good understanding  for the structure of the SM gauge sector. However, a fundamental understanding for the structure of the flavor sector, which contains the dominant fraction of the SM parameters, is still absent. The experimentally measured masses and mixing parameters of quarks and leptons provide a clue for new physics beyond standard model (SM). Motivated by the large lepton mixing angles, the non-Abelian discrete flavor symmetry has been extensively utilized to explain the lepton mixing patterns observed by neutrino oscillation experiments. The discrete flavor symmetry group should generally be broken along different directions in flavor space in the neutrino and charged lepton sectors, and the mismatch in the breaking of the flavor symmetry group gives rise to certain lepton mixing patterns, see Refs.~\cite{King:2017guk,Xing:2020ijf,Feruglio:2019ybq,Ding:2024ozt} for recent reviews. The dynamics breaking the flavor symmetry typically requires additional degree of freedom, additional symmetry and introduce more free parameters so that and resulting models are quite complicated and the predictive power of flavor symmetry is reduced.

The modular invariance as flavor symmetry was suggested to avoid the ad hoc flavor symmetry breaking sector~\cite{Feruglio:2017spp}. In the simplest modular invariant models, the vacuum expectation value (VEV) of the complex modulus $\tau$ is the unique source of flavor symmetry breaking, and flavons in traditional flavor symmetry models are unnecessary so that the complications of vacuum alignment is evaded\footnote{The dynamical determination of the VEV of $\tau$ is still elusive. In the bottom-up models with modular flavor symmetry, the VEV of $\tau$ is usually treated as a random complex parameter freely varying in the fundamental domain to optimize the agreement between the model predictions and the experimental data.}. The approach of modular flavor symmetry has been intensively studied from both bottom-up and top-down perspectives in the past few years, see Refs.~\cite{Kobayashi:2023zzc,Ding:2023htn} for recent review and an extensive list of papers on modular symmetry therein. The modular flavor symmetry replies on the level $N$ modular forms which are holomorphic functions of $\tau$ and can be arranged into multiplets of the inhomogeneous (or homogeneous) finite modular groups $\Gamma_N$ (or $\Gamma'_N$). The matter fields are assumed to transform in representations of $\Gamma_N$ (or $\Gamma'_N$) up to the automorphy factor. So far the modular symmetry is implemented in the framework of global (or rigid) supersymmetry (SUSY), thus the superpotential is a holomorphic function. As a consequence, modular invariance fixes the Yukawa couplings and fermion mass matrices to be modular forms of level $N$. Here the holomorphicity protected by SUSY plays a crucial role to enhances the predictive power of modular invariance. It is notable that the modular invariance allows to predict the masses and flavor mixing of both quarks and leptons in terms of a small number of free parameters. All three charged lepton masses, three neutrino masses, three lepton mixing angles and three CP violation phases for Majorana neutrinos can be described in terms of only 6 real parameters in the minimal phenomenologically viable model based on modular symmetry~\cite{Ding:2022nzn,Ding:2023ydy}, and extension to quark sector requires 8 additional parameters~\cite{Ding:2023ydy}.

SUSY is a well-motivated candidate for physics beyond SM, the experimental data from flavor physics, high-precision electroweak observables and astrophysics impose strong constraints on the allowed SUSY parameter space.  However, the signature of low energy SUSY has not been experimentally observed~\cite{ParticleDataGroup:2022pth}, and we still don't know whether and how SUSY is realized in Nature. Hence it is intriguing to implement the modular invariance without SUSY. Moreover, it was claimed that the low energy SUSY possibly is not required for the Yukawa couplings to be modular forms~\cite{Cremades:2004wa,Almumin:2021fbk}, and the amount of supersymemtry preserved at low energies depends on the geometrical details of the compactification. It was suggested that the framework of automorphic forms provides a non-supersymmetric realization of the modular flavor symmetry~\cite{Ding:2020zxw}, the assumption of holomorphicity is replaced by the Laplacian condition. For the case of single complex modulus, the automorphic forms coincide with the harmonic Maa{\ss} forms which are non-holomorphic modular functions satisfying the Laplacian equation.

From the view of top-down, the coefficients of the effective interactions in the low energy expansion of the four-supergraviton amplitude are non-holomorphic automorphic functions which satisfy Laplacian eigenvalue equations~\cite{Green:1997tv,Green:1997me,Pioline:1998mn,Green:1998by,deHaro:2002vk,Green:2010wi,Basu:2011he,Peeters:2000qj,Sinha:2002zr}. Moreover, the coefficient functions in the low energy expansion of string theory compactified on higher dimensional torus are generally automorphic functions invariant under the corresponding duality group~\cite{Green:2010wi}. The holomorphic part of the automorphic form is  the so-called mock modular form, and it could be related to the degeneracies of quantum black holes in string theories~\cite{Dabholkar:2012nd}.

This motivates us to discuss the non-holomorphic modular flavor symmetry in the framework of harmonic Maa{\ss} forms which have both holomorphic and non-holomorphic parts, modular invariance constrains the Yukawa couplings to be harmonic Maa{\ss} forms  of level $N$. Imposing moderate growth condition, we consider the so-called polyharmonic Maa{\ss} forms of level $N$ which span a linear space of finite dimension. Hence the Yukawa interactions are restricted and a number of fermion mass terms are allowed. Furthermore, we find that the integer weight polyharmonic Maa{\ss} forms of level $N$ can be decomposed into multiplets of the homogeneous finite modular group $\Gamma'_N$ up to the automorphy factor, and the even weight polyharmonic Maa{\ss} forms of level $N$ can be arranged into multiplets of the inhomogeneous finite modular group $\Gamma_N$. Here $\Gamma'_N$ (or $\Gamma_N$) plays the role of flavor symmetry, and the transformation properties of matter fields are characterized by the modular weights and transformations under $\Gamma'_N$ (or $\Gamma_N$). The non-holomorphic modular flavor symmetry extends the original modular invariance approach due to the presence of negative weight polyharmonic Maa{\ss} forms, while the weight of modular form must be non-negative.

The remaining of this paper is organized as follows. We give an introduction to the automorphic form and harmonic Maa{\ss} form in section~\ref{sec:automorphic-form-maass-form}. We show that the (even) integer weight polyharmonic Maa{\ss} forms at level $N$ can be arranged into multiplets of ($\Gamma_N$) $\Gamma'_N$, and their explicit expressions can be obtained from the known modular forms. The formalism of the non-holomorphic modular flavor symmetry is presented in section~\ref{sec:non-SUSY-modular-symmetry}. For illustration, the formalism is applied to the lepton sector and three benchmark models based on the $A_4$ modular symmetry are given in section~\ref{sec:benchmark-models}, two models for Majorana neutrinos and one for Dirac neutrinos. Section~\ref{sec:summary-discussions} is devoted to the summary and discussions. The Fourier expansion of a polyharmonic Maa{\ss} form of weight $k$ and level $N$ is presented in Appendix~\ref{app:Fourier-expansion}. The $D$ operator and $\xi$ operator and their action on the polyharmonic Maa{\ss} form  are given in Appendix~\ref{app:differential-operators}. We provide the multiplets of even weight polyharmonic Maa{\ss} forms at levels $N=2, 3, 4, 5$ in Appendix~\ref{app:polyharmonic-Maass-form-multiplets}.

%%%%%%%%%%%%%%%%%%%%%%%%%%%%%%%%%%%%%%%%%%%%%%%%%%%%%%%%%%%%%%%%%%%%%%%%%%%%%
\section{\label{sec:automorphic-form-maass-form}Automorphic forms and harmonic Maa{\ss} forms }
%%%%%%%%%%%%%%%%%%%%%%%%%%%%%%%%%%%%%%%%%%%%%%%%%%%%%%%%%%%%%%%%%%%%%%%%%%%%%

The special linear group $SL(2, \mathbb{R})$ is the group of the $2\times 2$ real matrices with unit determinant:
\begin{equation}
SL(2, \mathbb{R})=\left\{
\begin{pmatrix}
a  ~&~ b \\
c  ~&~ d
\end{pmatrix}\Bigg|
ad-bc=1,~~a, b,c,d\in \mathbb{R}
\right\}\,.
\end{equation}
The elements of $SL(2, \mathbb{R})$ act on the complex upper half-plane
$\mathcal{H}=\left\{\tau\in\mathbb{C}|\text{Im}\tau>0\right\}$ by M\"obius transformation:
\begin{equation}
\tau\mapsto g\tau=\frac{a\tau+b}{c\tau+d},~~~~g=\begin{pmatrix}
a ~&~ b \\
c ~&~ d
\end{pmatrix}\in SL(2, R)\,.
\end{equation}
One sees that the known $SO(2)$ group is a compact subgroup of $SL(2, \mathbb{R})$, the coset decomposition of $SL(2, \mathbb{R})$ with respect to $SO(2)$ implies that a generic element of $SL(2, \mathbb{R})$ can be decomposed as~\cite{Feruglio:2017spp}
\begin{equation}
\left(
\begin{array}{cc}
\sqrt{y} ~&~ x/\sqrt{y}\\
0 ~&~ 1/\sqrt{y}
\end{array}
\right)h\,,\qquad y>0\,,
\end{equation}
with
\begin{equation}
h=\left(\begin{matrix}
\cos\theta ~& -\sin\theta\\
\sin\theta ~& \cos\theta
\end{matrix}\right)\in SO(2)\,.
\end{equation}
The self dual modulus $\tau_0=i$ is invariant under the action of $SO(2)$, and a generic modulus $\tau=x+iy$ can always be related to $\tau_0$ by a $SL(2, \mathbb{R})$ transformation
\begin{equation}
g\tau_0=\left(
\begin{array}{cc}
\sqrt{y}~&~ x/\sqrt{y}\\
0~&~1/\sqrt{y}
\end{array}
\right)\cdot i=\tau\,.
\label{gi}
\end{equation}
Hence the modulus $\tau$ can be parameterized as the coset space $SL(2, \mathbb{R})/SO(2)$. In the same manner, $SL(2, \mathbb{R})$ and $SO(2)$ are generalized to a Lie group $G$ and its compact subgroup $K$, thus single complex modulus is naturally generalized to high dimensional moduli space descried by the coset $G/K$~\cite{Ding:2020zxw}. The modular invariance would require the Yukawa couplings to be certain modular functions $Y(\tau)$ which is closely related to the automorphic forms. For the case with single modulus, $Y(\tau)$ should fulfill the following conditions~\cite{Ding:2020zxw}:
\begin{subequations}
\begin{eqnarray}
\label{eq:modularity-def} Y(\gamma \tau) &=& j^k(\gamma,\tau) Y(\tau),~~~\gamma\in G_d\,, \\
\label{eq:autform-def}\Delta_k Y(\tau)&=&0\,,
\end{eqnarray}
\end{subequations}
where $j(\gamma, \tau)=c\tau+d$ is the automorphy factor, $k$ is the modular weight, $G_d$ is a discrete subgroup of $G$, and $\Delta_k$ is the weight $k$ hyperbolic Laplacian operator\footnote{Equivalently $\Delta_k$ can be compactly written as follow,
\begin{equation*}
\Delta_k=-4y^{2-k}\partial_{\tau} y^{k}\partial_{\taubar} \,.
\end{equation*}}
\begin{eqnarray}
\Delta_k = -y^2\left( \dfrac{\partial^2}{\partial x^2} + \dfrac{\partial^2}{\partial y^2} \right) + iky \left( \dfrac{\partial}{\partial x} + i\dfrac{\partial}{\partial y} \right)
=-4y^2\dfrac{\partial}{\partial \tau} \dfrac{\partial}{\partial \taubar} +2iky\dfrac{\partial}{\partial \taubar}\,,
\end{eqnarray}
The modular function $Y(\tau)$ satisfying Eq.~\eqref{eq:autform-def} is generally a non-holomorphic function of $\tau$ although any holomorphic function is the solution of Eq.~\eqref{eq:autform-def}. Hence the Yukawa couplings could have non-holomorphic part if Supersymmetry is not imposed. Moreover, the modular function $Y(\tau)$ should fulfill suitable growth conditions~\cite{Borel,Borel2}. In the present work, we are concerned with the polyharmonic Maa{\ss} forms, the moderate growth condition is
\begin{eqnarray}
\label{eq:growth-cond}Y(\tau)=\mathcal{O}(y^\alpha) ~~~\text{as $y\rightarrow +\infty$, uniformly in $x$}
\end{eqnarray}
for some $\alpha$, and similar condition holds at all cusps of $G_d$~\cite{Lagarias2015poly}. In the present work, we shall be concerned with the case of $G_d=\Gamma(N)$ which is the so-called principal congruence subgroup of level $N$ defined as
\begin{equation}
\Gamma(N)=\left\{\begin{pmatrix}
a  ~& b \\
c  ~& d
\end{pmatrix}\in SL(2,\mathbb{Z})\Bigg|a-1=d-1=b=c=0\;(\texttt{mod}\; N)
\right\}
\end{equation}
with $N=1, 2, 3, \ldots$ being a positive integer. It is a normal subgroup of $SL(2, \mathbb{Z})$, in particularly the element $T^{N}\in\Gamma(N)$. The modularity condition in Eq.~\eqref{eq:modularity-def} implies $Y(\tau+N)=Y(\tau)$. As shown in Appendix~\ref{app:Fourier-expansion}, after taking into account the harmonic condition of Eq.~\eqref{eq:autform-def}, the Fourier expansion of $Y(\tau)$ is determined to be:
\begin{eqnarray}
\label{eq:Fourier-expansion-weak-masss}Y(\tau)=\sum_{\substack{n\in\frac{1}{N}\mathbb{Z} }} c^+(n)q^n + c^-(0)y^{1-k} + \sum_{\substack{n\in\frac{1}{N}\mathbb{Z}\backslash\{0\} }} c^-(n)\Gamma(1-k,-4\pi n y)q^n\,,~~~q\equiv e^{2\pi i\tau} \,.~~
\end{eqnarray}
If $k=1$, the term $y^{1-k}$ should be replaced by $\ln y$. Here $\Gamma(s,z)$ is the incomplete gamma function given by~\cite{book:Ono}
\begin{eqnarray}
\label{eq:incomplete-gamma}\Gamma(s,z)=\int_z^{+\infty} e^{-t}t^{s-1}\,dt\,.
\end{eqnarray}
The incomplete gamma function have the asymptotic behavior for $z\in\mathbb{R}$,
\begin{eqnarray}
\Gamma(s, z)\sim z^{s-1}\,e^{-z}~~~\text{ as~~ $|z|\rightarrow +\infty$}\,.
\end{eqnarray}
Moreover, integration by parts gives the following recurssion formula
\begin{eqnarray}
\label{eq:Gamma-recursion}\Gamma(s+1,z)=s\Gamma(s,z) + z^s\,e^{-z}
\end{eqnarray}
Furthermore, the growth condition in Eq.~\eqref{eq:growth-cond} restricts the range of values of $n$, i.e.
\begin{eqnarray}
c^+(n)=0,~~c^-(-n)=0,~~ \text{for }~  n< 0\,.
\end{eqnarray}
Consequently the Fourier expansion of a level $N$ and weight $k$ polyharmonic Maa{\ss} form is of the following form
\begin{eqnarray}
\label{eq:poly-maass-form}Y(\tau)=\sum_{\substack{n\in\frac{1}{N}\mathbb{Z} \\ n\geq0}} c^+(n)q^n + c^-(0)y^{1-k}+ \sum_{\substack{n\in\frac{1}{N}\mathbb{Z} \\ n<0}} c^-(n)\Gamma(1-k,-4\pi n y)q^n \,.
\end{eqnarray}
Notice that the terms proportional to $c^-(0)$ and $c^-(n)$ are non-holomorphic, and they are vanishing for the holomorphic modular form. Obviously the level $N$ and weight $k$ modular forms are polyharmonic Maa{\ss} forms, nevertheless polyharmonic Maa{\ss} forms contain more modular functions than the modular forms. It is known that the product of two modular forms of level $N$ and weights $k$, $k'$ is a modular form of level $N$ and weight $k+k'$. However, the product of two polyharmonic Maa{\ss} forms of level $N$ and weights $k$, $k'$ generally is not a polyharmonic Maa{\ss} form of level $N$ and weight $k+k'$, because the harmonic condition could be spoiled for $k, k'<0$.

It is known that the modified weight $2$ Eisenstein series $\widehat{E}_2(\tau)$ is a polyharmonic Maa{\ss} form of $SL(2, Z)$ and $\widehat{E}_2(\tau)$ is defined as~\cite{Bruinier2008The},
\begin{eqnarray}
\widehat{E}_2(\tau)=E_2(\tau)-\dfrac{3}{\pi y}=1-\dfrac{3}{\pi y}-24\sum_{n=1}^\infty \sigma_1(n)q^n\,,
\end{eqnarray}
where $E_2(\tau)=1-24\sum_{n=1}^\infty \sigma_1(n)q^n$ is the weight 2 Eisenstein series and $\sigma_1(n) =\sum_{d|n} d$ is the sum of the divisors of $n$. The explicit $q$-expansion expression of $\widehat{E}_2(\tau)$ reads
\begin{eqnarray}
\widehat{E}_2(\tau) = 1 - \dfrac{3}{\pi y} - 24 q - 72 q^2 - 96 q^3 - 168 q^4 - 144 q^5 - \cdots \,.
\end{eqnarray}
Obviously $\widehat{E}_2(\tau)$ is a non-holomorphic function of the modulus $\tau$. Notice that $E_2(\tau)$ is not a modular form of weight $2$, and it fulfills the following transformation formula
\begin{equation}
E_2(\gamma\tau)=(c\tau+d)^2 E_2(\tau) -\frac{6i}{\pi} c(c\tau+d)\,,
\end{equation}
where the last term violates the modularity. Due to the presence of the term $-\dfrac{3}{\pi y}$, the modularity condition of Eq.~\eqref{eq:modularity-def} is satisfied, i.e. $\widehat{E}_2(\gamma\tau)=(c\tau+d)^2\widehat{E}_2(\tau)$ for any $\gamma\in SL(2, \mathbb{Z})$. Consequently $\widehat{E}_2(\tau)$ is also a polyharmonic Maa{\ss} form of weight 2 and any positive level $N$.

\subsection{Decomposition of polyharmonic Maa{\ss} forms }

The polyharmonic Maa{\ss} forms of integer weight $k$ and level $N$ span a linear space whose dimension is denoted as $\texttt{dim}\,\mathcal{PH}_{k}(\Gamma(N))$. One could choose a set of linearly independent basis denoted as $f_{i}(\tau)$ with $i=1, 2, \ldots, \texttt{dim}\,\mathcal{PH}_{k}(\Gamma(N))$. We shall demonstrate that $f_{i}(\tau)$ can be arranged into irreducible multiplets of the finite modular group $\Gamma'_N=SL(2, \mathbb{Z})/\Gamma(N)$ for general integer $k$ and multiplets of  $\Gamma_N=SL(2, \mathbb{Z})/\pm\Gamma(N)$ for even modular weight $k$. For a generic element $\gamma\in \mathrm{SL}(2,\mathbb{Z})$, we construct the following modular function
\begin{eqnarray}
F_{i\gamma}(\tau)=(c\tau + d)^{-k}f_i(\gamma\tau)=j^{-k}(\gamma,\tau) f_i(\gamma\tau)\,.
\end{eqnarray}
It is straightforward to check that the harmonic condition $\Delta_k F_{i\gamma}(\tau)=0$ is fulfilled. Under the action of any modular transformation $g\in\Gamma(N)$, we have
\begin{eqnarray}
F_{i\gamma}(g\tau)=j^{-k}(\gamma, g\tau)f_{i}(\gamma g\tau)=j^{-k}(\gamma, g\tau)f_{i}(\gamma g\gamma^{-1}\gamma\tau)=j^{k}(g, \tau)F_{i\gamma}(\tau)\,,
\end{eqnarray}
where we have used the fact that the element $\gamma g \gamma ^{-1}$ belongs to $\Gamma(N)$, because $\Gamma(N)$ is a normal subgroup of $SL(2, \mathbb{Z})$. Therefore $F_{i\gamma}(\tau)$ are also polyharmonic Maa{\ss} forms of level $N$ and weight $k$, and they are linear combinations of $f_i$ as follows,
\begin{eqnarray}
F_{i\gamma}(\tau)=\rho_{ij}(\gamma)f_j(\tau)\,,
\end{eqnarray}
which gives
\begin{equation}
\label{eq:PHMF-irrep}f_i(\gamma\tau)=(c\tau+d)^{k}\rho_{ij}(\gamma)f_j(\tau)\,.
\end{equation}
It is straightforwardly to check that $\rho$ forms a linear representation of $\mathrm{SL}(2,\mathbb{Z})$, and the following relation is satisfied,
\begin{eqnarray}
\rho(\gamma_1\gamma_2)&=&\rho(\gamma_1)\rho(\gamma_2)\,.
\end{eqnarray}
Furthermore, $f_{i}(\tau)$ are polyharmonic Maa{\ss} forms of level $N$ and weight $k$, by definition they should fulfill
\begin{equation}
\label{eq:PHMF-ID-G}f_i(\gamma\tau)=(c\tau+d)^{k}f_i(\tau),~~~~~\gamma\in\Gamma(N)\,.
\end{equation}
Thus we have
\begin{equation}
\rho(\gamma)=1,~~~~~\gamma\in \Gamma(N)\,.
\end{equation}
Hence $\rho$ is a representation of the homogeneous finite modular group  $\Gamma'_N=\mathrm{SL}(2,\mathbb{Z})/\Gamma(N)$ for generic integer $k$.
The representation $\rho$ is generally a reducible representation of $\Gamma'_N$. It is known that each reducible representation of a finite group can be decomposed into a direct sum of irreducible unitary representations~\cite{rao2006linear}. As a consequence, by properly choosing basis, $\rho$ can be written into a block diagonal form,
\begin{equation}
\label{dim_relation}\rho \sim \rho_{\bm{r_1}} \oplus \rho_{\bm{r_2}} \oplus \dots \,,~~~\text{with}~~~ \sum_i\texttt{dim}\,\rho_{\bm{r}_i} = \texttt{dim}\mathcal{PH}_{k}(\Gamma(N))\,,
\end{equation}
where $\rho_{\bm{r}_i}$ stand for irreducible unitary representations of $\Gamma'_N$. Hence the weight $k$ polyharmonic Maa{\ss} forms of level $N$ can be arranged into some multiplets $Y^{(k)}_{\bm{r_i}}$ which transform in the irreducible representation $\bm{r}_i$ of $\Gamma'_N$ up to the automorphy factor, i.e.
\begin{equation}
\label{eq:PHMF-irrep-decomp}Y^{(k)}_{\bm{r_i}}(\gamma\tau)=(c\tau+d)^{k}\rho_{\bm{r_i}}(\gamma)Y^{(k)}_{\bm{r_i}}(\tau),~~~\gamma\in SL(2, \mathbb{Z})\,.
\end{equation}
Applying Eq.~\eqref{eq:PHMF-irrep-decomp} to $\gamma=S^2=R$, we obtain
\begin{equation}
Y^{(k)}_{\bm{r_i}}(\gamma\tau)(R\tau)=Y^{(k)}_{\bm{r_i}}(\tau)=(-1)^{k}\rho_{\bm{r_i}}(R)Y^{(k)}_{\bm{r_i}}(\tau)\,,
\end{equation}
which implies
\begin{equation}
\rho_{\bm{r_i}}(R)=(-1)^k
\end{equation}
If $k$ is even, we obtain $\rho_{\bm{r_i}}(R)=1$. Hence the even weight polyharmonic Maa{\ss} forms of level $N$ can be organized into multiplets of
$\Gamma_N=\mathrm{SL}(2,\mathbb{Z})/\pm\Gamma(N)$. For the level $N=3$, the even weight polyharmonic Maa{\ss} forms of level 3 can be arranged into multiplets of $\Gamma_3\cong A_4$, while the generic integer weight polyharmonic Maa{\ss} forms of level 3 can be arranged into multiplets of $\Gamma'_3\cong T'$. Note that the representations of $T'$ with $R=1$ coincide with those of $A_4$, consequently $T'$ and $A_4$ can not be distinguished by these representations.

\subsection{\label{subsec:MF-PHMF}Lifting modular forms to polyharmonic Maa{\ss} forms}

It is remarkable that the polyharmonic Maa{\ss} form can be related to the modular forms through the $D$ operator and the $\xi$ operator. To be more specific, for any polyharmonic Maa{\ss} form $Y(\tau)$ of level $N$ and weight $k$, both $D^{1-k} Y(\tau)$ and $\xi_k Y(\tau)$ are modular forms of weight $2-k$ at level $N$~\cite{book:Ono}, as shown in the Appendix~\ref{app:differential-operators}. For a multiplet $Y^{(k)}_{\bm{r}}(\tau)$ of polyharmonic Maa{\ss} form in the irreducible representation $\bm{r}$ of $\Gamma'_N$ (or $\Gamma_N$), using Eqs.~(\ref{eq:D-1-k-mod}, \ref{eq:xi-modularity}) we find that both $\xi_k Y^{(k)}_{\bm{r}}$ and $D^{1-k}Y^{(k)}_{\bm{r}}$ are level $N$ modular form multiplets of weight $2-k$ in the representations $\bm{r}^{*}$ and $\bm{r}$ respectively, i.e.,
\begin{subequations}
\begin{eqnarray}
\label{eq:xi-oper-modf-a}(\xi_k Y^{(k)}_{\bm{r}})(\gamma \tau)&=&(c\tau+d)^{2-k}\rho^{*}_{\bm{r}}(\gamma)\left(\xi_k Y^{(k)}_{\bm{r}}\right)(\tau)\,, \\
\label{eq:D-oper-modf-b}(D^{1-k}Y^{(k)}_{\bm{r}})(\gamma \tau)&=&(c\tau+d)^{2-k}\rho_{\bm{r}}(\gamma)\left(D^{1-k}Y^{(k)}_{\bm{r}}\right)(\tau)\,,
\end{eqnarray}
\end{subequations}
where
\begin{equation}
\gamma=\begin{pmatrix}
a  ~&~  b \\
c  ~&~ d
\end{pmatrix}
\end{equation}
is a representative element of $\Gamma'_N$ (or $\Gamma_N$).
Here $\rho^{*}_{\bm{r}}$ stands for the conjugate representation of $\rho_{\bm{r}}$. Using Eqs.~(\ref{eq:xi-oper-modf-a}, \ref{eq:D-oper-modf-b}), one can lift modular form to polyharmonic Maa{\ss} forms. Note that the second equation Eq.~\eqref{eq:D-oper-modf-b} is automatically satisfied for weight $k=1$, and the polyharmonic Maa{\ss} form of weight 1 can not be fixed by Eq.~\eqref{eq:xi-oper-modf-a} alone. Generally the conjugate representation $\rho^{*}_{\bm{r}}$ is equivalent to another representation $\rho_{\bm{r}'}$ of $\Gamma'_N$ (or $\Gamma_N$)\footnote{If $\rho_{\bm{r}}$ is a real or pseudo-real representation, then $\rho^{*}_{\bm{r}}$ is equivalent to $\rho_{\bm{r}}$ so that $\rho_{\bm{r}'}$ is exactly $\rho_{\bm{r}}$. If $\rho_{\bm{r}}$ is a complex representation, then $\rho^{*}_{\bm{r}}$ is inequivalent to $\rho_{\bm{r}}$ and consequently $\rho_{\bm{r}'}$ is different from $\rho_{\bm{r}}$.}, and they are related by certain similarity transformation $\Omega$,
\begin{equation}
\rho^{*}_{\bm{r}}(\gamma)=\Omega\rho_{\bm{r}'}(\gamma)\Omega^\dag\,.
\end{equation}
It is convenient to work in the $T$ diagonal basis and then the representation matrices $\rho_{\bm{r}}(T)$ and $\rho_{\bm{r}'}(T)$ can be parameterized as
\begin{eqnarray}
\rho_{\bm{r}}(T)={\rm diag}\{1,e^{2\pi i r_2}\,\cdots \,e^{2\pi i r_d}\}\,,~~~\rho_{\bm{r}'}(T)={\rm diag}\{1,e^{2\pi i r'_2}\,\cdots \,e^{2\pi i r'_d}\},~~~0\leq r_i<1~~\,.
\end{eqnarray}
with $r_i=1-r'_j$ for certain $i$ and $j$. In the $T$ diagonal basis, the unitary transformation $\Omega$ can be chosen to be permutation matrix. From Eq.~\eqref{eq:xi-oper-modf-a} we see that $\Omega^{\dag}\xi_k Y^{(k)}_{\bm{r}}$ is a weight $2-k$ modular form in representation $\rho_{\bm{r}'}$, i.e.
\begin{eqnarray}
\Omega^\dag \xi_k Y^{(k)}_{\bm{r}}(\gamma\tau) = (c\tau+d)^{2-k}\rho_{\bm{r}'}(\gamma)\Omega^{\dag} \xi_k Y^{(k)}_{\bm{r}}(\tau)\,.
\end{eqnarray}
It is clear that both the imaginary part $y$ and $q^{n}$ is invariant under the action of $T$ for any integer $n$, and $q^{r}$ for rational $r$ transforms under $T$ as
\begin{equation}
q^{r}\stackrel{T}{\rightarrow} e^{2\pi i(\tau+1) r}=e^{2r\pi i}q^r\,.
\end{equation}
Therefore the Fourier expansion of a level $N$ weight $k$ polyharmonic Maa{\ss} form $Y^{(k)}_{\bm{r}}(\tau)=\left(Y^{(k)}_{\bm{r},1}(\tau), Y^{(k)}_{\bm{r},2}(\tau),\ldots, Y^{(k)}_{\bm{r},d}(\tau)\right)^T$ can be written as
\begin{eqnarray}
Y^{(k)}_{\bm{r},1}(\tau)&=&c_{1}^+(0)+ c_{1}^-(0)y^{1-k}+\sum_{n=1}^{\infty} c_{1}^+(n)q^n+c_{1}^-(-n)\Gamma(1-k,4\pi n y)q^{-n}\,, \\
Y^{(k)}_{\bm{r},i}(\tau)&=&q^{r_i}\sum_{n=0}^{\infty} c_{i}^+(n)q^n+q^{r_i}\sum_{n=1}^{\infty}c_{i}^-(-n)\Gamma(1-k,4\pi (n-r_i)y)q^{-n} \,,~~i\geq 2\,.
\end{eqnarray}
If there exist unique modular multiplets $Y^{(2-k)}_{\bm{r}'}(\tau)$ and $Y^{(2-k)}_{\bm{r}}(\tau)$ at level $N$, weight $2-k$ and representations $\bm{r}$ and $\bm{r}'$, then proportionality follows\footnote{Here the representation $\bm{r}'$ is equivalent to $\bm{r}^{*}$. Therefore $\Omega Y^{(2-k)}_{\bm{r}'}(\tau)$ is a modular multiplet in the representation $\rho^{*}$, and it fulfills $\Omega Y^{(2-k)}_{\bm{r}'}(\gamma\tau)=(c\tau+d)^{2-k}\rho^{*}_{\bm{r}}(\gamma)\,\Omega Y^{(2-k)}_{\bm{r}'}(\tau)$.}:
\begin{eqnarray}\label{eq:lift-cond}
\xi_k Y^{(k)}_{\bm{r}}(\tau) = \alpha \Omega Y^{(2-k)}_{\bm{r}'}(\tau)\,,~~~D^{1-k}Y^{(k)}_{\bm{r}}(\tau) =\beta Y^{(2-k)}_{\bm{r}}(\tau)\,.
\end{eqnarray}
Here $\alpha$ and $\beta$ are complex constants which are fixed later, The weight $2-k$ modular forms $Y_{\bm{r}}^{(2-k)}(\tau)$ and $\Omega\, Y^{(2-k)}_{\bm{r}'}(\tau)$ at level $N$ furnish the representations $\rho_{\bm{r}}$ and $\rho^{*}_{\bm{r}}$ respectively, and their $q$-expansions are denoted as
\begin{eqnarray}
Y_{\bm{r}, 1}^{(2-k)}(\tau)&=&\sum_{n=0}^{\infty}a_{1}(n) q^n,~~~~~\left(\Omega Y_{\bm{r}'}^{(2-k)}(\tau)\right)_{1}=\sum_{n=0}^{\infty}b_{1}(n) q^n\,, \\
Y_{\bm{r},i}^{(2-k)}(\tau)&=&q^{r_i}\sum_{n=0}^{\infty}a_{i}(n) q^n ,~~~\left(\Omega Y_{\bm{r}'}^{(2-k)}(\tau)\right)_{i}=q^{-r_i}\sum_{n=1}^{\infty}b_{i}(n) q^n,~~~ i>1\,.
\end{eqnarray}
From Eqs.~\eqref{eq:Bol-F} and \eqref{eq:xi-Y}, we can obtain
\begin{eqnarray}
\xi_k Y^{(k)}_{\bm{r},1}(\tau) &=& (1-k)\overline{c_{1}^-(0)}-\sum_{n=1}^{\infty}(4\pi n)^{1-k}\,\overline{c_{1}^-(-n)}\,q^n \\
\xi_k Y^{(k)}_{\bm{r},i}(\tau) &=& -q^{-r_i} \sum_{n=1}^{\infty}(4\pi (n-r_i))^{1-k}\,\overline{c_{i}^-(-n)}\,q^n\,, \\
D^{1-k} Y^{(k)}_{\bm{r},1}(\tau)&=&(-4\pi)^{k-1} (1-k)! \, c_{1}^-(0)+\sum_{n=1}^{\infty} n^{1-k}c_{i}^+(n) q^n\\
D^{1-k} Y^{(k)}_{\bm{r},i}(\tau)&=&q^{r_i}\sum_{n=0}^{\infty} (n+r_i)^{1-k}c_{i}^+(n) q^n\,,
\end{eqnarray}
with $i\geq2$. Comparing the expressions of $q$-expansion on both sides of Eq.~\eqref{eq:lift-cond}, we find that the coefficients of the Fourier expansion of the polyharmonic Maa{\ss} form multiplet $Y^{(k)}_{\bm{r}}(\tau)$ satisfy the following constraints
\begin{eqnarray}
\nonumber&& (1-k)\,\overline{c_{1}^-(0)}=\alpha b_{1}(0)\,,~~~(-4\pi)^{k-1} (1-k)! \, c_{1}^-(0)=\beta a_{1}(0)\,, \\
\nonumber&&-(4\pi n)^{1-k}\overline{c_{1}^-(-n)}=\alpha b_1(n) \,, ~~~ n^{1-k}c_{1}^+(n)=\beta a_{1}(n)\,, \\
&&-(4\pi (n-r_i))^{1-k}\overline{c_{i}^-(-n)} =\alpha b_{i}(n)\,,~~~ (n+r_i)^{1-k}c_{i}^+(n) = \beta a_{i}(n)\,.
\end{eqnarray}
Without loss of generality, we could choose the normalization factor $\alpha=1$, then we have
\begin{eqnarray}
\nonumber&& \beta= (-4\pi)^{k-1} (-k)!\,\overline{b_{1}(0)}/a_{1}(0) \,,~~~c_{1}^-(0)=\dfrac{\overline{b_1(0)}}{1-k}\,,\\
\nonumber&&c_{1}^-(-n)=-(4\pi n)^{k-1}\overline{b_1(n)}\,,~~~c_{1}^+(n)=(-4\pi n)^{k-1} (-k)!\,\overline{b_{1}(0)}\,\dfrac{a_1(n)}{a_{1}(0)},~~\text{for}~~n\neq0\,,\\
\label{eq:coefficients-values}&&c_{i}^-(-n)= -(4\pi (n-r_i))^{k-1}\,\overline{b_i(n)}\,,~~~
c_{i}^+(n)= (-4\pi (n+r_i))^{k-1} (-k)!\,\overline{b_{1}(0)}\dfrac{a_{i}(n)}{a_{1}(0)}\,.~~~~
\end{eqnarray}
Thus all the Fourier expansion coefficients except $c_{1}^+(0)$ of weight $k$ polyharmonic Maa{\ss} form multiplet $Y^{(k)}_{\bm{r}}(\tau)$ at level $N$ can be expressed in terms those of the weight $2-k$ and level $N$ modular forms. The value of $c_{1}^+(0)$ can be fixed from the decomposition equation in Eq.~\eqref{eq:PHMF-irrep-decomp}. In short, one can lift the level $N$ and weight $k$ modular forms to the level $N$ and weight $k$ polyharmonic Maa{\ss} form multiplet via the following formula
\begin{eqnarray}
\nonumber Y^{(k)}_{\bm{r},1}(\tau)&=&c_{1}^+(0)+ \dfrac{\overline{b_1(0)}}{1-k}y^{1-k}-\sum_{n=1}^{\infty} (4\pi n)^{k-1}\Big[(-1)^k (-k)!\,\overline{b_{1}(0)}\,\dfrac{a_1(n)}{a_{1}(0)}q^n+\overline{b_1(n)}\;\Gamma(1-k,4\pi n y)q^{-n}\Big] \\
\nonumber Y^{(k)}_{\bm{r},i}(\tau)&=& q^{r_i}\sum_{n=0}^{\infty} (-4\pi (n+r_i))^{k-1} (-k)!\,\overline{b_{1}(0)}\dfrac{a_{i}(n)}{a_{1}(0)}q^n\\
\label{eq:polyHarmonic-expr-main}&&~~ - q^{r_i}\sum_{n=1}^{\infty}(4\pi (n-r_i))^{k-1}\;\overline{b_i(n)}\Gamma(1-k,4\pi (n-r_i)y)q^{-n} \,,~~i\geq 2\,.
\end{eqnarray}
Notice that the product of two polyharmonic Maa{\ss} forms of weights $k$, $k'$ at level $N$ usually is not a polyharmonic Maa{\ss} form of weight $k+k'$, since the Laplacian condition of Eq.~\eqref{eq:autform-def} could be spoiled particularly if $k<0$ or $k'<0$. Moreover, the polyharmonic Maa{\ss} forms coincide with the modular forms at level $N$ for the modular weight $k\geq3$, because there is no non-zero negative weight modular forms.

\section{\label{sec:non-SUSY-modular-symmetry}Non-holomorphic modular flavor symmetry }

Supersymmetry is assumed in the modular flavor symmetry~\cite{Feruglio:2017spp},  the framework of automorphic form opens the way to non-supersymmetric realization of modular flavor symmetry, as suggested in Ref.~\cite{Ding:2020zxw}. The polyharmonic Maa{\ss} form is the automorphic form of single modulus $\tau$. In this section, we generalize the modular flavor symmetry of~\cite{Feruglio:2017spp} to non-supersymmetric case with the polyharmonic Maa{\ss} form. The level $N$ and the finite modular group $\Gamma'_N$ (or $\Gamma_N$) are kept fixed. The generic matter fields are denoted by $\psi_i$ and $\psi^c_i$, their transformations under the modular group are specified by the modular weights $-k_{\psi}$, $-k_{\psi^c}$ and the irreducible representations $\rho_{\psi}$, $\rho_{\psi^c}$ of the finite modular group $\Gamma'_N$ (or $\Gamma_N$), i.e.,
\begin{eqnarray}
\nonumber&& \qquad\qquad ~~ \tau\rightarrow\gamma\tau=\frac{a\tau+b}{c\tau+d},~~~\gamma=\begin{pmatrix}
a ~&~ b \\
c ~&~ d
\end{pmatrix} \in SL(2, \mathbb{Z})\,,\\
&&\psi_i(x)\rightarrow (c\tau+d)^{-k_{\psi}}\left[\rho_{\psi}(\gamma)\right]_{ij} \psi_j(x),~~~\psi^c_i(x)\rightarrow (c\tau+d)^{-k_{\psi^c}}\left[\rho_{\psi^c}(\gamma)\right]_{ij} \psi^c_j(x)\,.
\end{eqnarray}
We are mainly interesting in the fermion mass terms which arise from the Yukawa interaction for Dirac fermions. The modular invariant Yukawa interaction can be written as
\begin{eqnarray}
\label{eq:LD-Y}\mathcal{L}^D_Y=Y^{(k_Y)}(\tau) \psi^c\psi H+\mathrm{h.c.}\,,
\end{eqnarray}
where we have adopted the two-component spinor notation for fermion fields, and $H$ refers to the Higgs field or its complex conjugate. The modular transformation of the Higgs field is
\begin{equation}
H(x)\rightarrow (c\tau+d)^{-k_{H}}\rho_{H}(\gamma)H(x)\,.
\end{equation}
Modular invariance requires $Y^{(k_Y)}(\tau)$ should be polyharmonic Maa{\ss} forms of weight $k_Y$ and level $N$, and it transforms in the representation $\rho_{Y}$ of $\Gamma_N$, i.e.,
\begin{equation}
Y^{(k_Y)}(\gamma\tau)=(c\tau+d)^{k_{Y}}\rho_{Y}(\gamma)Y^{(k_Y)}(\tau)\,.
\end{equation}
Each term in the Yukawa interactions $\mathcal{L}^D_Y$ has to be invariant under the finite modular group $\Gamma'_N$ (or $\Gamma_N$) and its total modular weight has to be vanishing. Hence $k_Y$ and $\rho_Y$ should satisfy the following conditions
\begin{eqnarray}
\nonumber && k_Y=k_{\psi^c}+k_{\psi}+k_H\,,\\
&&\rho_{Y}\otimes\rho_{\psi^c}\otimes\rho_{\psi}\otimes\rho_H
\ni\mathbf{1}\,,
\end{eqnarray}
where $\mathbf{1}$ refers to the trivial singlet of $\Gamma'_N$ (or $\Gamma_N$). For Majorana fermions denoted by $\psi^c$, the corresponding mass terms read as\footnote{The general form of the modular invariant Majorana neutrino mass term described by Weinberg operator is given by
\begin{eqnarray*}
\mathcal{L}^{M}_{\nu}=\dfrac{1}{2\Lambda} LLH^2 Y^{(k_Y)}(\tau)+\text{h.c.}\,.
\end{eqnarray*}}:
\begin{eqnarray}
\label{eq:LM-Y}\mathcal{L}^M=Y^{(k_Y)}(\tau) \psi^c\psi^c +\mathrm{h.c.}\,,
\end{eqnarray}
Similarly the conditions of modular invariance are
\begin{equation}
k_Y=2k_{\psi^c},~~~\rho_{Y}\otimes\rho_{\psi^c}\otimes\rho_{\psi^c}\ni\mathbf{1}\,.
\end{equation}
In comparison with the supersymmetric modular flavor symmetry, here the modular weight $k_Y$ of the polyharmonic Maa{\ss} form in Eqs.~(\ref{eq:LD-Y}, \ref{eq:LM-Y}) can be negative. Moreover, after the modulus $\tau$ acquires a VEV, the kinetic terms invariant under the modular symmetry read\footnote{The modular transformations of $-i\tau+i\bar{\tau}$ and $\partial_{\mu}\tau$ are
\begin{eqnarray*}
-i\tau+i\bar{\tau}\rightarrow \frac{-i\tau+i\bar{\tau}}{|c\tau+d|^2},~~~\partial_{\mu}\tau\rightarrow\frac{\partial_{\mu}\tau}{(c\tau+d)^2}\,.
\end{eqnarray*}
The kinetic term of the complex modulus $\tau$ is given by
\begin{eqnarray*}
\frac{1}{\langle-i\tau+i\bar{\tau}\rangle^{2}}\partial^{\mu}\bar{\tau}\partial_{\mu}\tau\,.
\end{eqnarray*}
Before the modulus $\tau$ acquires a VEV, the modular invariant kinetic terms for the fermion fields read as
\begin{eqnarray*}\label{eq:Lag-kinetic-inv}
\mathcal{L}^{\text{inv}}_K=\frac{1}{2}\left(-i\tau+i\bar{\tau}\right)^{-k_{\psi}}\,i\,\psi^{\dagger} \,\overline{\sigma}^{\mu}D_{\mu}\psi+\frac{1}{2}\left(-i\tau+i\bar{\tau}\right)^{-k_{\psi^c}}\,i\,\psi^{c\dagger} \, \overline{\sigma}^{\mu}D_{\mu}\psi^c+\text{h.c.}\,,
\end{eqnarray*}
where $D_{\mu}$ is the modular covariant derivative depending on the weight $k$,
\begin{eqnarray*}
D_\mu = \partial_\mu + \dfrac{k\pi i}{6} E_2(\tau) \partial_\mu \tau\,.
\end{eqnarray*}
Then one can see that $D_{\mu}\psi$ and $\psi$ transform in the same way under modular symmetry,
\begin{eqnarray*}
D_{\mu}\psi\rightarrow(c\tau+d)^{-k_{\psi}}\rho(\gamma)D_{\mu}\psi\,.
\end{eqnarray*}
After the modulus $\tau$ gets a VEV, the kinetic terms in Eq.~\eqref{eq:L-kinetic} can be produced.}:
\begin{eqnarray}
\label{eq:L-kinetic}\mathcal{L}_K=\langle-i\tau+i\bar{\tau}\rangle^{-k_{\psi}}\,i\,\psi^{\dagger} \, \overline{\sigma}^{\mu}\partial_{\mu}\psi+\langle-i\tau+i\bar{\tau}\rangle^{-k_{\psi^c}}\,i\,\psi^{c\dagger} \, \overline{\sigma}^{\mu}\partial_{\mu}\psi^c+\ldots\,,
\end{eqnarray}
where the dots stand for other terms compatible with modular invariance. It is known that the modular symmetry does not fix the form of the kinetic terms uniquely~\cite{Chen:2019ewa}, the full kinetic terms beyond the one given in Eq.~\eqref{eq:L-kinetic} can be written as
\begin{eqnarray}
\nonumber \Delta \mathcal{L}_K&=&\sum_{n,\bm{r_1}, \bm{r_2}} c_{n,\bm{r_1}, \bm{r_2}}\langle-i\tau+i\bar{\tau}\rangle^{n-k_{\psi}}\left(Y^{{(n)}\dagger}_{\bm{r_1}}Y^{(n)}_{\bm{r_2}}\,i\,\psi^{\dagger} \, \overline{\sigma}^{\mu}\partial_{\mu}\psi\right)_{\mathbf{1}}\\
\nonumber &&+ c'_{n,\bm{r_1}, \bm{r_2}}\langle-i\tau+i\bar{\tau}\rangle^{n-k_{\psi^c}}\left(Y^{{(n)}\dagger}_{\bm{r_1}}Y^{(n)}_{\bm{r_2}}\,i\,\psi^{c\dagger} \, \overline{\sigma}^{\mu}\partial_{\mu}\psi^c\right)_{\mathbf{1}}+\text{h.c.}\\
\nonumber&&+\sum_{n,\bm{r_1}, \bm{r_2}} d_{n,\bm{r_1}, \bm{r_2}}\langle-i\tau+i\bar{\tau}\rangle^{-k_{\psi}}\left(Y^{{(-n)}}_{\bm{r_1}}Y^{(n)}_{\bm{r_2}}\,i\,\psi^{\dagger} \, \overline{\sigma}^{\mu}\partial_{\mu}\psi\right)_{\mathbf{1}}\\
&&+ d'_{n,\bm{r_1}, \bm{r_2}}\langle-i\tau+i\bar{\tau}\rangle^{-k_{\psi^c}}\left(Y^{{(-n)}}_{\bm{r_1}}Y^{(n)}_{\bm{r_2}}\,i\,\psi^{c\dagger} \,\overline{\sigma}^{\mu}\partial_{\mu}\psi^c\right)_{\mathbf{1}}+\text{h.c.}\,,
\end{eqnarray}
where one should sum over all singlet contractions. Note that each term in above can multiply any power of the combination $(-i\tau+i\bar{\tau})|\eta(\tau)|^4$ which is modular invariant. These terms are on the same foot as the leading terms in Eq.~\eqref{eq:L-kinetic}, the presence of the couplings $c_{n,\bm{r_1}, \bm{r_2}}$, $c'_{n,\bm{r_1}, \bm{r_2}}$, $d_{n,\bm{r_1}, \bm{r_2}}$ and $d'_{n,\bm{r_1}, \bm{r_2}}$ reduces the predictive power of modular symmetry~\cite{Chen:2019ewa}. It is known that the kinetic terms could be well controlled in the paradigm of eclectic flavor group in which the modular flavor symmetry is extended by the traditional flavor symmetry~\cite{Nilles:2020nnc,Nilles:2020kgo}. In the present work, we shall focus on the minimal kinetic terms in Eq.~\eqref{eq:L-kinetic}. In order to match the canonical form of kinetic terms, the spinor multiplets $\psi$ and $\psi^c$ should be rescaled as \begin{equation}\label{eq:rescaling}
\psi\rightarrow \langle-i\tau+i\bar{\tau}\rangle^{k_{\psi}/2}\psi,~~~~ \psi^c\rightarrow \langle-i\tau+i\bar{\tau}\rangle^{k_{\psi^c}/2}\psi^c\,.
\end{equation}
This effect of rescaling can be absorbed into the unknown couplings of
the Yukawa interactions.

It has been established that the generalized CP symmetry (gCP) can be consistently incorporated in the modular symmetry, and the gCP transformation of the modulus $\tau$ is uniquely fixed to be~\cite{Acharya:1995ag,Dent:2001cc,Giedt:2002ns,Baur:2019kwi,Novichkov:2019sqv}
\begin{equation}
\tau \xrightarrow{\mathcal{CP}} -\tau^*\,,
\label{eq:tauCP}
\end{equation}
up to a modular transformation. For a field multiplet $\varphi$ in the representation $\rho_{\bm{r}}$ of $\Gamma_N$, the action of gCP symmetry is
\begin{equation}
\varphi\xrightarrow{\mathcal{CP}} X_{\bm{r}}\varphi^{*}\,,
\end{equation}
where the obvious action of CP on the spinor indices has been omitted for the case of $\varphi$ being a spinor. The gCP transformation $X_{\bm{r}}$ is a unitary matrix in flavor space and it has to satisfy the following consistency condition~\cite{Novichkov:2019sqv}:
\begin{equation}
\label{eq:consistency-cond}X_{\bm{r}}\rho^{*}_{\bm{r}}(S)X^{-1}_{\bm{r}}=\rho^{-1}_{\bm{r}}(S),~~~X_{\bm{r}}\rho^{*}_{\bm{r}}(T)X^{-1}_{\bm{r}}=\rho^{-1}_{\bm{r}}(T)\,,
\end{equation}
If both modular generators $S$ and $T$ are represented by unitary and symmetric matrices in all irreducible presentations, the conditions of Eq.~\eqref{eq:consistency-cond} are always solved by $X_{\bm{r}}=1$. Thus the gCP symmetry would reduce to traditional CP in such choice of basis, the representation bases of $\Gamma_2\cong S_3$, $\Gamma_3\cong A_4$, $\Gamma_4\cong S_4$ and $\Gamma_5\cong A_5$ in the Appendix~\ref{app:polyharmonic-Maass-form-multiplets} all enjoy this property.

Similar to the case of modular forms~\cite{Novichkov:2019sqv,Ding:2021iqp}, one can show that the action of the CP transformation on the multiplet $Y^{(k_Y)}_{\bm{r}}(\tau)$ of polyharmonic Maa{\ss} form is as follow
\begin{equation}
Y^{(k_Y)}_{\bm{r}}(\tau)\stackrel{\mathcal{CP}}{\longrightarrow} Y^{(k_Y)}_{\bm{r}}(-\tau^{*})=X_{\mathbf{r}}Y^{(k_Y)*}_{\bm{r}}(\tau)\,.
\end{equation}
Hence the gCP symmetry would enforce the coupling constants to be real in the $S$ and $T$ symmetric basis, if all the Clebsch-Gordan coefficients of $\Gamma'_N$ (or $\Gamma_N$) are real, as shown in the Appendix~\ref{app:polyharmonic-Maass-form-multiplets}. Thus the VEV of $\tau$ is the unique source breaking both modular symmetry and gCP symmetry.

\begin{table}[t!]
\centering
\begin{tabular}{|c|c|}
\hline  \hline

Weight $k_Y$ & Polyharmonic Maa{\ss} forms $Y^{(k_Y)}_{\mathbf{r}}$ \\ \hline

$k_Y=-4$ & $Y^{(-4)}_{\mathbf{1}}$,\; $Y^{(-4)}_{\mathbf{3}}$\\ \hline

$k_Y=-2$ & $Y^{(-2)}_{\mathbf{1}}$,\; $Y^{(-2)}_{\mathbf{3}}$\\ \hline

$k_Y=0$ & $Y^{(0)}_{\mathbf{1}}$,\; $Y^{(0)}_{\mathbf{3}}$\\ \hline

$k_Y=2$ & $Y^{(2)}_{\mathbf{1}}$,\; $Y^{(2)}_{\mathbf{3}}$\\ \hline

$k_Y=4$ & $Y^{(4)}_{\mathbf{1}}$,\; $Y^{(4)}_{\mathbf{1}'}$,\; $Y^{(4)}_{\mathbf{3}}$\\ \hline

$k_Y=6$ & $Y^{(6)}_{\mathbf{1}}$, \; $Y^{(6)}_{\mathbf{3}I}$,\; $Y^{(6)}_{\mathbf{3}II}$\\
\hline \hline
\end{tabular}
\caption{\label{tab:PMF-level3}Summary of polyharmonic Maa{\ss} forms of weight $k_Y=-4, -2, 0, 2, 4, 6$ at level $N=3$, the subscript $\bm{r}$ denote the transformation property under $A_4$ modular symmetry. Here $Y^{(6)}_{\mathbf{3}I}$ and $Y^{(6)}_{\mathbf{3}II}$ stand for two independent weight 6 modular forms transforming as triplet $\mathbf{3}$ of $A_4$. }
\end{table}

\section{\label{sec:benchmark-models}Benchmark models for lepton masses and mixing}

In the following, we apply the formalism outlined above to construct models of lepton masses and flavor mixing with the polyharmonic Maa{\ss} forms of level $N=3$. The corresponding finite modular group is $\Gamma_3\cong A_4$, the polyharmonic Maa{\ss} forms of level 3 are arranged into multiplets of $A_4$, as is summarized in table~\ref{tab:PMF-level3}, see Appendix~\ref{sec:app-N3} for detailed results. It is notable that the polyharmonic Maa{\ss} forms coincide with the known modular forms at level 3 for weight $k_Y\geq4$, and the polyharmonic Maa{\ss} forms of weight $k_Y\leq 2$ can always be arranged into a singlet $Y^{(k_Y)}_{\bm{1}}$ and a triplet $Y^{(k_Y)}_{\bm{3}}$of $A_4$. In the following, we shall present three benchmark models which differ in the neutrino mass generation mechanism. No other flavon besides the modulus $\tau$ is introduced in these models.

\subsection{\label{subsec:model-Weinberg}Neutrino masses from Weinberg operator}

The light neutrino masses are described by the effective Weinberg operator in this model. We assume that the three generations of left-handed lepton doublets $L$ transform as a triplet $\bm{3}$ under $A_4$, and the right-handed charged leptons $E^c_{1,2,3}$ are assigned to singlet representations of $A_4$. The modular weight and representation assignments of lepton fields are summarized as follow,
\begin{eqnarray}
\rho_{E_1^c}=\bm{1}\,,\,\rho_{E_2^c}=\bm{1}''\,,\,\rho_{E_3^c}=\bm{1}'\,,\,\rho_L=\bm{3}\,,~~~k_{E_1^c}=0\,,\,k_{E_2^c}=2\,,\,k_{E_3^c}=2\,,\,k_{L}=-2\,.
\end{eqnarray}
The Higgs field is invariant under $A_4$ with zero weight. Then we can read out the modular invariant Lagrangian for the charged lepton Yukawa interaction and the Weinberg operator,
\begin{eqnarray}
\nonumber -\mathcal{L}_e &=& \alpha (E^c_1 L Y^{(-2)}_{\bm{3}}H^*)_{\bm{1}}  + \beta (E^c_2 L Y^{(0)}_{\bm{3}}H^*)_{\bm{1}}  + \gamma (E^c_3 L Y^{(0)}_{\bm{3}}H^*)_{\bm{1}}  + \text{h.c.}\,, \\
\label{eq:Lag-Weinberg}-\mathcal{L}_\nu &=& \dfrac{1}{2\Lambda} (LLH^2 Y^{(-4)}_{\bm{3}})_{\bm{1}} + \dfrac{g}{2\Lambda} (LL H^2 Y^{(-4)}_{\bm{1}})_{\bm{1}} + \text{h.c.}\,,
\end{eqnarray}
where the phases of the parameters $\alpha$, $\beta$, $\gamma$ and $\Lambda$ can be absorbed by redefining the lepton fields and consequently they can be taken real without loss of generality, while the coupling $g$ is generally complex\footnote{If gCP symmetry is imposed, the parameter $g$ would be constrained to be real. However, the experimental data of lepton masses and mixing angles can not be accommodated.}. Then we can read out the charged lepton and neutrino mass matrices as follows:
\begin{eqnarray}
\nonumber M_e&=&\begin{pmatrix}
\alpha Y_{\bm{3},1}^{(-2)} ~& \alpha Y_{\bm{3},3}^{(-2)} ~& \alpha Y_{\bm{3},2}^{(-2)} \\
\beta Y_{\bm{3},2}^{(0)} ~& \beta Y_{\bm{3},1}^{(0)} ~& \beta Y_{\bm{3},3}^{(0)} \\
\gamma Y_{\bm{3},3}^{(0)} ~& \gamma Y_{\bm{3},2}^{(0)} ~& \gamma Y_{\bm{3},1}^{(0)}
\end{pmatrix}v\,,\\
M_{\nu}&=&\begin{pmatrix}
2 Y_{\bm{3},1}^{(-4)} + g Y_{\bm{1}}^{(-4)} ~& -Y_{\bm{3},3}^{(-4)} ~& -Y_{\bm{3},2}^{(-4)} \\
-Y_{\bm{3},3}^{(-4)} ~& 2Y_{\bm{3},2}^{(-4)} ~& -Y_{\bm{3},1}^{(-4)} + g Y_{\bm{1}}^{(-4)} \\
-Y_{\bm{3},2}^{(-4)} ~& -Y_{\bm{3},1}^{(-4)} + g Y_{\bm{1}}^{(-4)} ~& 2Y_{\bm{3},3}^{(-4)}
\end{pmatrix}\dfrac{v^2}{\Lambda}\,,
\end{eqnarray}
where $v=\langle H^{0}\rangle$ is the VEV of the Higgs field. The charged lepton mass matrix $M_e$ involves three real couplings $\alpha$, $\beta$, $\gamma$ which can be adjusted to reproduce the charged lepton masses. The neutrino mass matrix $M_{\nu}$ depends on the complex coupling $g$ and an overall scale factor $v^2/\Lambda$ besides the complex modulus $\tau$. As a measure of goodness of fit, we perform a conventional $\chi^2$ analysis to determine whether this model can accommodate the experimental data on lepton masses and mixing angles. The contribution of the Dirac CP phase to the $\chi^2$ function is dropped, since it is less constrained by the present data. We search for the minimum of the $\chi^2$ function built with the data in table~\ref{tab:lepton-data}. The best fit values of the input parameters and lepton flavor observables for NO mass spectrum are found to be
\begin{eqnarray}
\nonumber &&\langle\tau\rangle = 0.3691 + 0.9611 i \,,\,\beta/\alpha = 125.74\,,\,\gamma/\alpha = 1574.04\,,\\
\nonumber &&g = 0.7531 + 0.3859 i\,,\, \alpha v= 3.6203\,{\rm MeV}\,,\, \dfrac{v^2}{\Lambda}= 97.390\,{\rm meV}\,, \\
\nonumber &&\sin^2\theta_{12}=0.307\,,~~\sin^2\theta_{13}=0.02224\,,~~\sin^2\theta_{23}=0.454\,, \\
\nonumber &&\delta_{CP}=1.225\pi\,,~~\alpha_{21}=0.667\pi\,,~~\alpha_{31}=1.385\pi\,,\\
&& m_1=21.092\,{\rm meV}\,,~~m_2=22.781\,{\rm meV}\,,~~m_3=54.313\,{\rm meV}\,,
\end{eqnarray}
with $\chi^2_{\text{min}}= 7.0\times 10^{-5}$. These predictions are in excellent agreement with experimental data. The central values of the charged lepton masses are exactly reproduced. Then we can determine the effective Majorana neutrino mass in neutrinoless double beta ($0\nu\beta\beta$) decay as $m_{\beta\beta}\simeq11.484$ meV, where $m_{\beta\beta}\equiv|\sum^{3}_{i=1}U^2_{ei}m_i|$ and $U_{ei}$ are the elements of the neutrino mixing matrix. This predicted value of $m_{\beta\beta}$ is below the most stringent upper bound $m_{\beta\beta}<(36\sim156)$ meV at $90\%$ C.L from the KamLAND-Zen collaboration~\cite{KamLAND-Zen:2022tow}, yet it is within the reach of the future tonne-scale neutrinoless double beta decay experiments such as LEGEND-1000~\cite{LEGEND:2021bnm} and nEXO~\cite{nEXO:2021ujk} whose sensitivities are expected to reach $(9\sim21)$ meV and $(4.7\sim20.3)$ meV respectively for 10 years of livetime.

\begin{table}[t!]
\centering
\begin{tabular}{| c | c | c |} \hline \hline
Observable & Central value and $1\sigma$ error & $3\sigma$ range \\ \hline
$m_e/m_\mu $ & $0.004737 $ & $-$  \\
$m_\mu/m_\tau$ & $0.05882$ & $-$  \\
$m_e/{\rm MeV}$ & $0.469652$ & $-$ \\
$\Delta m_{21}^2 / 10^{-5}\text{eV}^2$ & $7.41^{+0.21}_{-0.20}$ & $6.81\rightarrow 8.03$   \\
$\Delta m_{31}^2 / 10^{-3}\text{eV}^2$(NO) & $2.505^{+0.024}_{-0.026}$ & $2.426\rightarrow 2.586$  \\
$\Delta m_{32}^2 / 10^{-3}\text{eV}^2$(IO) & $-2.487^{+0.027}_{-0.024}$ & $-2.566\rightarrow -2.407$     \\ \hline
$\delta_{CP}/\pi$(NO) & $1.289^{+0.217}_{-0.139}$ & $0.772\rightarrow 1.944$  \\
$\delta_{CP}/\pi$(IO) & $1.517^{+0.144}_{-0.133}$ & $1.083\rightarrow 1.900$  \\
$\sin^2\theta_{12}$ $(\text{NO} \;\&\; \text{IO})$ & $0.307^{+0.012}_{-0.011}$ & $0.275\rightarrow 0.344$   \\
$\sin^2\theta_{13}$(NO) & $0.02224^{+0.00056}_{-0.00057}$ & $0.02047\rightarrow 0.02397$  \\
$\sin^2\theta_{13}$(IO) & $0.02222^{+0.00069}_{-0.00057}$ & $0.02049\rightarrow 0.02420$   \\
$\sin^2\theta_{23}$(NO) & $0.454^{+0.019}_{-0.016}$ & $0.411\rightarrow 0.606$  \\
$\sin^2\theta_{23}$(IO) & $0.568^{+0.016}_{-0.021}$ & $0.412\rightarrow 0.611$  \\ \hline \hline
\end{tabular}
\caption{\label{tab:lepton-data}The central values and the $1\sigma$ errors of the mass ratios and mixing angles and CP violation phases in lepton sector. The charged lepton mass ratios are taken from~\cite{Xing:2007fb} where the uncertainties are too small. We set the uncertainties of the charged lepton mass ratios to be $0.1\%$ of their central value in scanning the parameter space of our models. We adopt the values of the lepton mixing parameters from NuFIT v5.3 with Super-Kamiokanda atmospheric data for normal ordering (NO) and inverted ordering (IO)~\cite{Esteban:2020cvm}. }
\end{table}

\begin{figure}[hptb]
\centering
\includegraphics[width=0.99\textwidth]{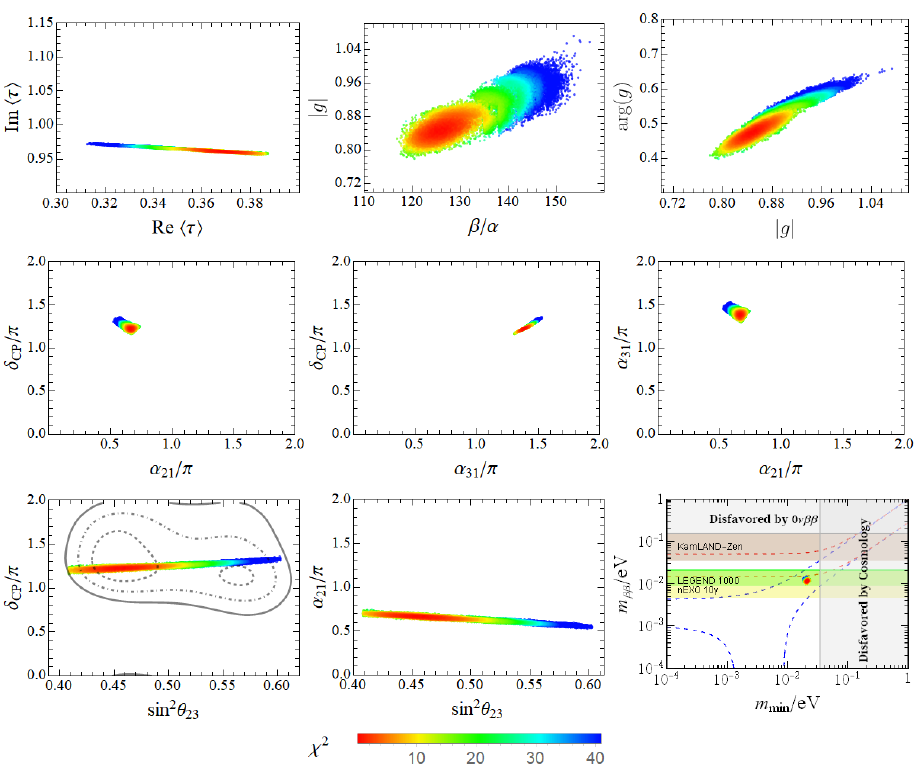}
\caption{\label{fig:model-Weinberg} The correlations among the input free parameters, neutrino mixing angles and CP violation phases for the model in section~\ref{subsec:model-Weinberg}, where the neutrino masses are described by Weinberg operator. In the plane $\delta_{CP}$ versus $\sin^2\theta_{23}$, the dashed, dash-dotted and solid contours
stand for the $1\sigma$, $2\sigma$ and $3\sigma$ allowed regions respectively~\cite{Esteban:2020cvm}. The last panel is the for the effective Majorana mass $m_{\beta\beta}$ with respect to the lightest neutrino mass $m_{\text{lightest}}$. The blue (red) dashed lines represent the most general allowed regions for normal ordering (inverted ordering) neutrino mass spectrum, where the neutrino oscillation parameter are varied within their $3\sigma$ ranges. Moreover, the vertical grey exclusion band denotes the most stringent upper bound on neutrino mass $\sum_i m_i<0.12$ eV from Planck~\cite{Planck:2018vyg}. The current experimental bound $m_{\beta\beta}<(36\sim156)\,$meV from KamLAND-Zen~\cite{KamLAND-Zen:2022tow} and the future sensitivity ranges $m_{\beta\beta}<(9\sim21)\,$ meV from LEGEND-1000~\cite{LEGEND:2021bnm} and $m_{\beta\beta}<(4.7\sim20.3)\,$ meV from nEXO~\cite{nEXO:2021ujk} are showed in light brown, green and light yellow horizontal bands respectively. }
\end{figure}

Furthermore, we perform a comprehensive numerical scan over the parameter space of the model. The complex modulus $\tau$ is taken as a random complex number in the fundamental domain $\mathcal{D}=\left\{\tau\in\mathbb{C}|\text{Im}\tau>0, |\text{Re}\tau|\leq 1/2, |\tau|\geq1\right\}$. Since the contribution of $\delta_{CP}$ to the $\chi^2$ function is not included, in the following we only show the results for $\text{Re}\tau>0$. The signs of the CP violation phases would be reversed while the lepton masses and mixing angles are unchanged, if the coupling constants are complex conjugated for $\text{Re}\tau<0$. We limit the absolute value of the coupling constants in the range $[0, 10^4]$ and the phase of $g$ freely varies in the region $[0, 2\pi]$. We require the mass ratios $m_e/m_{\mu}$, $m_{\mu}/m_{\tau}$, $\Delta m^2_{21}/\Delta m^2_{31}$ and the three lepton mixing angels $\theta_{12}$, $\theta_{13}$, $\theta_{23}$ are in the experimentally favored $3\sigma$ regions. The overall scales of the charged lepton and neutrino mass matrices are fixed by the electron mass and solar neutrino mass squared difference $\Delta m^2_{21}$ respectively. The correlations among the input parameters and flavor mixing obervables are plotted in figure~\ref{fig:model-Weinberg}. We see that the atmospheric mixing angle $\theta_{23}$ and the CP violation phase $\delta_{\text{CP}}$ are strongly correlated, and $\delta_{\text{CP}}$ is in the range of $[1.166\pi, 1.358\pi]$. These predictions for $\delta_{\text{CP}}$ and $\theta_{23}$ could be tested in future long baseline neutrino experiments DUNE~\cite{DUNE:2020ypp} and Hyper-Kamiokande (HK)~\cite{Hyper-Kamiokande:2018ofw}, which are capable of measuring $\theta_{23}$ and $\delta_{\rm CP}$ with very good precision. In the last panel of figure~\ref{fig:model-Weinberg}, we show the effective Majorana neutrino mass $m_{\beta\beta}$ with respect to the lightest neutrino mass. It is found the effective mass $m_{\beta\beta}$ lies in the region of $[9.94{\rm meV}, 14.08{\rm meV}]$ which is within the sensitivities of next generation $0\nu\beta\beta$ decay experiments.

This model can also accommodate inverted ordering neutrino masses. The corresponding best fit values of the free parameters and mixing parameters at the best fit point are determined to be
\begin{eqnarray}
\nonumber &&\langle\tau\rangle = 0.2429 + 1.1586 i \,,\,\beta/\alpha = 27.073\,,\,\gamma/\alpha = 1.0061 \times 10^{-2}\,,\\
\nonumber &&g = -0.5290 - 1.2049 i\,,\, \alpha v= 150.50\,{\rm MeV}\,,\, \dfrac{v^2}{\Lambda}= 65.693\,{\rm meV}\,, \\
\nonumber &&\sin^2\theta_{12}=0.307\,,~~\sin^2\theta_{13}=0.02222\,,~~\sin^2\theta_{23}=0.568\,, \\
\nonumber &&\delta_{CP}=1.453\pi\,,~~\alpha_{21}=1.532\pi\,,~~\alpha_{31}=1.695\pi\,, ~~ m_1=63.962\,{\rm meV}\,, \\
&&m_2=64.538\,{\rm meV}\,,~~m_3=40.963\,{\rm meV},~~ \chi^2_{\text{min}}= 2.5\times 10^{-4}\,.
\end{eqnarray}
The effective Majorana neutrino mass is $m_{\beta\beta}\simeq48.312$ meV. Analogously the predictions for $\theta_{23}$, $\delta_{CP}$ and $m_{\beta\beta}$ can be tested by future neutrino facilities.

\subsection{\label{subsec:model-seesaw}Neutrino masses from type-I seesaw mechanism }

\begin{figure}[t!]
\centering
\includegraphics[width=0.99\textwidth]{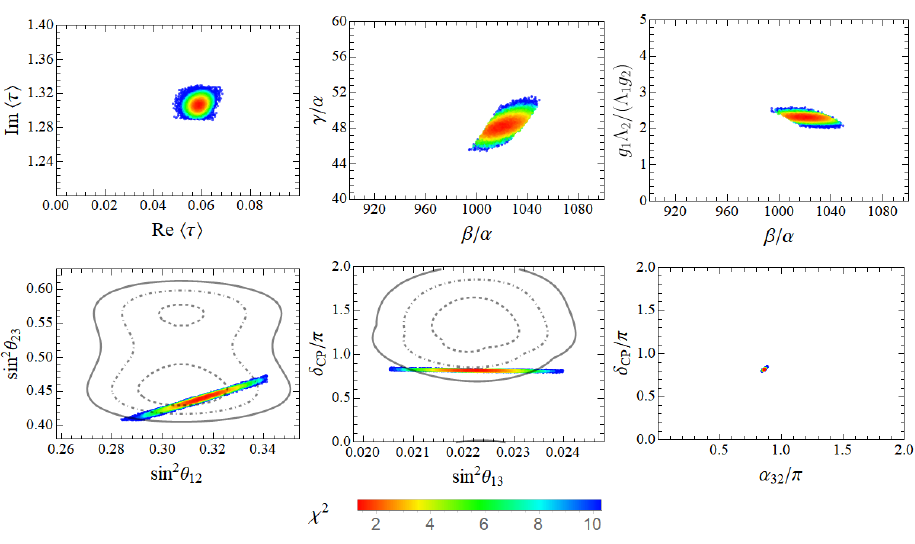}
\caption{\label{fig:model-seesaw} The correlations among the input free parameters, neutrino mixing angles and CP violation phases for the model in section~\ref{subsec:model-seesaw}, where neutrino masses are generated by type I seesaw mechanism. Here we adopt the same convention as
figure~\ref{fig:model-Weinberg}. }
\end{figure}

In this model, the neutrino masses are generated by the type-I seesaw mechanism and only two right-handed neutrinos $N^c_{1,2}$ are introduced for simplicity. The left-handed lepton fields are assigned to be $A_4$ triplet, both right-handed charged leptons and right-handed neutrinos are $A_4$ singlets, i.e.,
\begin{eqnarray}
\nonumber && \rho_{E^c_1}=\bm{1}''\,,~~ \rho_{E^c_2}=\bm{1}''\,,~~ \rho_{E^c_3}=\bm{1} \,,~~ \rho_L=\bm{3} \,,~~ \rho_{N^c_1}=\bm{1}''\,,~~ \rho_{N^c_2}=\bm{1}\,,\\
&& k_{E^c_1}=1 \,,~~k_{E^c_2}=3 \,,~ k_{E^c_3}=9 \,, ~~ k_L=-5\,, ~~ k_{N^c_1}=1\,, ~~ k_{N^c_2}= 3\,.
\end{eqnarray}
The Higgs field is invariant singlet of $A_4$ and its modular weight is vanishing. The modular invariant Lagrangian for the lepton masses is given by
\begin{eqnarray}
\nonumber -\mathcal{L}_e &=& \alpha (E^c_1 L Y^{(-4)}_{\bm{3}}H^*)_{\bm{1}}  + \beta (E^c_2 L Y^{(-2)}_{\bm{3}}H^*)_{\bm{1}}  + \gamma (E^c_3 L Y^{(4)}_{\bm{3}}H^*)_{\bm{1}}  + \text{h.c.}\,, \\
-\mathcal{L}_\nu &=& g_1 (N^c_1 L H Y^{(-4)}_{\bm{3}})_{\bm{1}}   + g_2 (N^c_2 L Y^{(-2)}_{\bm{3}}H)_{\bm{1}}  + \Lambda_1 N^c_1 N^c_2 Y^{(4)}_{\bm{1}'} + \dfrac{1}{2} \Lambda_2 N^c_2 N^c_2 Y^{(6)}_{(\bm{1})} +\text{h.c.}~\,.
\end{eqnarray}
We include gCP symmetry in this model so that all the couplings are constrained to be real, as explained in section~\ref{sec:non-SUSY-modular-symmetry}. The charged lepton mass matrix, the Dirac neutrino mass matrix and the heavy Majorana neutrino mass matrix read as
\begin{eqnarray}
\nonumber M_e &=& \begin{pmatrix}
\alpha Y^{(-4)}_{\bm{3},2} ~& \alpha Y^{(-4)}_{\bm{3},1} ~& \alpha Y^{(-4)}_{\bm{3},3} \\
\beta Y^{(-2)}_{\bm{3},2} ~& \beta Y^{(-2)}_{\bm{3},1} ~& \beta Y^{(-2)}_{\bm{3},3} \\
\gamma Y^{(4)}_{\bm{3},1} ~& \gamma Y^{(4)}_{\bm{3},3} ~& \gamma Y^{(4)}_{\bm{3},2}
\end{pmatrix}v \,,  ~~
M_N =\begin{pmatrix}
0 ~& \Lambda_1 Y^{(4)}_{\bm{1}'} \\
\Lambda_1 Y^{(4)}_{\bm{1}'} ~& \Lambda_2 Y^{(6)}_{\bm{1}}
\end{pmatrix}\,, \\
M_D &=& \begin{pmatrix}
g_1 Y^{(-4)}_{\bm{3},2} ~&~ g_1 Y^{(-4)}_{\bm{3},1} ~& g_1 Y^{(-4)}_{\bm{3},3}  \\
g_2 Y^{(-2)}_{\bm{3},1} ~&~ g_2 Y^{(-2)}_{\bm{3},3} ~& g_2 Y^{(-2)}_{\bm{3},2}
\end{pmatrix} v\,.
\end{eqnarray}
The light neutrino mass matrix is given by the seesaw formula
\begin{small}
\begin{eqnarray}
\nonumber M_\nu &=& - M_D^T M_N^{-1} M_D  \\
\nonumber &=& - \dfrac{g_1 g_2 v^2}{\Lambda_1}\dfrac{1}{Y^{(4)}_{\bm{1}'}}
\begin{pmatrix}
2 Y^{(-2)}_{\bm{3},1} Y^{(-4)}_{\bm{3},2} ~& Y^{(-2)}_{\bm{3},1} Y^{(-4)}_{\bm{3},1} + Y^{(-2)}_{\bm{3},3} Y^{(-4)}_{\bm{3},2} ~& Y^{(-2)}_{\bm{3},2} Y^{(-4)}_{\bm{3},2} + Y^{(-2)}_{\bm{3},1} Y^{(-4)}_{\bm{3},3}  \\
Y^{(-2)}_{\bm{3},1} Y^{(-4)}_{\bm{3},1} + Y^{(-2)}_{\bm{3},3} Y^{(-4)}_{\bm{3},2} ~& 2 Y^{(-2)}_{\bm{3},3} Y^{(-4)}_{\bm{3},1} ~& Y^{(-2)}_{\bm{3},2} Y^{(-4)}_{\bm{3},1} + Y^{(-2)}_{\bm{3},3} Y^{(-4)}_{\bm{3},3} \\
Y^{(-2)}_{\bm{3},2} Y^{(-4)}_{\bm{3},2} + Y^{(-2)}_{\bm{3},1} Y^{(-4)}_{\bm{3},3} ~& Y^{(-2)}_{\bm{3},2} Y^{(-4)}_{\bm{3},1} + Y^{(-2)}_{\bm{3},3} Y^{(-4)}_{\bm{3},3} ~& 2 Y^{(-2)}_{\bm{3},2} Y^{(-4)}_{\bm{3},3}
\end{pmatrix}   \\
&& + \dfrac{\Lambda_2 g_1^2 v^2}{\Lambda_1^2}\dfrac{Y^{(6)}_{\bm{1}}}{Y^{(4)}_{\bm{1}'}}
\begin{pmatrix}
(Y^{(-4)}_{\bm{3},2})^2 ~& Y^{(-4)}_{\bm{3},1} Y^{(-4)}_{\bm{3},2} ~& Y^{(-4)}_{\bm{3},2} Y^{(-4)}_{\bm{3},3} \\
Y^{(-4)}_{\bm{3},1} Y^{(-4)}_{\bm{3},2} ~& (Y^{(-4)}_{\bm{3},1})^2 ~& Y^{(-4)}_{\bm{3},1} Y^{(-4)}_{\bm{3},3} \\
Y^{(-4)}_{\bm{3},2} Y^{(-4)}_{\bm{3},3} ~& Y^{(-4)}_{\bm{3},1} Y^{(-4)}_{\bm{3},3} ~& (Y^{(-4)}_{\bm{3},3})^2
\end{pmatrix}\,.
\end{eqnarray}
\end{small}
Similar to the model of section~\ref{subsec:model-Weinberg}, the masses of electron, muon and tau are in a one-to-one correspondence with the parameters $\alpha$, $\beta$ and $\gamma$. We see that the light neutrino mass matrix $M_{\nu}$ depends on the combination $g_1\Lambda_2/(g_2\Lambda_1)$ and the overall mass scale $g_1 g_2 v^2 /\Lambda_1$ as well as the modulus $\tau$. This model describes all the lepton masses and mixing parameters in terms of seven real parameters including $\text{Re}\tau$ and $\text{Im}\tau$. The agreement between predictions and experimental data can be achieved for NO. The best fit values of the input parameters and the lepton flavor observables are determined to be
\begin{eqnarray}
\nonumber && \langle \tau \rangle = 0.0588 + 1.3067 i\,,~\beta/\alpha = 1019.5\,, ~\gamma/\alpha = 48.204\,,  \\
\nonumber && g_1\Lambda_2/ (g_2 \Lambda_1) = 2.3125 \,,~\alpha v = 1.9059\,{\rm MeV}\,, ~ g_1 g_2 v^2/\Lambda_1 = 24.779\,{\rm meV}\,, \\
\nonumber && \sin^2\theta_{12}=0.315 \,,~ \sin^2\theta_{13}=0.02225\,,~ \sin^2\theta_{23}=0.440\,, \\
\nonumber && \delta_{CP}=0.818 \pi\,,~ \alpha_{31}=1.138\pi\,, m_1 = 0\,{\rm meV}\,, \\
&& m_2 = 8.608\,{\rm meV}\,,~ m_2 = 49.941\,{\rm meV}\,,~ \chi^2_{\text{min}}=1.27\,.
\end{eqnarray}
The modulus $\tau$ is very close to the pure imaginary axis, its real part is very small. Since only two right-handed neutrinos are introduced, the lightest neutrino mass $m_1$ is zero and the Majorana phase $\alpha_{21}$ is unphysical. The effective Majorana neutrino mass is found to be $m_{\beta\beta}\simeq2.881$ meV which is too small to be detectable by next generation tonne-scale $0\nu\beta\beta$ experiments. The correlations among the free parameters and the flavor mixing parameters are shown in figure~\ref{fig:model-seesaw}. The atmospheric angle $\theta_{23}$ in the first octant is favored with $0.411<\sin^2\theta_{23}<0.474$ and the Dirac CP violation phase $\delta_{CP}$ is in the range of $[0.796\pi, 0.851\pi]$ which can be tested at future long baseline neutrino oscillation experiments DUNE~\cite{DUNE:2020ypp} and HK~\cite{Hyper-Kamiokande:2018ofw}. Moreover, the atmospheric angle $\theta_{23}$ are strongly correlated with the solar angle $\theta_{12}$ which lie in the interval $0.284<\sin^2\theta_{12}<0.341$. The reactor experiment JUNO has a unique sensitivity to the solar oscillation parameters $\sin^2\theta_{12}$ and $\Delta m^2_{21}$, and the error of $\sin^2\theta_{12}$ can be reduced to the world-leading level of $0.7\%$~\cite{JUNO:2015zny}. Hence JUNO can also test this model by precise measurement of the solar mixing parameters.

\subsection{\label{subsec:model-Dirac}Dirac neutrinos}

\begin{figure}[t!]
\centering
\includegraphics[width=0.99\textwidth]{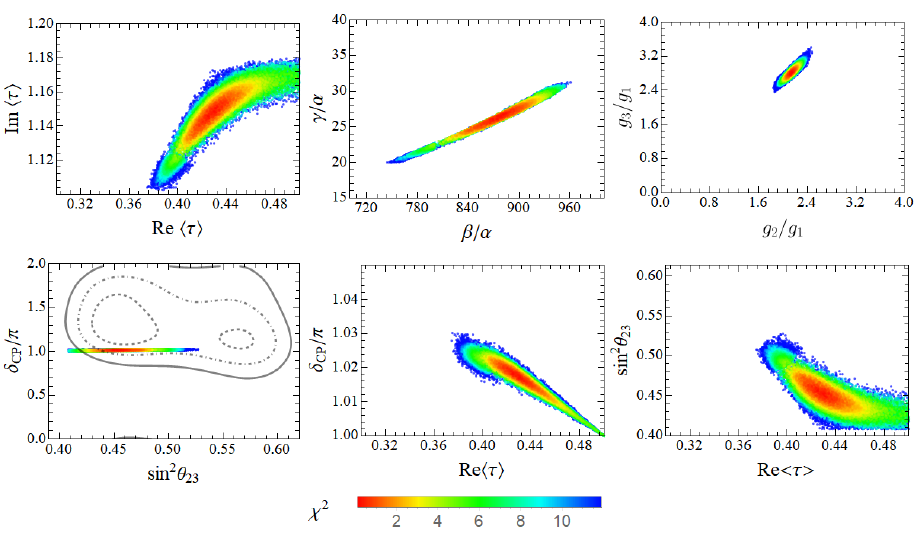}
\caption{\label{fig:model-Dirac} The correlations among the input free parameters, neutrino mixing angles and CP violation phases for the Dirac neutrino mass model in section~\ref{subsec:model-Dirac}. Here we adopt the same convention as figure~\ref{fig:model-Weinberg}.}
\end{figure}

Neutrinos are assumed to be Dirac fermions in this model. Usually the symmetry $U(1)_L$ or $U(1)_{B-L}$ is imposed to guarantee Diracness of neutrinos. Here it is remarkable that modular invariance can forbid the Majorana mass term of the right-handed neutrinos, consequently neutrinos are naturally Dirac particles without $U(1)_L$ or $U(1)_{B-L}$ symmetry. Similar to previous models, the Higgs field transforms as $\mathbf{1}$ under $A_4$ and its modular weight is zero, the left-handed lepton fields are embedded into a $A_4$ triplet while all right-handed lepton fields are $A_4$ singlets,
\begin{eqnarray}
\nonumber && \rho_{E^c_1} = \rho_{E^c_2} = \rho_{N^c_1} = \rho_{N^c_2} = \rho_{N^c_3} = \bm{1}''\,,~ \rho_{E^c_3} = \bm{1} \,,~ \rho_L = \bm{3}\,, \\
\nonumber && k_{E^c_1} = -4-k \,,~ k_{E^c_2} = -2-k \,,~ k_{E^c_3} = 4-k\,,~ k_{L}=k\,, \\
&& k_{N^c_1} = -2-k\,,~ k_{N^c_2} = -k\,, ~ k_{N^c_3} = 2-k\,.
\end{eqnarray}
In this case, we can take $k$ to be a generic integer with $k\geq -1$ so that the modular weights of $N^c_{1,2,3}$ are less than $4$. Modular invariance requires that the Majorana mass terms of right-handed neutrinos $N^c_{1,2,3}$ couple with polyharmonic Maa{\ss} form transforming as $\bm{1}''$ under $A_4$, however there is no such polyharmonic Maa{\ss} form for weight less than $8$, as can be seen from table~\ref{tab:PMF-level3}. Hence the Majorana neutrino mass terms of heavy neutrinos are forbidden by modular symmetry in this model. The Lagrangian for lepton Yukawa interactions invariant under the modular symmetry reads as
\begin{eqnarray}
\nonumber -\mathcal{L}_e &=& \alpha (E^c_1 L Y^{(-4)}_{\bm{3}}H^*)_{\bm{1}}  + \beta (E^c_2 L Y^{(-2)}_{\bm{3}}H^*)_{\bm{1}}  + \gamma (E^c_3 L Y^{(4)}_{\bm{3}}H^*)_{\bm{1}}  + \text{h.c.}\,,  \\
-\mathcal{L}_\nu &=& g_1 (N^c_1 LH Y^{(-2)}_{\bm{3}})_{\bm{1}}   + g_2 (N^c_2 LH Y^{(0)}_{\bm{3}})_{\bm{1}}  + g_3 (N^c_3 LH Y^{(2)}_{\bm{3}})_{\bm{1}} + \text{h.c.}\,,
\end{eqnarray}
where the phases of the couplings $\alpha$, $\beta$, $\gamma$, $g_1$, $g_2$ and $g_3$ can be absorbed into the right-handed lepton fields, and consequently they can be taken real no matter whether gCP symmetry is included or not. As a consequence, the VEV of $\tau$ is the unique source of CP violation. The charged lepton mass matrix and the Dirac neutrino mass matrix are given by
\begin{eqnarray}
M_e &=& \begin{pmatrix}
\alpha Y^{(-4)}_{\bm{3},2} ~& \alpha Y^{(-4)}_{\bm{3},1} ~& \alpha Y^{(-4)}_{\bm{3},3} \\
\beta Y^{(-2)}_{\bm{3},2} ~& \beta Y^{(-2)}_{\bm{3},1} ~& \beta Y^{(-2)}_{\bm{3},3} \\
\gamma Y^{(4)}_{\bm{3},1} ~& \gamma Y^{(4)}_{\bm{3},3} ~& \gamma Y^{(4)}_{\bm{3},2}
\end{pmatrix}v \,, ~~~
M_{\nu_D} = \begin{pmatrix}
g_1 Y^{(-2)}_{\bm{3},2} ~& g_1 Y^{(-2)}_{\bm{3},1} ~& g_1 Y^{(-2)}_{\bm{3},3} \\
g_2 Y^{(0)}_{\bm{3},2} ~& g_2 Y^{(0)}_{\bm{3},1} ~& g_2 Y^{(0)}_{\bm{3},3} \\
g_3 Y^{(2)}_{\bm{3},2} ~& g_3 Y^{(2)}_{\bm{3},1} ~& g_3 Y^{(2)}_{\bm{3},3}
\end{pmatrix}v \,.
\end{eqnarray}
We find that this model can be in excellent agreement with experimental data for NO neutrino mass spectrum, and the best fit values of the free parameters are
\begin{eqnarray}
\nonumber && \langle \tau \rangle = 0.4277 + 1.1477 i \,,~ \beta/\alpha = 876.29\,, ~ \gamma/\alpha = 26.390 \,, \\
&& g_2/g_1 = 2.1371\,, ~ g_3/g_1 = 2.8154\,,~ \alpha v = 3.0021\,{\rm MeV}\,,~ g_1 v = 15.277\,{\rm meV} \,.
\end{eqnarray}
The neutrino masses and mixing parameters at the above best fit point are determined to be
\begin{eqnarray}
\nonumber && \sin^2 \theta_{12} = 0.307\,,~ \sin^2 \theta_{13} = 0.02224\,,~ \sin^2 \theta_{23} = 0.454\,,  \delta_{CP} = 1.018 \pi \\
&&m_1 = 5.275\,{\rm meV} \,,~ m_2 = 10.096\,{\rm meV}\,,~ m_3 = 50.327\,{\rm meV}\,,~\chi^2_{\text{min}} = 1.7\times 10^{-4}\,.~~
\end{eqnarray}
We see that the Dirac CP phase approximately takes the CP conserved value $\delta_{CP}\approx\pi$, this is because the VEV of $\tau$ is close to the right vertical boundary $\text{Re}\tau=1/2$ which preserve the CP symmetry $T\circ \mathcal{CP}$~\cite{Novichkov:2019sqv,Ding:2023htn}. Furthermore, we display the correlations between the free parameters and lepton mixing parameters in figure~\ref{fig:model-Dirac}. It is notable that atmospheric angle and Dirac CP phase are limited in the narrow regions $0.411\leq\sin^2\theta_{23}\leq0.527$ and $\pi\leq\delta_{CP}\leq1.030\pi$ respectively, this prediction is compatible with the latest measurements of NOvA~\cite{NOvA:2023iam}. If the signal of CP violation would be detected by DUNE and HK in future, our construction will be ruled out.

\section{\label{sec:summary-discussions}Summary and discussions }

Modular flavor symmetry is an elegant and promising approach to understand the observed patterns of masses and mixing in both quark and lepton sectors.
In the limit of exact supersymmetry, modular invariance determines the Yukawa couplings to be level $N$ modular forms which are holomorphic functions of the complex modulus $\tau$. Usually supersymmetry is assumed to ensure that only the holomorphic modular forms are involved in the Yukawa couplings, and thus modular flavor symmetry becomes a predictive framework addressing the flavor structure of SM. However, the evidence of low energy supersymmetry has not been experimentally observed so far.

It was found that the coefficients of the effective interactions describing the four supergraviton scattering amplitude are non-holomorphic automorphic functions which satisfy Laplace eigenvalue equations~\cite{Green:1997tv,Green:1997me,Pioline:1998mn,Green:1998by,deHaro:2002vk,Green:2010wi,Basu:2011he,Peeters:2000qj,Sinha:2002zr}.
Motivated by this insight from top-down and the bottom-up approach based on automorphic forms~\cite{Ding:2020zxw}, we have investigated the non-holomorphic modular flavor symmetry in the framework of harmonic Maa{\ss} forms. The growth condition of Eq.~\eqref{eq:growth-cond} is imposed, the resulting harmonic Maa{\ss} forms are the so-called polyharmonic Maa{\ss} forms. The linearly independent polyharmonic Maa{\ss} forms at each given weight and level span a linear space of finite dimension, so that only a finite number of terms in the Yukawa couplings would be allowed and the predictive power of modular invariance is not lost. In comparison with the original modular flavor symmetry, the Yukawa couplings are the polyharmonic Maa{\ss} forms and  the assumption of holomorphicity is replaced by the Laplacian condition in Eq.~\eqref{eq:autform-def}, while the modularity is kept.

In this paradigm, the finite modular group $\Gamma'_N=\Gamma/\Gamma(N)$ or $\Gamma_N=\Gamma/\pm\Gamma(N)$ still plays the role of flavor symmetry. The transformation of the matter field under the modular symmetry is specified by their modular weight and representation assignment of the adopted finite modular group. It is notable that the polyharmonic Maa{\ss} forms of integer weight $k$ and level $N$ can be decomposed into irreducible multiplets of $\Gamma'_N$, and the even weight polyharmonic Maa{\ss} forms can be organized into multiplets of $\Gamma_N$. The polyharmonic Maa{\ss} forms can be lifted to the modular forms through the action of $D$ operator and $\xi$ operator in Eq.~\eqref{eq:lift-cond}. Thus one can determine the expressions of the polyharmonic Maa{\ss} forms from the already known modular forms, as shown in Eq.~\eqref{eq:polyHarmonic-expr-main}. Although there is no nontrivial modular form of negative weights, the modular weight of polyharmonic Maa{\ss} forms can be negative, and the number of the linearly independent polyharmonic Maa{\ss} forms at weight $k<0$ and level $N$ doesn't change with $k$. The polyharmonic Maa{\ss} forms of negative weights might provide an interesting opportunity for fermion mass models with modular symmetry. In Appendix~\ref{app:polyharmonic-Maass-form-multiplets}, we provide the multiplets of polyharmonic Maa{\ss} forms of weights $k=-4, -2, 0, 2, 4, 6$ at levels $N=2, 3, 4, 5$ together with the irreducible representations and the Clebsch-Gordan coefficients of the inhomogeneous finite modular groups $\Gamma_2\cong S_3$, $\Gamma_3\cong A_4$, $\Gamma_4\cong S_4$, $\Gamma_5\cong A_5$. These information should be useful when constructing concrete flavor models with polyharmonic Maa{\ss} forms.

Moreover, we show that the non-holomorphic modular flavor symmetry can be consistently combined with the gCP symmetry, the admissible CP transformation of the modulus $\tau$ is still $\tau \xrightarrow{\mathcal{CP}} -\tau^*$ up to modular transformations. The consistency condition between the modular symmetry and gCP symmetry is given in Eq.~\eqref{eq:consistency-cond}. In the basis where both modular generators $S$ and $T$ are represented by unitary and symmetric matrices given in the Appendix~\ref{app:polyharmonic-Maass-form-multiplets}, the gCP transformation would take the canonical form $X_{\bm{r}}=1$, and gCP invariance would require all the coupling constants to be real.

In order to illustrate how the formalism of non-holomorphic modular flavor symmetry can be applied to the flavor problem, we present three example models for leptons based on the $A_4$ modular symmetry. The three models don't introduce any other flavon except $\tau$, and they are different in the neutrino mass generation mechanism. In the model of section~\ref{subsec:model-Weinberg}, the light neutrino masses are generated by the Weinberg operator and the lepton sector depends on 8 real parameters, and the neutrino mass spectrum can be either NO or IO. The model of section~\ref{subsec:model-seesaw} depends on 7 real parameters, the neutrino masses are generated by the type-I seesaw mechanism with two right-handed neutrinos, and the gCP symmetry is included for a higher predictive power. The data of lepton masses and mixing angles can be well accommodated for NO spectrum. Neutrinos are Dirac particles in the model of section~\ref{subsec:model-Dirac} which involves 8 real parameters, the Majorana mass terms of right-handed neutrinos are forbidden by the modular symmetry. One can test these example models by confronting the predicited values of $\theta_{23}$ and $\delta_{CP}$ with measurements of future long baseline experiments DUNE and HK. Moreover, the predictions for the effective mass $m_{\beta\beta}$ can be tested at the next generation tonne-scale $0\nu\beta\beta$ decay experiments such as LEGEND-1000 and nEXO. In the present work, we have ignored the effects of the renormalization group (RG) running between the modular symmetry scale, usually assumed to be large, and the electroweak scale. It is known that the RG effects are rather small for hierarchical neutrino mass spectrum~\cite{Chankowski:2001mx,Antusch:2005gp,Xing:2020ijf}. Therefore we expect that the RG corrections could be negligible for the NO case in these benchmark models.

From a more theoretical point of view, the original modular flavor symmetry requires supersymmetry to naturally ensure that only the holomorphic modular forms enter into the Yukawa couplings. The assumption of holomorphicity is superseded by the Laplacian condition of Eq.~\eqref{eq:autform-def} in the present framework of non-holomorphic modular flavor symmetry. However, the fundamental origin of this Laplacian condition and the possible related symmetry is obscure, as far as we know. It has been argued that the low-energy supersymmetry may be unnecessary for the Yukawa couplings to be modular forms or metaplectic forms~\cite{Cremades:2004wa,Almumin:2021fbk}. It is interesting to explore whether the top-down approach can give rise to Yukawa couplings as polyharmonic Maa{\ss} forms in the absence of supersymmetry.

In summary, the formalism of the non-holomorphic modular flavor symmetry provides a new possibility for understanding the flavor structure of SM, and it is worthy of further in-depth study from both bottom-up and top-down.

%%%%%%%%%%%%%%%%%%%%%%%%%%%%%%%%%%%%%%%%%%%%%
\section*{Acknowledgements}

We are grateful to professor Ferruccio Feruglio for sharing his note on non-holomorphic Yukawa couplings and insightful comments and questions. GJD would like to acknowledge all the participants of the workshop ``Modular Invariance Approach to the Lepton and Quark Flavour Problems: from Bottom-up to Top-down'' for stimulating discussions and helpful suggestions. GJD is grateful to the Mainz Institute for Theoretical Physics (MITP) of the Cluster of Excellence PRISMA+ (Project ID 390831469), for its hospitality and its partial support during the completion of this work. This work is supported by the National Natural Science Foundation of China under Grant No.~12375104.

\section*{Appendix}

\setcounter{equation}{0}
\renewcommand{\theequation}{\thesection.\arabic{equation}}

\begin{appendix}

\section{\label{app:Fourier-expansion} Fourier expansion of polyharmonic Maa{\ss} forms }

From the condition in Eq.~\eqref{eq:modularity-def} for $\gamma=T^{N}$, it follows that
\begin{eqnarray}
Y(\tau+N)=Y(\tau)\,.
\end{eqnarray}
Hence the Fourier expansion of the polyharmonic Maa{\ss} form $Y(\tau)$ has the following form:
\begin{eqnarray}
Y(\tau)=\sum_{n\in \frac{1}{N}\mathbb{Z}}a_n(y)q^n,~~~q\equiv e^{2\pi i\tau}\,.
\end{eqnarray}
The second condition in Eq.~\eqref{eq:modularity-def} requires the coefficients $a_n(y)$ fulfill the following identity,
\begin{eqnarray}
\dfrac{d^2a_n(y)}{dy^2}=\left( 4\pi n -\dfrac{k}{y} \right)\dfrac{da_n(y)}{dy}\,,
\end{eqnarray}
from which $a_n(y)$ is determined to be
\begin{eqnarray}
\nonumber a_n(y)&=&c^+(n)+c^-(n)\Gamma(1-k,-4\pi n y),~~\text{for}~~n\neq0\,,\\
a_0(y)&=&\left\{\begin{array}{cc}
c^+(0)+c^-(0)y^{1-k}\,,  ~~&~~ k\neq1 \\ [0.1in]
c^+(0) + c^-(0)\ln y\,, ~~&~~ k=1
\end{array}
\right.\,,
\end{eqnarray}
where $c^+(n)$, $c^-(n)$, $c^+(0)$ and $c^-(0)$ are constants, and the incomplete gamma function $\Gamma(1-k,-4\pi n y)$ is defined in Eq.~\eqref{eq:incomplete-gamma}. From the recursion relation of Eq.~\eqref{eq:Gamma-recursion}, one can obtain the following expressions of the incomplete gamma functions,
\begin{eqnarray}
\nonumber \Gamma(1, z)&=&e^{-z} \\
\nonumber\Gamma(2,z)&=&(z+1)\,e^{-z} \\
\nonumber\Gamma(3,z)&=&(z^2+2z+2)\,e^{-z} \\
\nonumber\Gamma(4,z)&=&(z^3+3z^2+6z+6)\,e^{-z} \\
\label{eq:Gamma-1-x}\Gamma(5,z)&=&(z^4+4z^3+12z^2+24z+24)\,e^{-z}\,.
\end{eqnarray}
Hence the Fourier expansion of $Y(\tau)$ is~\cite{book:Ono}
\begin{eqnarray}
\label{eq:Fourier-expansion-weak-masss2}Y(\tau)=\sum_{\substack{n\in\frac{1}{N}\mathbb{Z} }} c^+(n)q^n + c^-(0)y^{1-k} + \sum_{\substack{n\in\frac{1}{N}\mathbb{Z}\backslash\{0\} }} c^-(n)\Gamma(1-k,-4\pi n y)q^n\,,
\end{eqnarray}
where the term $y^{1-k}$ should be $\ln y$ for $k=1$. Moreover, for a polyharmonic Maa{\ss} form, there exists a constant $\alpha$ such that $Y(\tau)=\mathcal{O}(y^\alpha)$ as $y\rightarrow \infty$. Therefore we have $c^+(n)=c^-(-n)=0~ \text{for}~ n< 0$. Hence the Fourier expansion of a polyharmonic Maa{\ss} form of level $N$ is of the form:
\begin{eqnarray}
\label{eq:poly-maass-form2}Y(\tau)=\sum_{\substack{n\in\frac{1}{N}\mathbb{Z} \\ n\geq0}} c^+(n)q^n + c^-(0)y^{1-k}+ \sum_{\substack{n\in\frac{1}{N}\mathbb{Z} \\ n<0}} c^-(n)\Gamma(1-k,-4\pi n y)q^n \,.
\end{eqnarray}

\section{\label{app:differential-operators}Differential operators: $D$ operator and $\xi$ operator }

The well-known $D$ differential operator is defined as~\cite{book:Ono,bruinier2004two}
\begin{eqnarray}
\label{eq:D-operator}D=\dfrac{1}{2\pi i}\dfrac{\partial}{\partial \tau}\,.
\end{eqnarray}
It is straightforward to determine the actions of the operator $D^{1-k}$ on $q^n$, $y^{1-k}$ and $\Gamma(1-k,4\pi n y)q^{-n}$ as follows
\begin{eqnarray}
\nonumber&&D^{1-k}(q^n)=n^{1-k}q^n \\
\nonumber&&D^{1-k}(y^{1-k})=(-4\pi)^{k-1}(1-k)! \\
\label{eq:D-act}&&D^{1-k}(\Gamma(1-k,4\pi n y)q^{-n})=0
\end{eqnarray}
for any integer $k\leq 0$. Acting on a polyharmonic Maa{\ss} form $Y(\tau)$ of weight $k$ and level $N$ with the operator $D^{1-k}$, then we have
\begin{eqnarray}\label{eq:Bol-F}
D^{1-k} Y(\tau)=(-4\pi)^{k-1}(1-k)! \, c^-(0)+\sum_{\substack{n\in\frac{1}{N}\mathbb{Z} \\ n>0}} n^{1-k}c^+(n) q^n\,,
\end{eqnarray}
which is a holomorphic function of $\tau$. Moreover, $D^{1-k}Y$ satisfies the modularity condition~\cite{book:Ono},
\begin{eqnarray}
\label{eq:D-1-k-mod}(D^{1-k}Y)(\gamma\tau)=(c\tau + d)^{2-k}D^{1-k}Y(\tau),~~~\forall\gamma=\begin{pmatrix}
a ~&~ b \\
c ~&~ d
\end{pmatrix}\in\Gamma(N)\,.
\end{eqnarray}
From Eq.~\eqref{eq:Bol-F} and Eq.~\eqref{eq:D-1-k-mod}, we conclude that $D^{1-k} Y(\tau)$ is a weight $2-k$ modular form of level $N$.

The $\xi$ operator is another differential operator and it is defined as~\cite{book:Ono}
\begin{eqnarray}
\xi_k=2iy^k\overline{\partialtaubar{}} \,.
\end{eqnarray}
We denote the complex conjugation with a bar or asterisk in this work. From this definition of the $\xi$ operator, we can obtain the following identities:
\begin{eqnarray}
\nonumber&&\xi_k(q^n)=0\,, \\
\nonumber&&\xi_k(y^{1-k})=1-k \,, \\
\nonumber&&\xi_k(\Gamma(1-k,4\pi n y))=-(4\pi n)^{1-k}\,e^{-4\pi n y} \,, \\
\label{eq:xi-act}&&\xi_k(\Gamma(1-k,4\pi n y)q^{-n})=-(4\pi n)^{1-k}\,q^{n}\,.
\end{eqnarray}
Hence the action of $\xi_k$ on the weight $k$ and level $N$ polyharmonic Maa{\ss} form $Y(\tau)$ gives
\begin{eqnarray}\label{eq:xi-Y}
\xi_k Y(\tau)=(1-k)\,\overline{c^-(0)}-\sum_{\substack{n\in\frac{1}{N}\mathbb{Z} \\ n>0}} (4\pi n)^{1-k}\,\overline{c^-(-n)}\,q^n\,,
\end{eqnarray}
which implies $\xi_2(\widehat{E}_2)=\dfrac{3}{\pi}$. Under a modular transformation, one can check that $(\xi_k Y)(\tau)$ transforms as
\begin{equation}
\label{eq:xi-modularity}(\xi_k Y)(\gamma\tau)=(c\tau +d)^{2-k}(\xi_k Y)(\tau),~~~\forall\gamma=\begin{pmatrix}
a ~&~ b \\
c ~&~ d
\end{pmatrix}\in\Gamma(N)\,.
\end{equation}
From Eqs.~(\ref{eq:xi-Y},\ref{eq:xi-modularity}), we see that $\xi_k Y(\tau)$ is a level $N$ modular form of weight $2-k$. Since there is no level $N$ modular forms of negative weights, the weight $k$ of the polyharmonic Maa{\ss} form should be less than or equal to 2 (i.e. $k\leq2$) otherwise they would coincide with level $N$ modular forms of weight $k$.

\section{\label{app:polyharmonic-Maass-form-multiplets}Polyharmonic Maa{\ss} form multiplets of level $N\leq5$ }

In the following, we shall present the explicit forms of the Fourier
expansion of the polyharmonic Maa{\ss} form multiplets of weight $k=-4, -2, 0, 2, 4, 6$ up to level 5. These expressions in \texttt{Mathematica} form can be downloaded from the link~\cite{Qu:2024supv2}. The irreducible representations and the Clebsch-Gordan coefficients of the inhomogeneous finite modular groups $\Gamma_2\cong S_3$, $\Gamma_3\cong A_4$, $\Gamma_4\cong S_4$, $\Gamma_5\cong A_5$ are also given for convenience of model building.

\subsection{$N=2$}

The polyharmonic Maa{\ss} forms of level $N=2$ can be arranged into irreducible multiplets of $\Gamma_2\cong S_3$. The $S_3$ group can be generated by two generators $S$ and $T$ satisfying the relations
\begin{equation}
S_3=\left\{S,T|S^2=T^2=(ST)^3=1\right\}\,.
\end{equation}
The $S_3$ group has a doublet representation $\mathbf{2}$ and two singlet representations $\mathbf{1}$, $\mathbf{1}'$, and the modular generators $S$ and $T$ are represented by
\begin{eqnarray}
\nonumber \mathbf{1}:~ & S=1,~&~T=1\,,\\
\nonumber \mathbf{1}':~ &S=-1,~&~T=-1\,,\\
 \mathbf{2}:~ &S=\dfrac{1}{2} \begin{pmatrix}
-1 ~&~ -\sqrt{3}  \\  -\sqrt{3} ~&~ 1
\end{pmatrix},~&~T=\begin{pmatrix}
1 ~&~ 0 \\ 0 ~&~ -1
\end{pmatrix}\,.
\end{eqnarray}
The product of two doublets is given by $\mathbf{2}\times \mathbf{2}=\mathbf{1}\oplus\mathbf{1}'\oplus\mathbf{2}$, and the contraction rules are given by
\begin{eqnarray}
\label{eq:S3-contraction}\begin{pmatrix}
\alpha_1\\
\alpha_2
\end{pmatrix}_{\mathbf{2}}\otimes\begin{pmatrix}
\beta_1\\
\beta_2
\end{pmatrix}_{\mathbf{2}}&=&\left(\alpha_1\beta_1+\alpha_2\beta_2\right)_{\mathbf{1}}\oplus\left(\alpha_1\beta_2-\alpha_2\beta_1\right)_{\mathbf{1}'}\oplus\begin{pmatrix}
\alpha_1\beta_1-\alpha_2\beta_2\\
-\alpha_1\beta_2-\alpha_2\beta_1
\end{pmatrix}_{\mathbf{2}} \,.
\end{eqnarray}
There are two linearly independent weight 2 modular forms at level 2~\cite{Kobayashi:2018vbk}:
\begin{eqnarray}
\nonumber Y_1(\tau)&=&\dfrac{2i}{\pi}\left[ \dfrac{\eta'(\tau/2)}{\eta(\pi/2)} + \dfrac{\eta'((\tau+1)/2)}{\eta((\tau+1)/2)}- \dfrac{8\eta'(2\tau)}{\eta(2\tau)} \right] \,, \\
Y_2(\tau)&=&\dfrac{2\sqrt{3}i}{\pi} \left[ \dfrac{\eta'(\tau/2)}{\eta(\tau/2)} - \dfrac{\eta'((\tau+1)/2)}{\eta((\tau+1)/2)} \right]\,,
\end{eqnarray}
where $\eta'(\tau)\equiv\dfrac{d\eta(\tau)}{d\tau}$ and $\eta(\tau)$ is the Dedekind eta-function defined as
\begin{equation}
\eta(\tau)=q^{1/24}\prod^{\infty}_{n=1}(1-q^n)\,.
\end{equation}
The weight $2$ modular forms of level 2 forms a doublet $\mathbf{2}$ of $S_3$~\cite{Kobayashi:2018vbk} and the $q$-expansion are given by
\begin{eqnarray}
Y_\mathbf{2}^{(2)}=\begin{pmatrix}
Y_1 \\ Y_2
\end{pmatrix}&=&
\begin{pmatrix}
1 + 24q + 24 q^2 + 96 q^3 + 24 q^4 + 144 q^5 + \cdots \\
8\sqrt{3}q^{1/2}( 1 + 4 q + 6 q^2 + 8 q^3 + 13 q^4 + 12 q^5 + \cdots )
\end{pmatrix}\,.
\end{eqnarray}
There are three polyharmonic Maa{\ss} forms of weight $2$ and level 2 including the modified Eisenstein series $\widehat{E}_2$ and the modular forms $Y_1$ and $Y_2$. $\widehat{E}_2$ is a singlet $\mathbf{1}$ of $S_3$ while the holomorphic $Y_1$ and $Y_2$ are a doublet $\mathbf{2}$ of $S_3$. Using the method of section~\ref{subsec:MF-PHMF}, one can obtain the explicit expressions of polyharmonic Maa{\ss} forms from the modular forms at level 2. The weight $k=0$ polyharmonic Maa{\ss} forms can be arranged into a doublet $Y^{(0)}_{\mathbf{2}}$ and a trivial singlet $Y^{(0)}_{\mathbf{1}}$ of $S_3$. Note that $Y^{(0)}_{\mathbf{1}}$ is a constant and without loss of generality we can take $Y^{(0)}_{\mathbf{1}}=1$. The master equation of Eq.~\eqref{eq:lift-cond} for the weight 0 polyharmonic Maa{\ss} form $Y^{(0)}_{\mathbf{2}}$ of level 2 is of the following
\begin{eqnarray}
\xi_0(Y^{(0)}_{\mathbf{2}})&=&  Y^{(2)}_{\mathbf{2}},~~~D(Y^{(0)}_{\mathbf{2}}) = -\dfrac{1}{4\pi} Y^{(2)}_{\mathbf{2}}\,,
\end{eqnarray}
where we have chosen the normalization factor $\alpha=1$, and thus $\beta=-\dfrac{1}{4\pi}$ given in Eq.~\eqref{eq:coefficients-values}. Using the general result of Eq.~\eqref{eq:polyHarmonic-expr-main}, we can obtain the $q$-expansion of $Y^{(0)}_{\mathbf{2}}=(Y_{\mathbf{2},1}^{(0)}, Y_{\mathbf{2},2}^{(0)})^{T}$ as
\begin{eqnarray}
\nonumber Y_{\mathbf{2},1}^{(0)}&=& y - \dfrac{6\,e^{-4\pi y}}{\pi q} - \dfrac{3\,e^{-8\pi y}}{\pi q^2} - \dfrac{8\,e^{-12\pi y}}{\pi q^3} - \dfrac{3\,e^{-16\pi y}}{2\pi q^4} + \cdots \\
\nonumber&& -\dfrac{4\log 2}{\pi} - \dfrac{6 q}{\pi} - \dfrac{3 q^2}{\pi} - \dfrac{8 q^3}{\pi} - \dfrac{3 q^4}{2\pi} - \dfrac{36 q^5}{5\pi} + \cdots \,,\\
\nonumber Y_{\mathbf{2},2}^{(0)}&=&-\dfrac{4\sqrt{3}q^{1/2}}{\pi}\left( \dfrac{e^{-2\pi y}}{q} + \dfrac{4\,e^{-6\pi y}}{3 q^2} + \dfrac{6\,e^{-10\pi y}}{5 q^3} + \dfrac{8\,e^{-14\pi y}}{7 q^4} + \dfrac{13\,e^{-18\pi y}}{9 q^5} + \cdots \right)  \\
&&-\dfrac{4\sqrt{3}\,q^{1/2}}{\pi}\left(1+\dfrac{4}{3}q+\dfrac{6}{4}q^2+\dfrac{8}{7}q^3+\dfrac{13}{9}q^4+\dfrac{12}{11}q^5+\cdots  \right)\,,
\end{eqnarray}
The exact expressions of $Y_{\mathbf{2}, 1}^{(0)}$ and $Y_{\mathbf{2}, 2}^{(0)}$ are
\begin{eqnarray}
\nonumber Y_{\mathbf{2},1}^{(0)}&=&y-\dfrac{4\log 2}{\pi} - \dfrac{4}{\pi}\sum_{n=1}^{+\infty}\left[ \dfrac{\sigma_1(2n)}{2n}\left( q^n + \dfrac{e^{-4\pi n y}}{q^n}\right) - \dfrac{\sigma_1(n)}{n}\left( q^{2n} + \dfrac{e^{-8\pi n y}}{q^{2n}} \right)  \right] \,, \\
Y_{\mathbf{2},2}^{(0)}&=&-\dfrac{4\sqrt{3}}{\pi}\sum_{n=0}^{+\infty} \dfrac{\sigma_1(2n+1)}{2n+1} \left[ q^{(2n+1)/2} + \dfrac{e^{-2\pi (2n+1) y}}{q^{(2n+1)/2}} \right]\,.
\end{eqnarray}
The weight $4$ modular forms of level 2 are $Y_{\mathbf{1}}^{(4)}$ and $Y_{\mathbf{2}}^{(4)}$ with
\begin{eqnarray}
\nonumber Y_{\mathbf{1}}^{(4)}&=& Y_1^2 + Y_2^2 \,, \\
Y_{\mathbf{2}}^{(4)}&=& \begin{pmatrix}
Y_1^2 - Y_2^2 \\
-2Y_1Y_2
\end{pmatrix}\,.
\end{eqnarray}
They can be lifted to weight $-2$ polyharmonic Maa{\ss} forms through  Eq.~\eqref{eq:lift-cond} as follow
\begin{eqnarray}
\nonumber \xi_{-2}(Y_{\mathbf{1}}^{(-2)}) &=& Y^{(4)}_{\mathbf{1}}\,,~~~~D^3(Y_{\mathbf{1}}^{(-2)})=-\dfrac{2}{(4\pi)^3} Y^{(4)}_{\mathbf{1}}\,, \\
\xi_{-2}(Y_{\mathbf{2}}^{(-2)}) &=& Y^{(4)}_{\mathbf{2}}\,,~~~~D^3(Y_{\mathbf{2}}^{(-2)})=-\dfrac{2}{(4\pi)^3} Y^{(4)}_{\mathbf{2}}\,.
\end{eqnarray}
Using the general result of Eq.~\eqref{eq:polyHarmonic-expr-main}, we find the weight $k=-2$ polyharmonic Maa{\ss} forms of level 2 are
\begin{eqnarray}
\nonumber Y_{\mathbf{1}}^{(-2)}&=& \dfrac{y^3}{3} - \dfrac{15\Gamma(3,4\pi y)}{4\pi^3 q} - \dfrac{135\Gamma(3,8\pi y)}{32\pi^3 q^2} - \dfrac{35\Gamma(3,12\pi y)}{9\pi^3 q^3} + \cdots \\
&&-\dfrac{\pi}{12}\dfrac{\zeta(3)}{\zeta(4)} - \dfrac{15 q}{2\pi^3} - \dfrac{135 q^2}{16\pi^3} - \dfrac{70 q^3}{q\pi^3} - \dfrac{1095 q^4}{128\pi^3} - \dfrac{189 q^5}{25\pi^3} - \dfrac{35 q^6}{4\pi^3} + \cdots \,, \\
\nonumber Y_{\mathbf{2},1}^{(-2)}&=& \dfrac{y^3}{3} + \dfrac{9\Gamma(3,4\pi y)}{4\pi^3 q} + \dfrac{57\Gamma(3,8\pi y)}{32\pi^3 q^2} + \dfrac{7\Gamma(3,12\pi y)}{3\pi^3 q^3} + \dfrac{441\Gamma(3,16\pi y)}{256\pi^3 q^4} + \cdots \\
&&+\dfrac{\pi\zeta(3)}{30\zeta(4)} + \dfrac{9 q}{2\pi^3} + \dfrac{57 q^2}{16\pi^3} + \dfrac{14 q^3}{3\pi^3} + \dfrac{441 q^4}{128\pi^3} + \dfrac{567 q^5}{125\pi^3} + \cdots \,, \\
\nonumber Y_{\mathbf{2},2}^{(-2)}&=&\dfrac{4\sqrt{3}q^{1/2}}{\pi^3}\left( \dfrac{\Gamma(3,2\pi y)}{2 q} + \dfrac{14\Gamma(3,6\pi y)}{27 q^2} + \dfrac{63 \Gamma(3,10\pi y)}{125 q^3} + \dfrac{172\Gamma(3,14\pi y)}{343 q^4} + \cdots \right) \\
&&+\dfrac{4\sqrt{3}q^{1/2}}{\pi^3}\left( 1 + \dfrac{28 q}{27} + \dfrac{126 q^2}{125} + \dfrac{344 q^3}{343} + \dfrac{757 q^4}{729} + \cdots \right)\,,
\end{eqnarray}
where $\zeta(s)$ is the Riemann zeta function of $s$. The weight $6$ modular forms of level 2 are $Y_{\mathbf{1}}^{(6)}$, $Y_{\mathbf{1}'}^{(6)}$ and $Y_{\mathbf{2}}^{(6)}$
\begin{eqnarray}
Y_{\mathbf{1}}^{(6)}&=& Y_1^3 - 3Y_1Y_2^2 \\
Y_{\mathbf{1}'}^{(6)}&=& Y_2^3 - 3Y_2Y_1^2 \\
Y_{\mathbf{2}}^{(6)}&=& \begin{pmatrix}
Y_1^3 + Y_1Y_2^2 \\
Y_2^3 + Y_2Y_1^2
\end{pmatrix}
\end{eqnarray}
We can see that $Y_{\mathbf{1}'}^{(6)}$ is a cusp form which can not be lifted as a polyharmonic Maa{\ss} form. The lifting equations are given by
\begin{eqnarray}
\xi_{-4}(Y_{\mathbf{1}}^{(-4)}) &=& Y^{(6)}_{\mathbf{1}}\,,~~~~D^5(Y_{\mathbf{1}}^{(-4)})=-\dfrac{24}{(4\pi)^5} Y^{(6)}_{\mathbf{1}} \\
\xi_{-4}(Y_{\mathbf{2}}^{(-2)}) &=& Y^{(6)}_{\mathbf{2}}\,,~~~~D^5(Y_{\mathbf{2}}^{(-4)})=-\dfrac{24}{(4\pi)^5} Y^{(6)}_{\mathbf{2}}
\end{eqnarray}
Using the general result of section~\ref{subsec:MF-PHMF},  the weight $k=-4$ polyharmonic Maa{\ss} forms are determined to be
\begin{eqnarray}
\nonumber Y_{\mathbf{1}}^{(-4)}&=& \dfrac{y^5}{5} + \dfrac{63\Gamma(5,4\pi y)}{128\pi^5 q} + \dfrac{2079\Gamma(5,8\pi y)}{4096\pi^5 q^2} + \dfrac{427\Gamma(5,12\pi y)}{864\pi^5 q^3} + \dfrac{66591\Gamma(5,16\pi y)}{131072 \pi^5 q^4} + \cdots   \\
\nonumber &&+\dfrac{\pi}{80}\dfrac{\zeta(5)}{\zeta(6)} + \dfrac{189 q}{16\pi^5} + \dfrac{6237 q^2}{512 \pi^5} + \dfrac{427 q^3}{36 \pi^5} + \dfrac{199773 q^4}{16384 \pi^5} + \cdots\,, \\
\nonumber Y_{\mathbf{2},1}^{(-4)}&=& \dfrac{y^5}{5} -  \dfrac{33\Gamma(5,4\pi y)}{128\pi^5 q} - \dfrac{993\Gamma(5,8\pi y)}{4096\pi^5 q^2} - \dfrac{671\Gamma(5,12\pi y)}{2592\pi^5 q^3} - \dfrac{31713 \Gamma(5,16\pi y)}{131072\pi^5 q^4} + \cdots \\
\nonumber &&-\dfrac{\pi}{168}\dfrac{\zeta(5)}{\zeta(6)} - \dfrac{99 q}{16\pi^5} - \dfrac{2979 q^2}{512\pi^5} - \dfrac{671 q^3}{108\pi^5} - \dfrac{95139 q^4}{16384 \pi^5} + \cdots \,, \\
\nonumber Y_{\mathbf{2},2}^{(-4)}&=& -\dfrac{\sqrt{3}q^{1/2}}{\pi^5}\left( \dfrac{\Gamma(5,2\pi y)}{4 q} + \dfrac{61\Gamma(5,6\pi y)}{243 q^2} + \dfrac{1563\Gamma(5,10\pi y)}{6250 q^3} + \dfrac{4202\Gamma(5,14\pi y)}{16807 q^4} + \cdots \right) \\
&&-\dfrac{6\sqrt{3}q^{1/2}}{\pi^5}\left( 1 + \dfrac{244 q}{243} + \dfrac{3126 q^2}{3125} + \dfrac{16808 q^3}{16807} + \dfrac{59293 q^4}{59049} + \dfrac{161052 q^5}{161051} + \cdots \right)\,.
\end{eqnarray}

\subsection{\label{sec:app-N3}$N=3$}

The inhomogeneous finite modular group $\Gamma_3$ is isomorphic to $A_4$, and it is the symmetry group of a regular tetrahedron. The group $\Gamma_3\cong A_4$ can be generated by $S$ and $T$ satisfying
\begin{equation}
A_4=\left\{S,T|S^2=T^3=(ST)^3=1\right\}\,.
\end{equation}
The $A_4$ group has three singlet representations $\mathbf{1}$, $\mathbf{1}'$, $\mathbf{1}''$  and a triplet representation $\mathbf{3}$. The generators $S$ and $T$ in irreducible representations of $A_4$ are represented by
\begin{eqnarray}
\nonumber \mathbf{1}:~ & S=1,~&~T=1\,,\\
\nonumber \mathbf{1}':~ &S=1,~&~T=\omega\,,\\
\nonumber \mathbf{1}'':~ &S=1,~&~T=\omega^2\,,\\
\label{eq:A4-irrep} \mathbf{3}:~ &S=\dfrac{1}{3}\begin{pmatrix}
-1 ~& 2 ~& 2 \\
2 ~& -1 ~& 2 \\
2 ~& 2 ~& -1
\end{pmatrix},~&~T=\begin{pmatrix}
1 ~& 0 ~& 0 \\
0 ~& \omega ~& 0 \\
0 ~& 0 ~& \omega^2
\end{pmatrix}\,,
\end{eqnarray}
with $\omega=e^{2\pi i/3}$. The triplet representation $\bm{3}$ is a real representation, and the complex conjugate of $\bm{3}$ is equivalent to $\bm{3}$, i.e.,
\begin{equation}
\rho_{\bm{3}}^*(T)=\Omega \rho_{\bm{3}}(T) \Omega^{\dag}\,,~~~\rho_{\bm{3}}^*(S)=\Omega \rho_{\bm{3}}(S) \Omega^{\dag}\,,
\end{equation}
where $\Omega$ is the similarity transformation,
\begin{eqnarray}
\Omega=\begin{pmatrix}
1 ~&~ 0 ~&~ 0 \\
0 ~&~ 0 ~&~ 1 \\
0 ~&~ 1 ~&~ 0
\end{pmatrix}\,.
\end{eqnarray}
For two generic $A_4$ triplets $\alpha=\left(\alpha_1, \alpha_2, \alpha_3\right)^T$ and $\beta=\left(\beta_1, \beta_2, \beta_3\right)^T$, the tensor product decomposition is $\bm{3}\otimes\bm{3}=\bm{1}\oplus \bm{1}'\oplus\bm{1}''\oplus\bm{3}_S\oplus\bm{3}_A$, and the contraction rules are given by
\begin{eqnarray}
\nonumber\begin{pmatrix}
\alpha_1\\
\alpha_2\\
\alpha_3
\end{pmatrix}_{\bm{3}}\otimes\begin{pmatrix}
\beta_1\\
\beta_2\\
\beta_3
\end{pmatrix}_{\bm{3}}&=&\left(\alpha_1\beta_1+\alpha_2\beta_3+\alpha_3\beta_2\right)_{\bm{1}}\oplus\left(\alpha_3\beta_3+\alpha_1\beta_2+\alpha_2\beta_1\right)_{\bm{1}'}\oplus\left(\alpha_2\beta_2+\alpha_1\beta_3+\alpha_3\beta_1\right)_{\bm{1}''}\\
\label{eq:A4-contraction}&&\oplus\begin{pmatrix}
2\alpha_1\beta_1-\alpha_2\beta_3-\alpha_3\beta_2 \\
2\alpha_3\beta_3-\alpha_1\beta_2-\alpha_2\beta_1\\
2\alpha_2\beta_2-\alpha_1\beta_3-\alpha_3\beta_1
\end{pmatrix}_{\bm{3}_S} \oplus \begin{pmatrix}
\alpha_2\beta_3-\alpha_3\beta_2\\
\alpha_1\beta_2-\alpha_2\beta_1\\
\alpha_3\beta_1-\alpha_1\beta_3
\end{pmatrix}_{\bm{3}_A}\,.
\end{eqnarray}
All the modular forms of weight $k$ and level $3$ can be expressed as the homogeneous polynomials of degree $k$ in the modular functions $\vartheta(\tau)$ and $\varepsilon(\tau)$ given by~\cite{Liu:2019khw,Lu:2019vgm,Ding:2022aoe}
\begin{eqnarray}
\vartheta(\tau) = 3\sqrt{2} \dfrac{\eta^3(3\tau)}{\eta(\tau)}\,,~~~\varepsilon(\tau)= - \dfrac{3\eta^3(3\tau) + \eta^3(\tau/3)}{\eta(\tau)} \,.
\end{eqnarray}
It has been shown that there are three modular forms of weight 2 and level 3, they form a irreducible triplet $\bm{3}$ of $A_4$~\cite{Feruglio:2017spp}, and they can be expressed in terms of $\vartheta(\tau)$ and $\varepsilon(\tau)$ as follow~\cite{Liu:2019khw,Lu:2019vgm,Ding:2022aoe},
\begin{eqnarray}
\label{eq:Yw2-level3}Y^{(2)}_{\bm{3}}=
\begin{pmatrix}
\varepsilon^2(\tau) \\ \sqrt{2} \vartheta(\tau) \varepsilon(\tau) \\ - \vartheta^2(\tau)
\end{pmatrix}\equiv \begin{pmatrix}Y_1(\tau)\\Y_2(\tau)\\Y_3(\tau)\end{pmatrix}\,,
\end{eqnarray}
with the $q$-expansions,
\begin{eqnarray}
\nonumber Y_1(\tau)&=&1 + 12q + 36q^2 + 12q^3 + 84q^4 + 72q^5 +\dots \\
\nonumber Y_2(\tau)&=&-6q^{1/3}(1 + 7q + 8q^2 + 18q^3 + 14q^4 +\dots) \\
 Y_3(\tau)&=&-18q^{2/3}(1 + 2q + 5q^2 + 4q^3 + 8q^4 +\dots)\,.
\end{eqnarray}
We can obtain five independent weight 4 modular forms from the tensor product of $Y^{(2)}_{\bm{3}}$ with itself,
\begin{eqnarray}
\nonumber Y^{(4)}_{\bm{3}}&=&(Y^{(2)}_{\bm{3}}Y^{(2)}_{\bm{3}})_{\bm{3}}=(Y_1^2-Y_2 Y_3,Y_3^2-Y_1 Y_2,Y_2^2-Y_1 Y_3)^{T}\,, \\
\nonumber Y^{(4)}_{\bm{1}}&=&(Y^{(2)}_{\bm{3}}Y^{(2)}_{\bm{3}})_{\bm{1}}=Y_1^2+2 Y_2 Y_3\,, \\
\label{eq:Yw4-level3}Y^{(4)}_{\bm{1}'}&=&(Y^{(2)}_{\bm{3}}Y^{(2)}_{\bm{3}})_{\bm{1}'}=Y_3^2+2 Y_1 Y_2\,.
\end{eqnarray}
Similarly there are seven modular forms of weight 6, and they decompose into a singlet $\bm{1}$ and two triplets $\bm{3}$ under $A_4$,
\begin{eqnarray}
Y^{(6)}_{\bm{1}}&=&(Y^{(2)}_{\bm{3}}Y^{(4)}_{\bm{3}})_{\bm{1}}=Y_1^3+Y_2^3+Y_3^3-3 Y_1 Y_2 Y_3\,,\nonumber\\
Y^{(6)}_{\bm{3} I}&=&Y^{(2)}_{\bm{3}}Y^{(4)}_{\bm{1}}=(Y_1^3+2 Y_1Y_2Y_3,Y_1^2Y_2+2Y_2^2Y_3,Y_1^2Y_3+2Y_3^2Y_2)^{T}\,,\nonumber\\
\label{eq:MF-w6-N3}Y^{(6)}_{\bm{3}II}&=&Y^{(2)}_{\bm{3}}Y^{(4)}_{\bm{1}'}=(Y_3^3+2 Y_1Y_2Y_3,Y_3^2Y_1+2Y_1^2Y_2,Y_3^2Y_2+2Y_2^2Y_1)^{T}\,.
\end{eqnarray}
For the integer weight $k>2$, the polyharmonic Maa{\ss} forms coincide with modular forms at level 3. Moreover, the weight 2 and level 3 polyharmonic Maa{\ss} forms include the modified Eisenstein series $\widehat{E}_2(\tau)$ and triplet modular form $Y^{(2)}_{\bm{3}}$ in Eq.~\eqref{eq:Yw2-level3}, and $\widehat{E}_2(\tau)$ is an $A_4$ trivial singlet. Consequently we denote the weight 2 polyharmonic Maa{\ss} form $Y^{(2)}_{\bm{1}}\equiv \widehat{E}_2(\tau)$.

The modular form $Y^{(2)}_{\bm{3}}$ implies the presence of weight 0 polyharmonic Maa{\ss} form $Y^{(0)}_{\bm{3}}$ besides the trivial one $Y^{(0)}_{\bm{1}}=1$, and the corresponding master equation of Eq.~\eqref{eq:lift-cond} is of the following form:
\begin{equation}
\xi_0(Y^{(0)}_{\bm{3}})= \Omega Y^{(2)}_{\bm{3}}\,,~~~D(Y^{(0)}_{\bm{3}}) = -\dfrac{1}{4\pi} Y^{(2)}_{\bm{3}}\,.
\end{equation}
Using the general results of Eq.~\eqref{eq:polyHarmonic-expr-main}, we find the $q$-expansion of $Y^{(0)}_{\bm{3}}$ is
\begin{eqnarray}
\nonumber Y_{\bm{3},1}^{(0)}&=&y-\dfrac{3\,e^{-4\pi y}}{\pi q} - \dfrac{9\,e^{-8\pi y}}{2\pi q^2} - \dfrac{e^{-12\pi y}}{\pi q^3} - \dfrac{21\,e^{-16\pi y}}{4\pi q^4} - \dfrac{18\,e^{-20\pi y}}{5\pi q^5} - \dfrac{3\,e^{-24\pi y}}{2\pi q^6} + \cdots \\
\nonumber&&-\dfrac{9\log 3}{4\pi} - \dfrac{3q}{\pi} - \dfrac{9q^2}{2\pi} - \dfrac{q^3}{\pi} - \dfrac{21q^4}{4\pi} - \dfrac{18q^5}{5\pi} - \dfrac{3q^6}{2\pi} + \cdots \,, \\
\nonumber Y_{\bm{3},2}^{(0)}&=&\dfrac{27 q^{1/3}e^{\pi y /3}}{\pi} \left( \dfrac{e^{-3\pi y}}{4 q} +\dfrac{e^{-7\pi y}}{5 q^2} + \dfrac{5\,e^{-11\pi y}}{16 q^3} + \dfrac{2\,e^{-15\pi y}}{11 q^4} + \dfrac{2\,e^{-19\pi y}}{7 q^5} + \dfrac{4\,e^{-23\pi y}}{17 q^6} + \cdots \right) \\
\nonumber&&+\dfrac{9q^{1/3}}{2\pi}\left( 1 + \dfrac{7 q}{4} + \dfrac{8 q^2}{7} + \dfrac{9 q^3}{5} + \dfrac{14 q^4}{13} + \dfrac{31 q^5}{16} + \dfrac{20 q^6}{19} + \cdots \right)\,, \\
\nonumber Y_{\bm{3},3}^{(0)}&=&\dfrac{9\,q^{2/3}e^{2\pi y /3}}{2\pi}\left( \dfrac{e^{-2\pi y}}{q} + \dfrac{7\,e^{-6\pi y}}{4 q^2} + \dfrac{8\,e^{-10\pi y}}{7 q^3} + \dfrac{9\,e^{-14\pi y}}{5 q^4} + \dfrac{14\,e^{-18\pi y}}{13 q^5} + \dfrac{31\,e^{-22\pi y}}{16 q^6} + \cdots \right) \\
&&+\dfrac{27 q^{2/3}}{\pi}\left( \dfrac{1}{4} + \dfrac{q}{5}  + \dfrac{5 q^2}{16} + \dfrac{2 q^3}{11} + \dfrac{2 q^4}{7} + \dfrac{9 q^5}{17} + \dfrac{21 q^6}{20} + \cdots \right)\,,
\end{eqnarray}
One can obtain the weight $-2$ polyharmonic Maa{\ss} form of level 3 from the weight $4$ modular forms $Y^{(4)}_{\bm{1}}$, $Y^{(4)}_{\bm{1'}}$ and $Y^{(4)}_{\bm{3}}$ shown in Eq.~\eqref{eq:Yw4-level3}. Notice that $Y^{(4)}_{\bm{1'}}$ is a cusp form in $\Gamma(3)$ and consequently it cannot be lifted to a polyharmonic Maa{\ss} form. From the general formula Eq.~\eqref{eq:lift-cond}, we can obtain
\begin{eqnarray}
\nonumber \xi_{-2}(Y_{\bm{1}}^{(-2)}) &=& Y^{(4)}_{\bm{1}}\,,~~~~D^3(Y_{\bm{1}}^{(-2)})=-\dfrac{2}{(4\pi)^3} Y^{(4)}_{\bm{1}} \,, \\
\xi_{-2}(Y_{\bm{3}}^{(-2)}) &=& \Omega Y^{(4)}_{\bm{3}}\,,~~~~D^3(Y_{\bm{3}}^{(-2)})=-\dfrac{2}{(4\pi)^3} Y^{(4)}_{\bm{3}}\,.
\end{eqnarray}
Using the general results of Eq.~\eqref{eq:polyHarmonic-expr-main}, we find the $q$-expansions of $Y_{\bm{1}}^{(-2)}$ and $Y_{\bm{3}}^{(-2)}$ are
\begin{eqnarray}
\nonumber Y_{\bm{1}}^{(-2)}(\tau)&=& \dfrac{y^3}{3} - \dfrac{15\Gamma(3,4\pi y)}{4\pi^3 q} - \dfrac{135\Gamma(3,8\pi y)}{32\pi^3 q^2} - \dfrac{35\Gamma(3,12\pi y)}{9\pi^3 q^3} + \cdots \\
&&-\dfrac{\pi}{12}\dfrac{\zeta(3)}{\zeta(4)} - \dfrac{15 q}{2\pi^3} - \dfrac{135 q^2}{16\pi^3} - \dfrac{70 q^3}{9\pi^3} - \dfrac{1095 q^4}{128\pi^3} - \dfrac{189 q^5}{25\pi^3} - \dfrac{35 q^6}{4\pi^3} + \cdots\,,
\end{eqnarray}
and
\begin{eqnarray}
\nonumber Y_{\bm{3},1}^{(-2)}(\tau)&=& \dfrac{y^3}{3} + \dfrac{21\Gamma(3,4\pi y)}{16\pi^3 q} + \dfrac{189\Gamma(3,8\pi y)}{128\pi^3 q^2} + \dfrac{169\Gamma(3,12\pi y)}{144\pi^3 q^3} + \dfrac{1533 \Gamma(3,16\pi y)}{1024\pi^3 q^4} + \cdots \\
\nonumber&&+\dfrac{\pi}{40}\dfrac{\zeta(3)}{\zeta(4)} + \dfrac{21 q}{8\pi^3} + \dfrac{189 q^2}{64\pi^3} + \dfrac{169 q^3}{72\pi^3} + \dfrac{1533 q^4}{512\pi^3} + \dfrac{1323 q^5}{500\pi^3} + \dfrac{169 q^6}{64\pi^3} + \cdots\,, \\
\nonumber Y_{\bm{3},2}^{(-2)}(\tau)&=&-\dfrac{729 q^{1/3}}{16\pi^3}\left( \dfrac{\Gamma(3,8\pi y/3)}{16 q} + \dfrac{7\Gamma(3,20\pi y/3)}{125 q^2} + \dfrac{65\Gamma(3,32\pi y/3)}{1024 q^3} + \dfrac{74 \Gamma(3,44\pi y/3)}{1331 q^4} + \cdots \right)  \\
\nonumber&&-\dfrac{81q^{1/3}}{16\pi^3}\left( 1 + \dfrac{73 q}{64} + \dfrac{344 q^2}{343} + \dfrac{567 q^3}{500} + \dfrac{20198 q^4}{2197} + \dfrac{4681 q^5}{4096} + \cdots \right) \,, \\
\nonumber Y_{\bm{3},3 }^{(-2)}(\tau)&=&-\dfrac{81 q^{2/3}}{32\pi^3} \left( \dfrac{\Gamma(3,4\pi y/3)}{q} + \dfrac{73 \Gamma(3,16\pi y/3)}{64 q^2} + \dfrac{344 \Gamma(3,28\pi y/3)}{343 q^3} + \dfrac{567 \Gamma(3,40\pi y/3)}{500 q^4} + \cdots \right)  \\
&&-\dfrac{729 q^{2/3}}{8\pi^3}\left( \dfrac{1}{16} + \dfrac{7 q}{125} + \dfrac{65 q^2}{1024} + \dfrac{74 q^3}{1331} + \cdots \right)\,.
\end{eqnarray}
The weight $6$ modular forms in $\Gamma (3)$ are $Y^{(6)}_{\bm{1}}$, $Y^{(6)}_{\bm{3}I}$ and $Y^{(6)}_{\bm{3}II}$, as shown in Eq.~\eqref{eq:MF-w6-N3}. Notice that $Y^{(6)}_{\bm{3}II}$ are the cusp forms in $\Gamma (3)$, and only $Y^{(6)}_{\bm{1}}$ and the triplet combination $Y^{(6)}_{\bm{3}}=Y^{(6)}_{\bm{3}I}-\frac{5}{13}Y^{(6)}_{\bm{3}II}$ can be lifted to weight $-4$ polyharmonic Maa{\ss} form of level 3 through the following identities:
\begin{eqnarray}
\nonumber \xi_{-4}(Y_{\bm{1}}^{(-4)}) &=& Y^{(6)}_{\bm{1}}\,,~~~~D^5(Y_{\bm{1}}^{(-4)})=-\dfrac{24}{(4\pi)^5} Y^{(6)}_{\bm{1}}\,, \\
\xi_{-4}(Y_{\bm{3}}^{(-2)}) &=& \Omega Y^{(6)}_{\bm{3}}\,,~~~~D^5(Y_{\bm{3}}^{(-4)})=-\dfrac{24}{(4\pi)^5} Y^{(6)}_{\bm{3}}\,,
\end{eqnarray}
which can be solved and gives
\begin{eqnarray}
\nonumber Y_{\bm{1}}^{(-4)}&=& \dfrac{y^5}{5} + \dfrac{63\Gamma(5,4\pi y)}{128\pi^5 q} + \dfrac{2079\Gamma(5,8\pi y)}{4096\pi^5 q^2} + \dfrac{427\Gamma(5,12\pi y)}{864\pi^5 q^3} + \dfrac{66591\Gamma(5,16\pi y)}{131072 \pi^5 q^4} + \cdots   \\
\nonumber&&+\dfrac{\pi}{80}\dfrac{\zeta(5)}{\zeta(6)} + \dfrac{189 q}{16\pi^5} + \dfrac{6237 q^2}{512 \pi^5} + \dfrac{427 q^3}{36 \pi^5} + \dfrac{199773 q^4}{16384 \pi^5} + \cdots \,, \\
\nonumber Y_{\bm{3},1}^{(-4)} &=& \dfrac{y^5}{5} - \dfrac{ 549 }{3328 \pi^5} \left( \dfrac{\Gamma(5, 4 \pi y)}{q}+ \dfrac{33 \Gamma(5, 8 \pi y)}{32 q^2}+ \dfrac{14641 \Gamma(5, 12 \pi y)}{14823 q^3}+ \dfrac{1057 \Gamma(5, 16 \pi y)}{1024 q^4} + \cdots \right) \\
\nonumber&&-\dfrac{3\pi}{728}\dfrac{\zeta(5)}{\zeta(6)} - \dfrac{1647 }{416 \pi^5} \left( q + \dfrac{33 q^2}{32}+ \dfrac{14641 q^3}{14823}+ \dfrac{1057 q^4}{1024}+ \dfrac{3126 q^5}{3125} + \cdots \right)\,, \\
\nonumber Y_{\bm{3},2}^{(-4)} &=& \dfrac{72171 q^{1/3}}{212992 \pi^5} \left( \dfrac{\Gamma(5, 8 \pi y/3)}{q}+ \dfrac{33344 \Gamma(5, 20 \pi y/3)}{34375 q^2}+ \dfrac{1025 \Gamma(5, 32 \pi y/3)}{1024 q^3} + \cdots  \right) \\
\nonumber&& + \dfrac{6561 q^{1/3}}{832 \pi^5} \left( 1 + \dfrac{1057 q}{1024}+ \dfrac{16808 q^2}{16807}+ \dfrac{51579 q^3}{50000}+ \dfrac{371294 q^4}{371293} + \cdots \right)\,, \\
\nonumber Y_{\bm{3},3}^{(-4)} &=& \dfrac{2187 q^{2/3}}{6656 \pi^5} \left( \dfrac{\Gamma(5, 4 \pi y/3)}{q}+ \dfrac{1057 \Gamma(5, 16 \pi y/3)}{1024 q^2}+ \dfrac{16808 \Gamma(5, 28 \pi y/3)}{16807 q^3} + \cdots  \right) \\
&& +\dfrac{216513 q^{2/3}}{26624 \pi^5} \left( 1 + \dfrac{33344 q}{34375}+ \dfrac{1025 q^2}{1024}+ \dfrac{1717888 q^3}{1771561}+ \dfrac{16808 q^4}{16807} + \cdots \right)\,.
\end{eqnarray}

\subsection{$N=4$}

The inhomogeneous finite modular group $\Gamma_4$ is isomorphic to $S_4$ whose defining relations are
\begin{equation}
S_4=\left\{S,T|S^2=T^4=(ST)^3=1\right\}\,.
\end{equation}
The $S_4$ group has two singlet representations $\mathbf{1}$, $\mathbf{1}'$, a doublet representation $\mathbf{2}$, and two triplet representations $\mathbf{3}$, $\mathbf{3}'$. The generators $S$ and $T$ are represented by
\begin{eqnarray}
\nonumber \mathbf{1}:~ & S=1,~&~T=1\,,\\
\nonumber \mathbf{1}':~ &S=-1,~&~T=-1\,,\\
\nonumber \mathbf{2}:~ &S=\dfrac{1}{2} \begin{pmatrix}
-1 ~& \sqrt{3}  \\  \sqrt{3} ~& 1
\end{pmatrix},~&~T=\begin{pmatrix}
1 ~& 0 \\ 0 ~& -1
\end{pmatrix} \,,\\
\nonumber \mathbf{3}:~ & S=\dfrac{1}{2}\begin{pmatrix}
0 ~& \sqrt{2} ~& \sqrt{2} \\
\sqrt{2} ~& -1 ~& 1 \\
\sqrt{2} ~& 1 ~& -1
\end{pmatrix},~&~T=\begin{pmatrix}
1 ~& 0 ~& 0 \\
0 ~& i ~& 0 \\
0 ~& 0 ~& -i
\end{pmatrix}\,,\\
\mathbf{3}':~ & S=-\dfrac{1}{2}\begin{pmatrix}
0 ~& \sqrt{2} ~& \sqrt{2} \\
\sqrt{2} ~& -1 ~& 1 \\
\sqrt{2} ~& 1 ~& -1
\end{pmatrix},~&~T=-\begin{pmatrix}
1 ~& 0 ~& 0 \\
0 ~& i ~& 0 \\
0 ~& 0 ~& -i
\end{pmatrix}\,. \label{eq:rep-basis-S4}
\end{eqnarray}
The tensor products of different $S_4$ multiplets are given by
\begin{eqnarray}
\nonumber \begin{pmatrix}
\alpha_1\\
\alpha_2
\end{pmatrix}_{\mathbf{2}}\otimes\begin{pmatrix}
\beta_1\\
\beta_2
\end{pmatrix}_{\mathbf{2}}&=&\left(\alpha_1\beta_1+\alpha_2\beta_2\right)_{\mathbf{1}}\oplus\left(\alpha_1\beta_2-\alpha_2\beta_1\right)_{\mathbf{1}'}\oplus\begin{pmatrix}
-\alpha_1\beta_1+\alpha_2\beta_2\\
\alpha_1\beta_2+\alpha_2\beta_1
\end{pmatrix}_{\mathbf{2}} \,,
\end{eqnarray}

\begin{eqnarray}
\nonumber\begin{pmatrix}
\alpha_1\\
\alpha_2
\end{pmatrix}_{\mathbf{2}}\otimes
\begin{pmatrix}
\beta_1 \\
\beta_2 \\
\beta_3
\end{pmatrix}_{\mathbf{3}(\mathbf{3}')} &=&
\begin{pmatrix}
2\alpha_1 \beta_1 \\
-\alpha_1 \beta_2 + \sqrt{3}\alpha_2 \beta_3 \\
-\alpha_1 \beta_3 + \sqrt{3}\alpha_2 \beta_2
\end{pmatrix}_{\mathbf{3}(\mathbf{3}')} \oplus
\begin{pmatrix}
-2\alpha_2 \beta_1 \\
\sqrt{3}\alpha_1 \beta_3 + \alpha_2 \beta_2 \\
\sqrt{3}\alpha_1 \beta_2 + \alpha_2 \beta_3
\end{pmatrix}_{\mathbf{3}'(\mathbf{3})}\,,
\end{eqnarray}
\begin{eqnarray}
\nonumber &&\begin{pmatrix}
\alpha_1 \\
\alpha_2 \\
\alpha_3
\end{pmatrix}_{\mathbf{3}}\otimes\begin{pmatrix}
\beta_1 \\
\beta_2 \\
\beta_3
\end{pmatrix}_{\mathbf{3}}
=\begin{pmatrix}
\alpha_1 \\
\alpha_2 \\
\alpha_3
\end{pmatrix}_{\mathbf{3}'}\otimes\begin{pmatrix}
\beta_1 \\
\beta_2 \\
\beta_3
\end{pmatrix}_{\mathbf{3}'}
=\left(\alpha_1 \beta_1 + \alpha_2 \beta_3 + \alpha_3 \beta_2\right)_{\mathbf{1}}\\
\nonumber && \oplus
\begin{pmatrix}
2\alpha_1 \beta_1 - \alpha_2 \beta_3 - \alpha_3 \beta_2 \\
\sqrt{3}\alpha_2 \beta_2 + \sqrt{3} \alpha_3 \beta_3
\end{pmatrix}_{\mathbf{2}} \oplus
\begin{pmatrix}
\alpha_2 \beta_3 - \alpha_3 \beta_2 \\
\alpha_1 \beta_2 - \alpha_2 \beta_1 \\
\alpha_3 \beta_1 - \alpha_1 \beta_3
\end{pmatrix}_{\mathbf{3}} \oplus
\begin{pmatrix}
\alpha_2 \beta_2 - \alpha_3 \beta_3 \\
-\alpha_1 \beta_3 - \alpha_3 \beta_1 \\
\alpha_1 \beta_2 + \alpha_2 \beta_1
\end{pmatrix}_{\mathbf{3}'}\,,
\end{eqnarray}

\begin{eqnarray}
\nonumber &&\begin{pmatrix}
\alpha_1 \\
\alpha_2 \\
\alpha_3
\end{pmatrix}_{\mathbf{3}}\otimes\begin{pmatrix}
\beta_1 \\
\beta_2 \\
\beta_3
\end{pmatrix}_{\mathbf{3}'}
=\begin{pmatrix}
\alpha_1 \\
\alpha_2 \\
\alpha_3
\end{pmatrix}_{\mathbf{3}'}\otimes\begin{pmatrix}
\beta_1 \\
\beta_2 \\
\beta_3
\end{pmatrix}_{\mathbf{3}'}
=\begin{pmatrix}
\alpha_1 \beta_1 + \alpha_2 \beta_3 + \alpha_3 \beta_2
\end{pmatrix}_{\mathbf{1}'} \\
&& \oplus
\begin{pmatrix}
\sqrt{3}\alpha_2 \beta_2 + \sqrt{3} \alpha_3 \beta_3  \\
- 2\alpha_1 \beta_1+\alpha_2 \beta_3 + \alpha_3 \beta_2  \\
\end{pmatrix}_{\mathbf{2}}\oplus
\begin{pmatrix}
\alpha_2 \beta_2 - \alpha_3 \beta_3 \\
-\alpha_1 \beta_3 - \alpha_3 \beta_1 \\
\alpha_1 \beta_2 + \alpha_2 \beta_1
\end{pmatrix}_{\mathbf{3}} \oplus
\begin{pmatrix}
\alpha_2 \beta_3 - \alpha_3 \beta_2 \\
\alpha_1 \beta_2 - \alpha_2 \beta_1 \\
\alpha_3 \beta_1 - \alpha_1 \beta_3
\end{pmatrix}_{\mathbf{3}'}\,.
\end{eqnarray}
The weight $k$ modular forms at level $4$ can be expressed as the homogeneous polynomials of degree $2k$ in the Jacobi theta functions $\vartheta_1$ and $\vartheta_2$~\cite{Liu:2020msy,Novichkov:2020eep},
\begin{eqnarray}
\nonumber \vartheta_1(\tau) &=& \sum_{m\in \mathbb{Z}} e^{2\pi i \tau m^2} = 1 + 2 q + 2 q^4 + 2 q^9 + 2 q^{16} + \cdots \,, \\
\vartheta_2(\tau) &=& - \sum_{m\in \mathbb{Z}} e^{2\pi i \tau (m+1/2)^2} = - 2 q^{1/4} ( 1+ q^2 + q^6 + q^{12} + \cdots ) \,.
\end{eqnarray}
In the representation basis of Eq.~\eqref{eq:rep-basis-S4}, the weight $2$ modular forms of level $4$ can be arranged into a doublet and a triplet of $S_4$~\cite{Penedo:2018nmg,Novichkov:2018ovf}:
\begin{eqnarray}
\nonumber Y_{\mathbf{2}}^{(2)}&=&\begin{pmatrix}
Y_1 \\ Y_2
\end{pmatrix}=
\begin{pmatrix}
\vartheta_1^4 + \vartheta_2^4  \\
-2\sqrt{3} \vartheta_1^2 \vartheta_2^2
\end{pmatrix} \,, \\
Y_{\mathbf{3}}^{(2)}&=&\begin{pmatrix}
Y_3 \\ Y_4  \\ Y_5
\end{pmatrix}=
\begin{pmatrix}
\vartheta_1^4 - \vartheta_2^4  \\
2\sqrt{2} \vartheta_1^3 \vartheta_2 \\
2\sqrt{2} \vartheta_1 \vartheta_2^3
\end{pmatrix}\,,
\end{eqnarray}
The $q$-expansion of the modular doublet $Y^{(2)}_{\mathbf{2}}$ and triplet $Y^{(2)}_{\mathbf{3}}$ reads as follow
\begin{eqnarray}
\nonumber Y_{\mathbf{2}}^{(2)}&=&\begin{pmatrix}
Y_1 \\ Y_2
\end{pmatrix}=\begin{pmatrix}
1 + 24 q + 24 q^2 + 96 q^3 + 24 q^4 + 144 q^5 \cdots \\
-8\sqrt{3}\, q^{1/2} (1 + 4 q + 6 q^2 +8 q^3 + 13 q^4 + 12 q^5 + \cdots)
\end{pmatrix}\,, \\
Y_{\mathbf{3}}^{(2)}&=&\begin{pmatrix}
Y_3 \\ Y_4  \\ Y_5
\end{pmatrix}=\begin{pmatrix}
1 - 8 q + 24 q^2 - 32 q^3 + 24 q^4 - 48 q^5 + \cdots  \\
-4\sqrt{2}\, q^{1/4} (1 + 6 q + 13 q^2 + 14 q^3 + 18 q^4 + 32 q^5 + \cdots) \\
-16\sqrt{2} \,q^{3/4} (1 + 2 q + 3 q^2 + 6 q^3 + 5 q^4 + 6 q^5 + \cdots)
\end{pmatrix}\,.
\end{eqnarray}
The modular forms of level 4 can also be constructed from the derivative of the Dedekind eta function ~\cite{Penedo:2018nmg,Novichkov:2018ovf} or the products of eta function~\cite{Ding:2019gof,Ding:2021zbg,Qu:2021jdy}, the $q$-expansion of the modular forms would be identical when going to the same representation basis of $S_4$.

The weight $2$ polyharmonic Maa{\ss} forms include the modified Eisenstein series $\widehat{E}_2$ which is an $S_4$ singlet and the modular forms $Y^{(2)}_{\mathbf{2}}$ and $Y^{(2)}_{\mathbf{3}}$. The level 4 and weight $0$ polyharmonic Maa{\ss} forms can be arranged into a doublet $Y^{(0)}_{\mathbf{2}}$ and a triplet $Y^{(0)}_{\mathbf{3}}$ of $S_4$ besides
the trivial singlet $Y^{(0)}_{\mathbf{1}}=1$. The master equation of Eq.~\eqref{eq:lift-cond} for the nontrivial weight 0 polyharmonic Maa{\ss} form of level $4$ is of the following form
\begin{eqnarray}
\nonumber \xi_0(Y^{(0)}_{\mathbf{2}})&=&  Y^{(2)}_{\mathbf{2}},~~~D(Y^{(0)}_{\mathbf{2}}) = -\dfrac{1}{4\pi} Y^{(2)}_{\mathbf{2}}\,,  \\
\xi_0(Y^{(0)}_{\mathbf{3}})&=& \Omega Y^{(2)}_{\mathbf{3}},~~~D(Y^{(0)}_{\mathbf{3}}) = -\dfrac{1}{4\pi} Y^{(2)}_{\mathbf{3}}\,,
\end{eqnarray}
where $\Omega$ is a permutation matrix
\begin{eqnarray}
\Omega=\begin{pmatrix}
1 ~&~ 0 ~&~ 0 \\
0 ~&~ 0 ~&~ 1 \\
0 ~&~ 1 ~&~ 0
\end{pmatrix}\,.
\end{eqnarray}
Then we can determine the expression of the doublet $Y^{(0)}_{\mathbf{2}}$ as \begin{eqnarray}
\nonumber Y^{(0)}_{\mathbf{2},1}&=& y - \dfrac{6\,e^{-4\pi y}}{\pi q}-\dfrac{3\,e^{-8\pi y}}{\pi q^2}-\dfrac{8\,e^{-12\pi y}}{\pi q^3}-\dfrac{3\,e^{-16\pi y}}{2\pi q^4}-\dfrac{36\,e^{-20\pi y}}{5\pi q^5} + \cdots \\
\nonumber&&-\dfrac{4\log2}{\pi}-\dfrac{6q}{\pi}-\dfrac{3q^2}{\pi}-\dfrac{8q^3}{\pi}-\dfrac{3q^4}{2\pi}-\dfrac{36q^5}{5\pi}+\cdots \,, \\
\nonumber Y^{(0)}_{\mathbf{2},2}&=&4\sqrt{3}\,q^{1/2}\left( \dfrac{e^{-2\pi y}}{\pi q}+\dfrac{4\,e^{-6\pi y}}{3\pi q^2}+\dfrac{6\,e^{-10\pi y}}{5\pi q^3}+\dfrac{8\,e^{-14\pi y}}{7\pi q^4}+\dfrac{13\,e^{-18\pi y}}{9\pi q^5}+\cdots \right) \\
&&+\dfrac{4\sqrt{3}\,q^{1/2}}{\pi}\left(1+\dfrac{4}{3}q+\dfrac{6}{5}q^2+\dfrac{8}{7}q^3+\dfrac{13}{9}q^4+\dfrac{12}{11}q^5+\cdots \right)\,,
\end{eqnarray}
Similarly the $q$-expansion of the triplet $Y^{(0)}_{\mathbf{3}}$ is determined to be
\begin{eqnarray}
\nonumber Y_{\mathbf{3},1}^{(0)}&=&y + \dfrac{2\,e^{-4\pi y}}{\pi q} - \dfrac{3\,e^{-8\pi y}}{\pi q^2} + \dfrac{8\, e^{-12\pi y}}{3\pi q^3} - \dfrac{3\,e^{-16\pi y}}{2\pi q^4} + \dfrac{12\,e^{-20\pi y}}{5\pi q^5} + \cdots \\
\nonumber&&-\dfrac{2\log2}{\pi}+\dfrac{2q}{\pi}-\dfrac{3q^2}{\pi}+\dfrac{8q^3}{3\pi}-\dfrac{3q^4}{2\pi}+\dfrac{12q^5}{5\pi}-\dfrac{4q^6}{\pi}+\cdots\,, \\
\nonumber Y_{\mathbf{3},2}^{(0)}&=&\dfrac{16\sqrt{2}q^{1/4}}{\pi}\left( \dfrac{e^{-3\pi y}}{3 q} + \dfrac{2\,e^{-7\pi y}}{7 q^2} + \dfrac{3\,e^{-11\pi y}}{11 q^3} + \dfrac{2\,e^{-15\pi y}}{5 q^4} + \dfrac{5\,e^{-19\pi y}}{19 q^5} + \cdots \right) \\
\nonumber &&+\dfrac{4\sqrt{2}q^{1/4}}{\pi}\left( 1 + \dfrac{6q}{5} + \dfrac{13q^2}{9} + \dfrac{14q^3}{13} + \dfrac{18q^4}{17} + \dfrac{32q^5}{21} + \dfrac{31 q^6}{25} + \cdots \right) \,, \\
\nonumber Y_{\mathbf{3},3}^{(0)}&=&\dfrac{4\sqrt{2}q^{3/4}}{\pi} \left( \dfrac{e^{-\pi y}}{q} + \dfrac{6\,e^{-5\pi y}}{5q^2} + \dfrac{13\, e^{-9\pi y}}{9q^3} + \dfrac{14\,e^{-13\pi y}}{13q^4} + \dfrac{18\,e^{-17\pi y}}{14q^5} + \cdots \right) \,, \\
&&+\dfrac{16\sqrt{2}q^{3/4}}{\pi} \left( \dfrac{1}{3} + \dfrac{2q}{7} + \dfrac{3q^2}{11} + \dfrac{2 q^3}{5} + \dfrac{5q^4}{19} + \dfrac{6q^5}{23} + \dfrac{10q^6}{27} + \cdots \right)\,.
\end{eqnarray}
The weight $4$ modular forms of level $4$ are $Y^{(4)}_{\mathbf{1}}$, $Y^{(4)}_{\mathbf{2}}$, $Y^{(4)}_{\mathbf{3}}$ and $Y^{(4)}_{\mathbf{3}'}$ which are the tensor products of weight 2 modular forms:
\begin{eqnarray}
\nonumber Y^{(4)}_{\mathbf{1}} &=& \left(Y_{\mathbf{2}}^{(2)}Y_{\mathbf{2}}^{(2)}\right)_{\mathbf{1}}=Y_1^2 + Y_2^2 \,, \\
\nonumber Y^{(4)}_{\mathbf{2}} &=&-\left(Y_{\mathbf{2}}^{(2)}Y_{\mathbf{2}}^{(2)}\right)_{\mathbf{2}}= \begin{pmatrix}
Y_1^2 - Y_2^2 \\
-2Y_1 Y_2
\end{pmatrix} \,, \\
\nonumber Y^{(4)}_{\mathbf{3}} &=& \frac{1}{2}\left(Y_{\mathbf{2}}^{(2)}Y_{\mathbf{3}}^{(2)}\right)_{\mathbf{3}}= \begin{pmatrix}
Y_1 Y_3 \\
-\dfrac{1}{2}Y_1 Y_4 + \dfrac{\sqrt{3}}{2}Y_2 Y_5 \\
-\dfrac{1}{2}Y_1 Y_5 + \dfrac{\sqrt{3}}{2}Y_2 Y_4
\end{pmatrix}\,, \\
Y^{(4)}_{\mathbf{3}'} &=& \frac{1}{2}\left(Y_{\mathbf{2}}^{(2)}Y_{\mathbf{3}}^{(2)}\right)_{\mathbf{3}'}= \begin{pmatrix}
-Y_2 Y_3 \\
\dfrac{\sqrt{3}}{2}Y_1 Y_5 + \dfrac{1}{2}Y_2 Y_4 \\
\dfrac{\sqrt{3}}{2}Y_1 Y_4 + \dfrac{1}{2}Y_2 Y_5
\end{pmatrix}\,.
\end{eqnarray}
Notice that $Y_{\mathbf{3}'}^{(4)}$ are cusp forms which cannot be lifted to the polyharmonic Maa{\ss} forms. Others can be lifted to weight $-2$ polyharmonic Maa{\ss} forms through  Eq.~\eqref{eq:lift-cond} as follow
\begin{eqnarray}
\nonumber \xi_{-2}(Y_{\mathbf{1}}^{(-2)}) &=& Y^{(4)}_{\mathbf{1}}\,,~~~~D^3(Y_{\mathbf{1}}^{(-2)})=-\dfrac{2}{(4\pi)^3} Y^{(4)}_{\mathbf{1}}\,, \\
\nonumber \xi_{-2}(Y_{\mathbf{2}}^{(-2)}) &=& Y^{(4)}_{\mathbf{2}}\,,~~~~D^3(Y_{\mathbf{2}}^{(-2)})=-\dfrac{2}{(4\pi)^3} Y^{(4)}_{\mathbf{2}} \,, \\
\xi_{-2}(Y_{\mathbf{3}}^{(-2)}) &=& \Omega Y^{(4)}_{\mathbf{3}}\,,~~~~D^3(Y_{\mathbf{3}}^{(-2)})=-\dfrac{2}{(4\pi)^3} Y^{(4)}_{\mathbf{3}} \,.
\end{eqnarray}
Using the general result of Eq.~\eqref{eq:polyHarmonic-expr-main}, we find the weight $k=-2$ polyharmonic Maa{\ss} forms of level $4$ are
\begin{eqnarray}
\nonumber Y_{\mathbf{1}}^{(-2)}&=& \dfrac{y^3}{3} - \dfrac{15\Gamma(3,4\pi y)}{4\pi^3 q} - \dfrac{135\Gamma(3,8\pi y)}{32\pi^3 q^2} - \dfrac{35\Gamma(3,12\pi y)}{9\pi^3 q^3} + \cdots \\
&&-\dfrac{\pi}{12}\dfrac{\zeta(3)}{\zeta(4)} - \dfrac{15 q}{2\pi^3} - \dfrac{135 q^2}{16\pi^3} - \dfrac{70 q^3}{q\pi^3} - \dfrac{1095 q^4}{128\pi^3} - \dfrac{189 q^5}{25\pi^3} - \dfrac{35 q^6}{4\pi^3} + \cdots \,, \\
\nonumber Y_{\mathbf{2},1}^{(-2)}&=& \dfrac{y^3}{3} + \dfrac{9 \Gamma(3, 4 \pi y)}{4 \pi^3 q}+ \dfrac{57 \Gamma(3, 8 \pi y)}{32 \pi^3 q^2}+ \dfrac{7 \Gamma(3, 12 \pi y)}{3 \pi^3 q^3}+ \dfrac{441 \Gamma(3, 16 \pi y)}{256 \pi^3 q^4} + \cdots  \\
\nonumber &&+\dfrac{\pi}{30} \dfrac{\zeta(3)}{\zeta(4)} + \dfrac{9 q}{2 \pi^3}+ \dfrac{57 q^2}{16 \pi^3}+ \dfrac{14 q^3}{3 \pi^3}+ \dfrac{441 q^4}{128 \pi^3}+ \dfrac{567 q^5}{125 \pi^3} + \cdots \,, \\
\nonumber Y_{\mathbf{2},2}^{(-2)}&=& -\dfrac{2\sqrt{3}q^{1/2}}{\pi^3}\left( \dfrac{\Gamma(3, 2 \pi y)}{q}+ \dfrac{28 \Gamma(3, 6 \pi y)}{27 q^2}+ \dfrac{126 \Gamma(3, 10 \pi y)}{125 q^3}+ \dfrac{344 \Gamma(3, 14 \pi y)}{343 q^4} + \cdots  \right) \\
\nonumber && -\dfrac{4\sqrt{3}q^{1/2}}{\pi^3} \left( 1 + \dfrac{28 q}{27}+ \dfrac{126 q^2}{125}+ \dfrac{344 q^3}{343}+ \dfrac{757 q^4}{729}+ \dfrac{1332 q^5}{1331} + \cdots \right) \,, \\
\nonumber Y_{\mathbf{3},1}^{(-2)}&=& \dfrac{y^3}{3} - \dfrac{ \Gamma(3, 4 \pi y)}{4 \pi^3 q}+ \dfrac{9 \Gamma(3, 8 \pi y)}{32 \pi^3 q^2}- \dfrac{ 7 \Gamma(3, 12 \pi y)}{27 \pi^3 q^3}+ \dfrac{57 \Gamma(3, 16 \pi y)}{256 \pi^3 q^4}- \dfrac{ 63 \Gamma(3, 20 \pi y)}{250 \pi^3 q^5} + \cdots  \\
\nonumber &&+ \dfrac{\pi}{240}\dfrac{\zeta(3)}{\zeta(4)} - \dfrac{q}{2 \pi^3}+ \dfrac{9 q^2}{16 \pi^3}- \dfrac{14 q^3}{27 \pi^3}+ \dfrac{57 q^4}{128 \pi^3}- \dfrac{63 q^5}{125 \pi^3} + \cdots \,, \\
\nonumber Y_{\mathbf{3},2}^{(-2)}&=&-\dfrac{56\sqrt{2} q^{1/4}}{27\pi^3} \left( \dfrac{\Gamma(3, 3 \pi y)}{q}+ \dfrac{2322 \Gamma(3, 7 \pi y)}{2401 q^2}+ \dfrac{8991 \Gamma(3, 11 \pi y)}{9317 q^3}+ \dfrac{126 \Gamma(3, 15 \pi y)}{125 q^4} + \cdots  \right) \\
\nonumber && - \dfrac{4\sqrt{2} q^{1/4}}{\pi^3} \left( 1 + \dfrac{126 q}{125}+ \dfrac{757 q^2}{729}+ \dfrac{2198 q^3}{2197}+ \dfrac{4914 q^4}{4913}+ \dfrac{1376 q^5}{1323} + \cdots \right) \,, \\
\nonumber Y_{\mathbf{3},3}^{(-2)}&=& -\dfrac{2\sqrt{2} q^{3/4}}{\pi^3} \left( \dfrac{\Gamma(3, \pi y)}{q}+ \dfrac{126 \Gamma(3, 5 \pi y)}{125 q^2}+ \dfrac{757 \Gamma(3, 9 \pi y)}{729 q^3}+ \dfrac{2198 \Gamma(3, 13 \pi y)}{2197 q^4} + \cdots  \right)  \\
&& -\dfrac{112\sqrt{2} q^{3/4}}{27\pi^3} \left( 1 + \dfrac{2322 q}{2401}+ \dfrac{8991 q^2}{9317}+ \dfrac{126 q^3}{125}+ \dfrac{6615 q^4}{6859}+ \dfrac{82134 q^5}{85169} + \cdots \right)\,.
\end{eqnarray}
The weight $6$ modular forms of level $4$ can be organized into six multiplets of $S_4$: $Y^{(6)}_{\mathbf{1}}$, $Y^{(6)}_{\mathbf{1}'}$, $Y^{(6)}_{\mathbf{2}}$, $Y^{(6)}_{\mathbf{3}I}$, $Y^{(6)}_{\mathbf{3}II}$ and $Y^{(6)}_{\mathbf{3}'}$ with
\begin{eqnarray}
\nonumber Y^{(6)}_{\mathbf{1}} &=&(Y_{\mathbf{2}}^{(2)}Y_{\mathbf{2}}^{(4)})_{\mathbf{1}} =Y_1^3 - 3Y_1 Y_2^2 \,, \\
\nonumber Y^{(6)}_{\mathbf{1}'} &=&(Y_{\mathbf{2}}^{(2)}Y_{\mathbf{2}}^{(4)})_{\mathbf{1}'} = Y_2^3 -3Y_1^2 Y_2\,, \\
\nonumber Y^{(6)}_{\mathbf{2}} &=&(Y_{\mathbf{2}}^{(2)}Y_{\mathbf{1}}^{(4)})_{\mathbf{2}} =\begin{pmatrix}
 Y_1(Y_1^2+Y_2^2)  \\
 Y_2 (Y_1^2+Y_2^2)
\end{pmatrix} \,, \\
\nonumber Y^{(6)}_{\mathbf{3}I} &=& \frac{1}{2} (Y_{\mathbf{3}}^{(2)}Y_{\mathbf{2}}^{(4)})_{\mathbf{3}} =\begin{pmatrix}
(Y_1^2-Y_2^2) Y_3  \\
\dfrac{1}{2}(Y_2^2-Y_1^2)Y_4 - \sqrt{3}Y_1 Y_2 Y_5 \\
\dfrac{1}{2}(Y_2^2-Y_1^2)Y_5 - \sqrt{3}Y_1 Y_2 Y_4 \\
\end{pmatrix} \,, \\
\nonumber Y^{(6)}_{\mathbf{3}II} &=&(Y_{\mathbf{3}}^{(2)}Y_{\mathbf{1}}^{(4)})_{\mathbf{3}} = \begin{pmatrix}
Y_3(Y_1^2+ Y_2^2)  \\
Y_4(Y_1^2+ Y_2^2) \\
Y_5(Y_1^2+ Y_2^2)
\end{pmatrix} \,, \\
Y^{(6)}_{\mathbf{3}'} &=& (Y_{\mathbf{3}}^{(2)}Y_{\mathbf{2}}^{(4)})_{\mathbf{3}'} =\begin{pmatrix}
4 Y_1 Y_2 Y_3 \\
\sqrt{3}(Y_1^2 - Y_2^2) Y_5 - 2 Y_1 Y_2 Y_4 \\
\sqrt{3}(Y_1^2 - Y_2^2) Y_4- 2 Y_1 Y_2 Y_5
\end{pmatrix}\,.
\end{eqnarray}
We can find that $Y^{(6)}_{\mathbf{1}'}$, $Y^{(6)}_{\mathbf{3}'}$ and $Y^{(6)}_{\mathbf{3}I}-Y^{(6)}_{\mathbf{3}II}$ are cusp forms. We can lift $Y^{(6)}_{\mathbf{1}}$, $Y^{(6)}_{\mathbf{2}}$ and the triplet modular form $Y^{(6)}_{\mathbf{3}}\equiv \dfrac{5}{8} Y^{(6)}_{\mathbf{3}I} + \dfrac{3}{8} Y^{(6)}_{\mathbf{3}II}$ into weight $-4$ polyharmonic Maa{\ss} forms through  Eq.~\eqref{eq:lift-cond} as follows
\begin{eqnarray}
\nonumber \xi_{-4}(Y_{\mathbf{1}}^{(-4)}) &=& Y^{(6)}_{\mathbf{1}}\,,~~~~D^5(Y_{\mathbf{1}}^{(-4)})= -\dfrac{24}{(4\pi)^5}} Y^{(6)}_{\mathbf{1}\,, \\
\nonumber \xi_{-4}(Y_{\mathbf{2}}^{(-4)}) &=& Y^{(6)}_{\mathbf{2}}\,,~~~~D^5(Y_{\mathbf{2}}^{(-4)})=-\dfrac{24}{(4\pi)^5}} Y^{(6)}_{\mathbf{2}\,, \\
\xi_{-4}(Y_{\mathbf{3}}^{(-4)}) &=& \Omega Y^{(6)}_{\mathbf{3}}\,,~~~~D^5(Y_{\mathbf{3}}^{(-4)})=-\dfrac{24}{(4\pi)^5} Y^{(6)}_{\mathbf{3}} \,.
\end{eqnarray}
Using the general result of Eq.~\eqref{eq:polyHarmonic-expr-main}, we find the weight $k=-4$ polyharmonic Maa{\ss} forms of level $4$ are given by
\begin{eqnarray}
\nonumber Y_{\mathbf{1}}^{(-4)}&=& \dfrac{y^5}{5} + \dfrac{63\Gamma(5,4\pi y)}{128\pi^5 q} + \dfrac{2079\Gamma(5,8\pi y)}{4096\pi^5 q^2} + \dfrac{427\Gamma(5,12\pi y)}{864\pi^5 q^3} + \dfrac{66591\Gamma(5,16\pi y)}{131072 \pi^5 q^4} + \cdots   \\
\nonumber &&+\dfrac{\pi}{80}\dfrac{\zeta(5)}{\zeta(6)} + \dfrac{189 q}{16\pi^5} + \dfrac{6237 q^2}{512 \pi^5} + \dfrac{427 q^3}{36 \pi^5} + \dfrac{199773 q^4}{16384 \pi^5} + \cdots \,, \\
\nonumber Y_{\mathbf{2},1}^{(-4)} &=& \dfrac{y^5}{5} - \dfrac{ 33 \Gamma(5, 4 \pi y)}{128 \pi^5 q}- \dfrac{ 993 \Gamma(5, 8 \pi y)}{4096 \pi^5 q^2}- \dfrac{ 671 \Gamma(5, 12 \pi y)}{2592 \pi^5 q^3}- \dfrac{ 31713 \Gamma(5, 16 \pi y)}{131072 \pi^5 q^4}+ \cdots  \\
\nonumber && - \dfrac{\pi}{168}\dfrac{\zeta(5)}{\zeta(6)} - \dfrac{99 q}{16 \pi^5}- \dfrac{2979 q^2}{512 \pi^5}- \dfrac{671 q^3}{108 \pi^5}- \dfrac{95139 q^4}{16384 \pi^5}- \dfrac{154737 q^5}{25000 \pi^5} + \cdots \,, \\
\nonumber Y_{\mathbf{2},2}^{(-4)} &=& \dfrac{\sqrt{3}q^{1/2}}{4\pi^5} \left( \dfrac{\Gamma(5, 2 \pi y)}{q}+ \dfrac{244 \Gamma(5, 6 \pi y)}{243 q^2}+ \dfrac{3126 \Gamma(5, 10 \pi y)}{3125 q^3}+ \dfrac{16808 \Gamma(5, 14 \pi y)}{16807 q^4} + \cdots  \right) \\
\nonumber && + \dfrac{6\sqrt{3}q^{1/2}}{\pi^5} \left( 1 + \dfrac{244 q}{243}+ \dfrac{3126 q^2}{3125}+ \dfrac{16808 q^3}{16807}+ \dfrac{59293 q^4}{59049}+ \dfrac{161052 q^5}{161051} + \cdots \right)\,, \\
\nonumber Y_{\mathbf{3},1}^{(-4)}&=& \dfrac{y^5}{5} + \dfrac{\Gamma(5, 4 \pi y)}{128 \pi^5 q}- \dfrac{ 33 \Gamma(5, 8 \pi y)}{4096 \pi^5 q^2}+ \dfrac{61 \Gamma(5, 12 \pi y)}{7776 \pi^5 q^3}- \dfrac{ 993 \Gamma(5, 16 \pi y)}{131072 \pi^5 q^4}+ \dfrac{1563 \Gamma(5, 20 \pi y)}{200000 \pi^5 q^5} + \cdots  \\
\nonumber && - \dfrac{\pi}{5376}\dfrac{\zeta(5)}{\zeta(6)}  + \dfrac{3 q}{16 \pi^5}- \dfrac{99 q^2}{512 \pi^5}+ \dfrac{61 q^3}{324 \pi^5}- \dfrac{2979 q^4}{16384 \pi^5}+ \dfrac{4689 q^5}{25000 \pi^5} + \cdots \,, \\
\nonumber Y_{\mathbf{3},2}^{(-4)}&=& \dfrac{61\sqrt{2}q^{1/4}}{243\pi^5} \left( \dfrac{\Gamma(5, 3 \pi y)}{q}+ \dfrac{1021086 \Gamma(5, 7 \pi y)}{1025227 q^2}+ \dfrac{9783909 \Gamma(5, 11 \pi y)}{9824111 q^3}+ \dfrac{3126 \Gamma(5, 15 \pi y)}{3125 q^4} + \cdots  \right)  \\
\nonumber && + \dfrac{6\sqrt{2}q^{1/4}}{\pi^5} \left( 1 + \dfrac{3126 q}{3125}+ \dfrac{59293 q^2}{59049}+ \dfrac{371294 q^3}{371293}+ \dfrac{1419858 q^4}{1419857}+ \dfrac{4101152 q^5}{4084101} + \cdots \right) \,, \\
\nonumber Y_{\mathbf{3},3}^{(-4)}&=& \dfrac{q^{3/4}}{2\sqrt{2}\pi^5} \left( \dfrac{\Gamma(5, \pi y)}{q}+ \dfrac{3126 \Gamma(5, 5 \pi y)}{3125 q^2}+ \dfrac{59293 \Gamma(5, 9 \pi y)}{59049 q^3}+ \dfrac{371294 \Gamma(5, 13 \pi y)}{371293 q^4} + \cdots  \right) \\
&&+\dfrac{488\sqrt{2}q^{3/4}}{81\pi^5} \left( 1 + \dfrac{1021086 q}{1025227}+ \dfrac{9783909 q^2}{9824111}+ \dfrac{3126 q^3}{3125}+ \dfrac{150423075 q^4}{151042039} + \cdots \right)\,.
\end{eqnarray}

\subsection{$N=5$}

The finite modular group $\Gamma_5$ is isomorphic to $A_5$ which is the symmetry group of the regular icosahedron, and the generators $S$ and $T$ satisfy the following relations:
\begin{equation}
A_5=\left\{S,T|S^2=T^5=(ST)^3=1\right\}\,.
\end{equation}
The $A_5$ group has one singlet representation $\mathbf{1}$, two triplet representations $\mathbf{3}$, $\mathbf{3}'$, one quartet representation $\mathbf{4}$ and one quintet representation $\mathbf{5}$. The generators $S$ and $T$ are represented by
\begin{eqnarray}
\nonumber \mathbf{1}:~ & S=1,~&~T=1\,,\\
\nonumber \mathbf{3}:~ &S=\dfrac{1}{\sqrt{5}}\begin{pmatrix}
1 ~& -\sqrt{2} ~& -\sqrt{2} \\
-\sqrt{2} ~& -\phi ~& \frac{1}{\phi} \\
-\sqrt{2} ~& \frac{1}{\phi} ~& -\phi
\end{pmatrix},~&~T=\begin{pmatrix}
1 ~& 0 ~& 0 \\
0 ~& \omega_5 ~& 0 \\
0 ~& 0 ~& \omega_5^4
\end{pmatrix}\,,\\
\nonumber \mathbf{3}':~ & S=\dfrac{1}{\sqrt{5}}\begin{pmatrix}
-1 ~& \sqrt{2} ~& \sqrt{2} \\
\sqrt{2} ~& -\frac{1}{\phi} ~& \phi \\
\sqrt{2} ~& \phi ~& -\frac{1}{\phi}
\end{pmatrix},~&~T=\begin{pmatrix}
1 ~& 0 ~& 0 \\
0 ~& \omega_5^2 ~& 0 \\
0 ~& 0 ~& \omega_5^3
\end{pmatrix}\,,\\
\nonumber \mathbf{4}:~ & S=\dfrac{1}{\sqrt{5}}\begin{pmatrix}
1 ~& \frac{1}{\phi} ~& \phi ~& -1  \\
\frac{1}{\phi} ~& -1 ~& 1 ~& \phi \\
\phi ~& 1 ~& -1 ~& \frac{1}{\phi} \\
-1 ~& \phi ~& \frac{1}{\phi} ~& 1
\end{pmatrix},~&~T=\begin{pmatrix}
\omega_5 ~& 0 ~& 0 ~& 0 \\
0 ~& \omega_5^2 ~& 0 ~& 0\\
0 ~& 0 ~& \omega_5^3 ~& 0 \\
0 ~& 0 ~& 0 ~& \omega_5^4
\end{pmatrix}\,,\\
\mathbf{5}:~ & S=\dfrac{1}{5}\begin{pmatrix}
-1 ~& \sqrt{6} ~& \sqrt{6} ~& \sqrt{6} ~& \sqrt{6} \\
\sqrt{6} ~& \frac{1}{\phi^2} ~& -2\phi ~& \frac{2}{\phi} ~& \phi^2 \\
\sqrt{6} ~& -2\phi ~& \phi^2 ~& \frac{1}{\phi^2} ~& \frac{2}{\phi} \\
\sqrt{6} ~& \frac{2}{\phi} ~& \frac{1}{\phi^2} ~& \phi^2 ~& -2\phi \\
\sqrt{6} ~& \phi^2 ~& \frac{2}{\phi} ~& -2\phi ~& \frac{1}{\phi^2}
\end{pmatrix},~&~T=\begin{pmatrix}
1 ~& 0 ~& 0 ~& 0 ~& 0  \\
0 ~& \omega_5 ~& 0 ~& 0 ~& 0 \\
0 ~& 0 ~& \omega_5^2 ~& 0 ~& 0\\
0 ~& 0 ~& 0 ~& \omega_5^3 ~& 0 \\
0 ~& 0 ~& 0 ~& 0 ~& \omega_5^4
\end{pmatrix}\,,
\end{eqnarray}
where $\omega_5=e^{2\pi i/5}$ is the quintic unit root and $\phi=(1+\sqrt{5})/2$ is the golden ratio. We adopt the same representation basis as that of~\cite{Ding:2011cm,Novichkov:2018nkm,Ding:2019xna}. The tensor products of different $A_5$ irreducible representations are as follows~\cite{Ding:2011cm},
\begin{eqnarray*}
\begin{pmatrix}
\alpha_1 \\
\alpha_2 \\
\alpha_3
\end{pmatrix}_{\mathbf{3}}\otimes\begin{pmatrix}
\beta_1 \\
\beta_2 \\
\beta_3
\end{pmatrix}_{\mathbf{3}}&=&(\alpha_1 \beta_1 + \alpha_2 \beta_3 + \alpha_3\beta_2)_{\mathbf{1}} \oplus \begin{pmatrix}
\alpha_2 \beta_3 - \alpha_3 \beta_2 \\
\alpha_1 \beta_2 - \alpha_2 \beta_1 \\
\alpha_3 \beta_1 - \alpha_1 \beta_3
\end{pmatrix}_{\mathbf{3}} \oplus
\begin{pmatrix}
2\alpha_1 \beta_1 - \alpha_2 \beta_3 - \alpha_3 \beta_2 \\
- \sqrt{3}\alpha_1 \beta_2 - \sqrt{3}\alpha_2 \beta_1 \\
\sqrt{6} \alpha_2 \beta_2 \\
\sqrt{6} \alpha_3 \beta_3 \\
- \sqrt{3}\alpha_1 \beta_3 - \sqrt{3}\alpha_3 \beta_1
\end{pmatrix}_{\mathbf{5}}\,,  \\
\begin{pmatrix}
\alpha_1 \\
\alpha_2 \\
\alpha_3
\end{pmatrix}_{\mathbf{3}}\otimes\begin{pmatrix}
\beta_1 \\
\beta_2 \\
\beta_3
\end{pmatrix}_{\mathbf{3}'}&=&
\begin{pmatrix}
\sqrt{2} \alpha_2 \beta_1 + \alpha_3 \beta_2 \\
-\sqrt{2} \alpha_1 \beta_2 - \alpha_3 \beta_3 \\
- \sqrt{2} \alpha_1 \beta_3 - \alpha_2 \beta_2  \\
\alpha_2 \beta_3 + \sqrt{2} \alpha_3 \beta_1
\end{pmatrix}_{\mathbf{4}} \oplus
\begin{pmatrix}
\sqrt{3} \alpha_1 \beta_1 \\
\alpha_2 \beta_1 - \sqrt{2} \alpha_3 \beta_2 \\
\alpha_1 \beta_2 - \sqrt{2} \alpha_3 \beta_3 \\
\alpha_1 \beta_3 - \sqrt{2} \alpha_2 \beta_2 \\
\alpha_3 \beta_1- \sqrt{2} \alpha_2 \beta_3
\end{pmatrix}_{\mathbf{5}}\,, \\
\begin{pmatrix}
\alpha_1 \\
\alpha_2 \\
\alpha_3
\end{pmatrix}_{\mathbf{3}'}\otimes
\begin{pmatrix}
\beta_1 \\
\beta_2 \\
\beta_3
\end{pmatrix}_{\mathbf{3}'}&=&
(\alpha_1 \beta_1 + \alpha_2 \beta_3 + \alpha_3\beta_2)_{\mathbf{1}} \oplus
\begin{pmatrix}
\alpha_2 \beta_3 - \alpha_3 \beta_2 \\
\alpha_1 \beta_2 - \alpha_2 \beta_1 \\
- \alpha_1 \beta_3 + \alpha_3 \beta_1
\end{pmatrix}_{\mathbf{3}} \oplus
\begin{pmatrix}
2 \alpha_1 \beta_1 - \alpha_2 \beta_3 - \alpha_3 \beta_2 \\
\sqrt{6} \alpha_3 \beta_3 \\
- \sqrt{3} \alpha_1 \beta_2 - \sqrt{3} \alpha_2 \beta_1  \\
- \sqrt{3} \alpha_1 \beta_3 - \sqrt{3} \alpha_3 \beta_1 \\
\sqrt{6} \alpha_2 \beta_2
\end{pmatrix}_{\mathbf{5}} \,, \\
\begin{pmatrix}
\alpha_1 \\
\alpha_2 \\
\alpha_3
\end{pmatrix}_{\mathbf{3}}\otimes
\begin{pmatrix}
\beta_1 \\
\beta_2 \\
\beta_3 \\
\beta_4
\end{pmatrix}_{\mathbf{4}} &=&
\begin{pmatrix}
- \sqrt{2} \alpha_2 \beta_4 - \sqrt{2} \alpha_3 \beta_1 \\
\sqrt{2} \alpha_1 \beta_2 - \alpha_2 \beta_1 + \alpha_3 \beta_3 \\
\sqrt{2} \alpha_1 \beta_3 + \alpha_2 \beta_2 - \alpha_3 \beta_4
\end{pmatrix}_{\mathbf{3}'} \oplus
\begin{pmatrix}
\alpha_1 \beta_1 - \sqrt{2} \alpha_3 \beta_2  \\
- \alpha_1 \beta_2 - \sqrt{2} \alpha_2 \beta_1  \\
\alpha_1 \beta_3 + \sqrt{2} \alpha_3 \beta_4 \\
- \alpha_1 \beta_4 + \sqrt{2} \alpha_2 \beta_3  \\
\end{pmatrix}_{\mathbf{4}} \\ &\oplus & \nonumber
\begin{pmatrix}
\sqrt{6} \alpha_2 \beta_4 - \sqrt{6} \alpha_3 \beta_1  \\
2\sqrt{2} \alpha_1 \beta_1 + 2 \alpha_3 \beta_2 \\
- \sqrt{2} \alpha_1 \beta_2 + \alpha_2 \beta_1 + 3 \alpha_3 \beta_3   \\
\sqrt{2} \alpha_1 \beta_3 - 3 \alpha_2 \beta_2 - \alpha_3 \beta_4 \\
- 2\sqrt{2} \alpha_1 \beta_4 - 2 \alpha_2 \beta_3
\end{pmatrix}_{\mathbf{5}}\,,  \\
\begin{pmatrix}
\alpha_1 \\
\alpha_2 \\
\alpha_3
\end{pmatrix}_{\mathbf{3}'}\otimes
\begin{pmatrix}
\beta_1 \\
\beta_2 \\
\beta_3 \\
\beta_4
\end{pmatrix}_{\mathbf{4}} &=&
\begin{pmatrix}
- \sqrt{2} \alpha_2 \beta_3 - \sqrt{2} \alpha_3 \beta_2 \\
\sqrt{2} \alpha_1 \beta_1 + \alpha_2 \beta_4 - \alpha_3 \beta_3 \\
\sqrt{2} \alpha_1 \beta_4 - \alpha_2 \beta_2 + \alpha_3 \beta_1
\end{pmatrix}_{\mathbf{3}} \oplus
\begin{pmatrix}
\alpha_1 \beta_1 + \sqrt{2} \alpha_3 \beta_3 \\
\alpha_1 \beta_2 - \sqrt{2} \alpha_3 \beta_4 \\
- \alpha_1 \beta_3 + \sqrt{2} \alpha_2 \beta_1  \\
- \alpha_1 \beta_4 - \sqrt{2} \alpha_2 \beta_2
\end{pmatrix}_{\mathbf{4}}  \\ &\oplus &
\begin{pmatrix}
\sqrt{6} \alpha_2 \beta_3 - \sqrt{6} \alpha_3 \beta_2 \\
\sqrt{2} \alpha_1 \beta_1 - 3 \alpha_2 \beta_4  - \alpha_3 \beta_3 \\
2\sqrt{2} \alpha_1 \beta_2 + 2 \alpha_3 \beta_4 \\
- 2\sqrt{2} \alpha_1 \beta_3 - 2 \alpha_2 \beta_1  \\
- \sqrt{2} \alpha_1 \beta_4 + \alpha_2 \beta_2 + 3 \alpha_3 \beta_1
\end{pmatrix}_{\mathbf{5}} \,, \\
\begin{pmatrix}
\alpha_1 \\
\alpha_2 \\
\alpha_3
\end{pmatrix}_{\mathbf{3}}\otimes
\begin{pmatrix}
\beta_1 \\
\beta_2 \\
\beta_3 \\
\beta_4 \\
\beta_5
\end{pmatrix}_{\mathbf{5}} &=&
\begin{pmatrix}
- 2 \alpha_1 \beta_1 + \sqrt{3} \alpha_2 \beta_5 + \sqrt{3} \alpha_3 \beta_2 \\
\sqrt{3} \alpha_1 \beta_2 + \alpha_2 \beta_1  - \sqrt{6} \alpha_3 \beta_3 \\
\sqrt{3} \alpha_1 \beta_5 - \sqrt{6} \alpha_2 \beta_4 + \alpha_3 \beta_1
\end{pmatrix}_{\mathbf{3}}  \oplus
\begin{pmatrix}
\sqrt{3} \alpha_1 \beta_1 + \alpha_2 \beta_5 + \alpha_3 \beta_2 \\
\alpha_1 \beta_3 - \sqrt{2} \alpha_2 \beta_2 - \sqrt{2} \alpha_3 \beta_4 \\
\alpha_1 \beta_4 - \sqrt{2} \alpha_2 \beta_3 - \sqrt{2} \alpha_3 \beta_5
\end{pmatrix}_{\mathbf{3}'}  \\ &\oplus &  \nonumber
\begin{pmatrix}
2\sqrt{2} \alpha_1 \beta_2 - \sqrt{6} \alpha_2 \beta_1 + \alpha_3 \beta_3 \\
- \sqrt{2} \alpha_1 \beta_3 + 2 \alpha_2 \beta_2 - 3 \alpha_3 \beta_4 \\
\sqrt{2} \alpha_1 \beta_4 + 3 \alpha_2 \beta_3 - 2\alpha_3 \beta_5 \\
- 2\sqrt{2} \alpha_1 \beta_5 - \alpha_2 \beta_4 + \sqrt{6} \alpha_3 \beta_1
\end{pmatrix}_{\mathbf{4}} \oplus
\begin{pmatrix}
\sqrt{3} \alpha_2 \beta_5 - \sqrt{3} \alpha_3 \beta_2 \\
- \alpha_1 \beta_2 - \sqrt{3} \alpha_2 \beta_1 - \sqrt{2} \alpha_3 \beta_3  \\
- 2 \alpha_1 \beta_3 - \sqrt{2} \alpha_2 \beta_2  \\
2 \alpha_1 \beta_4 + \sqrt{2} \alpha_3 \beta_5 \\
\alpha_1 \beta_5 + \sqrt{2} \alpha_2 \beta_4 + \sqrt{3} \alpha_3 \beta_1
\end{pmatrix}_{\mathbf{5}} \,, \\
\begin{pmatrix}
\alpha_1 \\
\alpha_2 \\
\alpha_3
\end{pmatrix}_{\mathbf{3}'}\otimes
\begin{pmatrix}
\beta_1 \\
\beta_2 \\
\beta_3 \\
\beta_4 \\
\beta_5
\end{pmatrix}_{\mathbf{5}} &=&
\begin{pmatrix}
\sqrt{3} \alpha_1 \beta_1 + \alpha_2 \beta_4 + \alpha_3 \beta_3 \\
\alpha_1 \beta_2 - \sqrt{2} \alpha_2 \beta_5 - \sqrt{2} \alpha_3 \beta_4  \\
\alpha_1 \beta_5 - \sqrt{2} \alpha_2 \beta_3 - \sqrt{2} \alpha_3 \beta_2
\end{pmatrix}_{\mathbf{3}} \oplus
\begin{pmatrix}
- 2 \alpha_1 \beta_1 + \sqrt{3} \alpha_2 \beta_4 + \sqrt{3} \alpha_3 \beta_3   \\
\sqrt{3} \alpha_1 \beta_3 + \alpha_2 \beta_1 - \sqrt{6} \alpha_3 \beta_5 \\
\sqrt{3} \alpha_1 \beta_4 - \sqrt{6} \alpha_2 \beta_2 + \alpha_3 \beta_1
\end{pmatrix}_{\mathbf{3}'}  \\ &\oplus & \nonumber
\begin{pmatrix}
\sqrt{2} \alpha_1 \beta_2 + 3 \alpha_2 \beta_5 - 2 \alpha_3 \beta_4 \\
2\sqrt{2} \alpha_1 \beta_3 - \sqrt{6} \alpha_2 \beta_1 + \alpha_3 \beta_5 \\
- 2\sqrt{2} \alpha_1 \beta_4 - \alpha_2 \beta_2 + \sqrt{6} \alpha_3 \beta_1   \\
- \sqrt{2} \alpha_1 \beta_5 + 2 \alpha_2 \beta_3 - 3 \alpha_3 \beta_2
\end{pmatrix}_{\mathbf{4}}   \oplus
\begin{pmatrix}
\sqrt{3} \alpha_2 \beta_4 - \sqrt{3} \alpha_3 \beta_3 \\
2\alpha_1 \beta_2 + \sqrt{2} \alpha_3 \beta_4 \\
- \alpha_1 \beta_3 - \sqrt{3} \alpha_2 \beta_1  - \sqrt{2} \alpha_3 \beta_5 \\
\alpha_1 \beta_4 + \sqrt{2} \alpha_2 \beta_2 + \sqrt{3} \alpha_3 \beta_1  \\
- 2 \alpha_1 \beta_5 - \sqrt{2} \alpha_2 \beta_3
\end{pmatrix}_{\mathbf{5}}  \,, \\
\begin{pmatrix}
\alpha_1 \\
\alpha_2 \\
\alpha_3 \\
\alpha_4
\end{pmatrix}_{\mathbf{4}}\otimes
\begin{pmatrix}
\beta_1 \\
\beta_2 \\
\beta_3 \\
\beta_4
\end{pmatrix}_{\mathbf{4}} &=& (\alpha_1 \beta_4 + \alpha_2 \beta_3 + \alpha_3 \beta_2 + \alpha_4 \beta_1)_{\mathbf{1}} \oplus
\begin{pmatrix}
-\alpha_1 \beta_4 + \alpha_2 \beta_3 - \alpha_3 \beta_2+\alpha_4 \beta_1 \\
\sqrt{2} \alpha_2 \beta_4 - \sqrt{2} \alpha_4 \beta_2 \\
\sqrt{2} \alpha_1 \beta_3 - \sqrt{2} \alpha_3 \beta_1
\end{pmatrix}_{\mathbf{3}} \\ &\oplus & \nonumber
\begin{pmatrix}
 \alpha_1 \beta_4 +\alpha_2 \beta_3- \alpha_3 \beta_2- \alpha_4 \beta_1 \\
\sqrt{2} \alpha_3 \beta_4 - \sqrt{2} \alpha_4 \beta_3 \\
\sqrt{2} \alpha_1 \beta_2 - \sqrt{2} \alpha_2 \beta_1
\end{pmatrix}_{\mathbf{3}'}  \oplus
\begin{pmatrix}
\alpha_2 \beta_4+\alpha_3 \beta_3 + \alpha_4 \beta_2  \\
\alpha_1 \beta_1 + \alpha_3 \beta_4+ \alpha_4 \beta_3  \\
\alpha_1 \beta_2+ \alpha_2 \beta_1 +\alpha_4 \beta_4 \\
\alpha_1 \beta_3+\alpha_2 \beta_2 + \alpha_3 \beta_1
\end{pmatrix}_{\mathbf{4}}  \\ &\oplus & \nonumber
\begin{pmatrix}
\sqrt{3} \alpha_1 \beta_4- \sqrt{3} \alpha_2 \beta_3 - \sqrt{3} \alpha_3 \beta_2+\sqrt{3} \alpha_4 \beta_1 \\
-\sqrt{2} \alpha_2 \beta_4 +2\sqrt{2} \alpha_3 \beta_3 - \sqrt{2} \alpha_4 \beta_2\\
- 2\sqrt{2} \alpha_1 \beta_1+ \sqrt{2} \alpha_3 \beta_4+\sqrt{2} \alpha_4 \beta_3\\
\sqrt{2} \alpha_1 \beta_2 +\sqrt{2} \alpha_2 \beta_1- 2\sqrt{2} \alpha_4 \beta_4 \\
-\sqrt{2} \alpha_1 \beta_3+2\sqrt{2} \alpha_2 \beta_2 - \sqrt{2} \alpha_3 \beta_1
\end{pmatrix}_{\mathbf{5}} \,, \\
\begin{pmatrix}
\alpha_1 \\
\alpha_2 \\
\alpha_3 \\
\alpha_4
\end{pmatrix}_{\mathbf{4}}\otimes
\begin{pmatrix}
\beta_1 \\
\beta_2 \\
\beta_3 \\
\beta_4 \\
\beta_5
\end{pmatrix}_{\mathbf{5}} &=&
\begin{pmatrix}
2\sqrt{2} \alpha_1 \beta_5 - \sqrt{2} \alpha_2 \beta_4 + \sqrt{2} \alpha_3 \beta_3 - 2\sqrt{2} \alpha_4 \beta_2  \\
- \sqrt{6} \alpha_1 \beta_1 + 2 \alpha_2 \beta_5 + 3 \alpha_3 \beta_4 - \alpha_4 \beta_3  \\
\alpha_1 \beta_4 - 3 \alpha_2 \beta_3 - 2 \alpha_3 \beta_2 + \sqrt{6} \alpha_4 \beta_1
\end{pmatrix}_{\mathbf{3}}  \\  &\oplus & \nonumber
\begin{pmatrix}
\sqrt{2} \alpha_1 \beta_5 + 2 \sqrt{2} \alpha_2 \beta_4 - 2\sqrt{2} \alpha_3 \beta_3 - \sqrt{2} \alpha_4 \beta_2 \\
3 \alpha_1 \beta_2 - \sqrt{6} \alpha_2 \beta_1 - \alpha_3 \beta_5 + 2 \alpha_4 \beta_4 \\
- 2 \alpha_1 \beta_3 + \alpha_2 \beta_2 + \sqrt{6} \alpha_3 \beta_1 - 3 \alpha_4 \beta_5
\end{pmatrix}_{\mathbf{3}'}   \\  &\oplus & \nonumber
\begin{pmatrix}
\sqrt{3} \alpha_1 \beta_1 - \sqrt{2} \alpha_2 \beta_5 + \sqrt{2} \alpha_3 \beta_4 - 2\sqrt{2} \alpha_4 \beta_3  \\
- \sqrt{2} \alpha_1 \beta_2 - \sqrt{3} \alpha_2 \beta_1 + 2\sqrt{2} \alpha_3 \beta_5 + \sqrt{2} \alpha_4 \beta_4  \\
\sqrt{2} \alpha_1 \beta_3 + 2\sqrt{2} \alpha_2 \beta_2  - \sqrt{3} \alpha_3 \beta_1 - \sqrt{2} \alpha_4 \beta_5 \\
- 2\sqrt{2} \alpha_1 \beta_4 + \sqrt{2} \alpha_2 \beta_3 - \sqrt{2} \alpha_3 \beta_2 + \sqrt{3} \alpha_4 \beta_1
\end{pmatrix}_{\mathbf{4}}  \\  &\oplus & \nonumber
\begin{pmatrix}
\sqrt{2} \alpha_1 \beta_5 - \sqrt{2} \alpha_2 \beta_4 - \sqrt{2} \alpha_3 \beta_3 + \sqrt{2} \alpha_4 \beta_2 \\
-\sqrt{2} \alpha_1 \beta_1 - \sqrt{3} \alpha_3 \beta_4 - \sqrt{3} \alpha_4 \beta_3 \\
\sqrt{3} \alpha_1 \beta_2 + \sqrt{2} \alpha_2 \beta_1 + \sqrt{3} \alpha_3 \beta_5  \\
\sqrt{3} \alpha_2 \beta_2 + \sqrt{2} \alpha_3 \beta_1  + \sqrt{3} \alpha_4 \beta_5 \\
- \sqrt{3} \alpha_1 \beta_4 - \sqrt{3} \alpha_2 \beta_3 - \sqrt{2} \alpha_4 \beta_1
\end{pmatrix}_{\mathbf{5}}  \\  &\oplus & \nonumber
\begin{pmatrix}
2 \alpha_1 \beta_5 + 4 \alpha_2 \beta_4 + 4 \alpha_3 \beta_3 + 2 \alpha_4 \beta_2  \\
4 \alpha_1 \beta_1 + 2\sqrt{6} \alpha_2 \beta_5 \\
- \sqrt{6} \alpha_1 \beta_2 + 2 \alpha_2 \beta_1 - \sqrt{6} \alpha_3 \beta_5 + 2\sqrt{6} \alpha_4 \beta_4   \\
2\sqrt{6} \alpha_1 \beta_3 - \sqrt{6} \alpha_2 \beta_2 + 2 \alpha_3 \beta_1 - \sqrt{6} \alpha_4 \beta_5 \\
2\sqrt{6} \alpha_3 \beta_2 + 4 \alpha_4 \beta_1
\end{pmatrix}_{\mathbf{5}} \,, \\
\begin{pmatrix}
\alpha_1 \\
\alpha_2 \\
\alpha_3 \\
\alpha_4 \\
\alpha_5
\end{pmatrix}_{\mathbf{5}}\otimes
\begin{pmatrix}
\beta_1 \\
\beta_2 \\
\beta_3 \\
\beta_4 \\
\beta_5
\end{pmatrix}_{\mathbf{5}} &=&
(\alpha_1 \beta_1 + \alpha_2 \beta_5 + \alpha_3 \beta_4 + \alpha_4 \beta_3 + \alpha_5 \beta_2)_{\mathbf{1}}  \\  &\oplus & \nonumber
\begin{pmatrix}
\alpha_2 \beta_5 + 2\alpha_3 \beta_4 - 2 \alpha_4 \beta_3  - \alpha_5 \beta_2 \\
- \sqrt{3} \alpha_1 \beta_2 + \sqrt{3} \alpha_2 \beta_1 + \sqrt{2} \alpha_3 \beta_5 - \sqrt{2} \alpha_5 \beta_3 \\
\sqrt{3} \alpha_1 \beta_5 + \sqrt{2} \alpha_2 \beta_4 - \sqrt{2} \alpha_4 \beta_2 - \sqrt{3} \alpha_5 \beta_1
\end{pmatrix}_{\mathbf{3}}  \\  &\oplus & \nonumber
\begin{pmatrix}
2 \alpha_2 \beta_5 - \alpha_3 \beta_4 + \alpha_4 \beta_3 - 2 \alpha_5 \beta_2 \\
\sqrt{3} \alpha_1 \beta_3 - \sqrt{3} \alpha_3 \beta_1 + \sqrt{2} \alpha_4 \beta_5 - \sqrt{2} \alpha_5 \beta_4 \\
- \sqrt{3} \alpha_1 \beta_4 + \sqrt{2} \alpha_2 \beta_3 - \sqrt{2} \alpha_3 \beta_2 + \sqrt{3} \alpha_4 \beta_1
\end{pmatrix}_{\mathbf{3}'}  \\  &\oplus & \nonumber
\begin{pmatrix}
\sqrt{6} \alpha_1 \beta_2 + \sqrt{6} \alpha_2 \beta_1 - \alpha_3 \beta_5 + 4 \alpha_4 \beta_4 - \alpha_5 \beta_3 \\
+ \sqrt{6} \alpha_1 \beta_3 + 4 \alpha_2 \beta_2 + \sqrt{6} \alpha_3 \beta_1 - \alpha_4 \beta_5 - \alpha_5 \beta_4 \\
+ \sqrt{6} \alpha_1 \beta_4 - \alpha_2 \beta_3 - \alpha_3 \beta_2 + \sqrt{6} \alpha_4 \beta_1 + 4 \alpha_5 \beta_5 \\
+ \sqrt{6} \alpha_1 \beta_5 - \alpha_2 \beta_4 + 4 \alpha_3 \beta_3 - \alpha_4 \beta_2 + \sqrt{6} \alpha_5  \beta_1
\end{pmatrix}_{\mathbf{4}}   \\  &\oplus & \nonumber
\begin{pmatrix}
\sqrt{2} \alpha_1 \beta_2 - \sqrt{2} \alpha_2 \beta_1 + \sqrt{3} \alpha_3 \beta_5 - \sqrt{3} \alpha_5 \beta_3 \\
- \sqrt{2} \alpha_1 \beta_3 + \sqrt{2} \alpha_3 \beta_1 + \sqrt{3} \alpha_4 \beta_5 - \sqrt{3} \alpha_5 \beta_4 \\
- \sqrt{2} \alpha_1 \beta_4 - \sqrt{3} \alpha_2 \beta_3 + \sqrt{3} \alpha_3 \beta_2 + \sqrt{2} \alpha_4 \beta_1 \\
\sqrt{2} \alpha_1 \beta_5 - \sqrt{3} \alpha_2 \beta_4 + \sqrt{3} \alpha_4 \beta_2 - \sqrt{2} \alpha_5 \beta_1
\end{pmatrix}_{\mathbf{4}}  \\  &\oplus & \nonumber
\begin{pmatrix}
2 \alpha_1 \beta_1 + \alpha_2 \beta_5 - 2\alpha_3 \beta_4 - 2 \alpha_4 \beta_3 + \alpha_5 \beta_2  \\
\alpha_1 \beta_2 + \alpha_2 \beta_1 + \sqrt{6} \alpha_3 \beta_5 + \sqrt{6} \alpha_5 \beta_3 \\
- 2 \alpha_1 \beta_3 + \sqrt{6} \alpha_2 \beta_2 - 2 \alpha_3 \beta_1  \\
- 2 \alpha_1 \beta_4 - 2 \alpha_4 \beta_1 + \sqrt{6} \alpha_5 \beta_5 \\
\alpha_1 \beta_5 + \sqrt{6} \alpha_2 \beta_4 + \sqrt{6} \alpha_4 \beta_2 + \alpha_5 \beta_1
\end{pmatrix}_{\mathbf{5}}   \\  &\oplus & \nonumber
\begin{pmatrix}
2 \alpha_1 \beta_1 - 2\alpha_2 \beta_5 + \alpha_3 \beta_4 + \alpha_4 \beta_3 - 2 \alpha_5 \beta_2  \\
- 2 \alpha_1 \beta_2 - 2 \alpha_2 \beta_1 + \sqrt{6} \alpha_4 \beta_4  \\
\alpha_1 \beta_3 + \alpha_3 \beta_1 + \sqrt{6} \alpha_4 \beta_5 + \sqrt{6} \alpha_5 \beta_4 \\
\alpha_1 \beta_4 + \sqrt{6} \alpha_2 \beta_3 + \sqrt{6} \alpha_3 \beta_2 + \alpha_4 \beta_1 \\
2 \alpha_1 \beta_5 + \sqrt{6} \alpha_3 \beta_3 - 2 \alpha_5 \beta_1  \\
\end{pmatrix}_{\mathbf{5}}\,.
\end{eqnarray*}
Modular forms of level 5 and weight $k$ form a linear space of dimension $5k+1$, and the explicit forms can be constructed through the Dedekind eta-function and Klein form as follow~\cite{schultz2015notes,Ding:2019xna}
\begin{eqnarray}
\nonumber \mathcal{M}_{k}(\Gamma(5))&=&\bigoplus_{a+b=5k,\,a,b\ge0} \mathbb{C} \frac{\eta^{15k}(5\tau)}{\eta^{3k}(\tau)} \mathfrak{k}^a_{\frac{1}{5},0}(5\tau)\mathfrak{k}^b_{\frac{2}{5},0}(5\tau) \\
&=&\bigoplus_{a+b=5k,\,a,b\ge0} \mathbb{C}
\left(\frac{\eta^{3}(5\tau)}{\eta^{3/5}(\tau)}\mathfrak{k}_{\frac{1}{5},0}(5\tau)\right)^a
\left(\frac{\eta^{3}(5\tau)}{\eta^{3/5}(\tau)}\mathfrak{k}_{\frac{2}{5},0}(5\tau)\right)^b\,,
\end{eqnarray}
where $\mathfrak{k}_{(r_1,r_2)}(\tau)$ is the Klein form defined as infinite product expansion
\begin{equation}\label{KleinForm}
\mathfrak{k}_{(r_1,r_2)}(\tau)=q^{(r_1-1)/2}_z(1-q_z)\prod_{n=1}^\infty(1-q^nq_z)(1-q^nq_z^{-1})(1-q^n)^{-2}\,,
\end{equation}
with $q_z=e^{2\pi iz}$ and $z=r_1\tau+r_2$. Hence $\mathcal{M}_{2k}(\Gamma(5))$ is spanned by polynomials of even degree $10k$ in two functions $F_1(\tau)$ and $F_2(\tau)$ defined as
\begin{equation}
F_1(\tau)=\frac{\eta^{3}(5\tau)}{\eta^{3/5}(\tau)}\mathfrak{k}_{\frac{2}{5},0}(5\tau)\,,~~~F_2(\tau)=\frac{\eta^{3}(5\tau)}{\eta^{3/5}(\tau)}\mathfrak{k}_{\frac{1}{5},0}(5\tau)\,,
\end{equation}
which turn out to be modular forms of weight $1/5$~\cite{Yao:2020zml}. $F_1(\tau)$ and $F_1(\tau)$ admit the following $q$-expansion~\cite{Yao:2020zml},
\begin{equation}
\label{eq:q-series_F1F2} F_1(\tau)\,\eta(\tau)^{3/5}=q^{1/40}\sum_{m\in \mathbb{Z}}(-1)^m q^{(5m^2+m)/2},~~~~F_2(\tau)\,\eta(\tau)^{3/5}=q^{9/40}\sum_{m\in \mathbb{Z}}(-1)^m q^{(5m^2+3m)/2}\,,
\end{equation}
The weight $2$ modular forms of level $5$ can be arranged into two triplets and a quintet of $A_5$~\cite{Novichkov:2018nkm,Ding:2019xna}:
\begin{eqnarray}
\nonumber Y^{(2)}_{\mathbf{3}}(\tau)&=&
\begin{pmatrix}
F_1^{10}-36 F_2^5 F_1^5-F_2^{10} \\
 5 \sqrt{2} F_1^4 F_2 \left(F_1^5-3 F_2^5\right) \\
 5 \sqrt{2} F_1 F_2^4 \left(3 F_1^5+F_2^5\right) \\
\end{pmatrix}\equiv \begin{pmatrix}
Y_1(\tau) \\ Y_2(\tau) \\ Y_3(\tau)
\end{pmatrix}\,, \\
\nonumber Y^{(2)}_{\mathbf{3'}}(\tau) &=&
\begin{pmatrix}
F_1^{10}+14 F_2^5 F_1^5-F_2^{10} \\
 -5 \sqrt{2} F_1^3 F_2^2 \left(F_1^5+2 F_2^5\right) \\
 -5 \sqrt{2} \left(2 F_1^7 F_2^3-F_1^2 F_2^8\right) \\
\end{pmatrix}\equiv
\begin{pmatrix}
Y'_1(\tau) \\ Y'_2(\tau) \\ Y'_3(\tau)
\end{pmatrix}\,, \\
Y^{(2)}_{\mathbf{5}}(\tau) &=&\begin{pmatrix}
F_1^{10}+F_2^{10} \\
-\sqrt{6}\, F_1^4 F_2 \left(F_1^5+7 F_2^5\right) \\
-\sqrt{6} F_1^3 F_2^2 \left(3 F_1^5-4 F_2^5\right) \\
-\sqrt{6}\left(4 F_1^7 F_2^3+3 F_1^2 F_2^8\right) \\
-\sqrt{6} F_1 F_2^4 \left(7 F_1^5-F_2^5\right) \\
\end{pmatrix}\equiv \begin{pmatrix}
\widetilde{Y}_1(\tau) \\ \widetilde{Y}_2(\tau) \\ \widetilde{Y}_3(\tau) \\ \widetilde{Y}_4(\tau) \\ \widetilde{Y}_5(\tau)
\end{pmatrix}\,.
\end{eqnarray}
The $q$-expansions of the weight 2 and level 5 modular forms are given by
\begin{eqnarray}
\nonumber Y^{(2)}_{\mathbf{3}}&=&
\begin{pmatrix}
1 - 30 q - 20 q^2 - 40 q^3 - 90 q^4 + \cdots \\
5\sqrt{2} q^{1/5} \left( 1 + 2 q + 12 q^2 + 11 q^3 + 12 q^4 + \cdots \right) \\
5\sqrt{2} q^{4/5} \left( 3 + 7 q + 6 q^2 + 20 q^3 + 10 q^4 + \cdots \right)
\end{pmatrix}\,,  \\
\nonumber Y^{(2)}_{\mathbf{3}'}&=&
\begin{pmatrix}
1 + 20 q + 30 q^2 + 60 q^3 + 60 q^4 + \cdots \\
-5\sqrt{2} q^{2/5} \left( 1 + 6 q + 6 q^2 + 16 q^3 + 12 q^4 + \cdots \right) \\
-5\sqrt{2} q^{3/5} \left( 2 + 5 q + 12 q^2 + 7 q^3 + 22 q^4 + \cdots \right)
\end{pmatrix} \,, \\
Y^{(2)}_{\mathbf{5}}&=&
\begin{pmatrix}
1 + 6 q + 18 q^2 + 24 q^3 + 42 q^4 \\
-\sqrt{6}q^{1/5} \left( 1 + 12 q + 12 q^2 + 31 q^3 + 32 q^4 + \cdots \right) \\
-\sqrt{6}q^{2/5} \left( 3 + 8 q + 28 q^2 + 18 q^3 + 36 q^4 + \cdots \right) \\
-\sqrt{6}q^{3/5} \left( 4 + 15 q + 14 q^2 + 39 q^3 + 24 q^4 + \cdots \right) \\
-\sqrt{6}q^{4/5} \left( 7 + 13 q + 24 q^2 + 20 q^3 + 60 q^4 + \cdots \right)
\end{pmatrix}\,.
\end{eqnarray}
Besides the above weight 2 modular multiplets $Y^{(2)}_{\mathbf{3}}$, $Y^{(2)}_{\mathbf{3}'}$ and $Y^{(2)}_{\mathbf{5}}$, the weight 2 polyharmonic Maa{\ss} form at level 5 includes the modified Eisenstein series $\widehat{E}_2(\tau)$ which is a singlet of $A_5$.

The weight $0$ polyharmonic Maa{\ss} forms can be arranged into two triplets $Y^{(0)}_{\mathbf{3}}$, $Y^{(0)}_{\mathbf{3}'}$ and a quintet $Y^{(0)}_{\mathbf{5}}$ of $A_5$ besides the weight 0 modular form $Y^{(0)}_{\mathbf{1}}=1$. The master equation of Eq.~\eqref{eq:lift-cond} for the weight 0 polyharmonic Maa{\ss} form of level $5$ is of the following form
\begin{eqnarray}
\nonumber \xi_0(Y^{(0)}_{\mathbf{3}})&=& \Omega_{\mathbf{3}}  Y^{(2)}_{\mathbf{3}},~~~D(Y^{(0)}_{\mathbf{3}}) = -\dfrac{1}{4\pi} Y^{(2)}_{\mathbf{3}} \,,  \\
\nonumber \xi_0(Y^{(0)}_{\mathbf{3}'})&=& \Omega_{\mathbf{3}'}  Y^{(2)}_{\mathbf{3}'},~~~D(Y^{(0)}_{\mathbf{3}'}) = -\dfrac{1}{4\pi} Y^{(2)}_{\mathbf{3}'} \,, \\
\xi_0(Y^{(0)}_{\mathbf{5}})&=& \Omega_{\mathbf{5}} Y^{(2)}_{\mathbf{5}},~~~D(Y^{(0)}_{\mathbf{5}}) = -\dfrac{1}{4\pi} Y^{(2)}_{\mathbf{5}}\,,
\end{eqnarray}
where the unitary transformations $\Omega_{\mathbf{3}}$, $\Omega_{\mathbf{3}'}$ and $\Omega_{\mathbf{5}}$ are permutation matrices
\begin{eqnarray}
\Omega_{\mathbf{3}}=\Omega_{\mathbf{3}'}=
\begin{pmatrix}
1 ~&~ 0 ~&~ 0 \\
0 ~&~ 0 ~&~ 1 \\
0 ~&~ 1 ~&~ 0
\end{pmatrix}\,,~~~
\Omega_{\mathbf{5}}=
\begin{pmatrix}
1 ~&~ 0 ~&~ 0 ~&~ 0 ~&~ 0 \\
0 ~&~ 0 ~&~ 0 ~&~ 0 ~&~ 1 \\
0 ~&~ 0 ~&~ 0 ~&~ 1 ~&~ 0 \\
0 ~&~ 0 ~&~ 1 ~&~ 0 ~&~ 0 \\
0 ~&~ 1 ~&~ 0 ~&~ 0 ~&~ 0
\end{pmatrix}\,.
\end{eqnarray}
The expression of weight $0$ polyharmonic Maa{\ss} forms are determined to be
\begin{eqnarray}
\nonumber Y_{\mathbf{3},1}^{(0)} &=& y + \dfrac{15 }{2 \pi}\left( \dfrac{e^{-4 \pi y}}{ q}+ \dfrac{e^{-8 \pi \
y}}{3 q^2}+ \dfrac{4\, e^{-12 \pi y}}{9 q^3}+ \dfrac{3\, e^{-16 \pi y}}{4 q^4} + \cdots  \right)  \\
\nonumber &&+\dfrac{5\sqrt{5}\log \phi}{2\pi} + \dfrac{15 }{2 \pi}\left( q + \dfrac{ q^2}{3}+ \dfrac{4 q^3}{9}+ \dfrac{3 q^4}{4}+ \dfrac{13 q^5}{15} + \cdots \right) \,,   \\
\nonumber Y_{\mathbf{3},2}^{(0)} &=& - \dfrac{75 q^{1/5}}{8 \sqrt{2} \pi}\left( \dfrac{e^{-16 \pi y/5}}{ q}+ \dfrac{28\, e^{-36 \pi y/5}}{27 q^2}+ \dfrac{4\, e^{-56 \pi y/5}}{7 q^3}+ \dfrac{80\, e^{-76 \pi y/5}}{57 q^4} + \cdots  \right)   \\
\nonumber &&- \dfrac{25 q^{1/5}}{2 \sqrt{2} \pi}\left( 1 + \dfrac{ q}{3}+ \dfrac{12 q^2}{11}+ \dfrac{11 q^3}{16}+ \dfrac{4 q^4}{7}+ \dfrac{6 q^5}{13} + \cdots \right) \,,   \\
\nonumber Y_{\mathbf{3},3}^{(0)} &=& - \dfrac{25 q^{4/5}}{2 \sqrt{2} \pi}\left( \dfrac{e^{-4 \pi y/5}}{ q}+ \dfrac{e^{-24 \pi y/5}}{3 q^2}+ \dfrac{12\, e^{-44 \pi y/5}}{11 q^3}+ \dfrac{11\, e^{-64 \pi y/5}}{16 q^4} + \cdots  \right)  \\
\nonumber &&- \dfrac{75 q^{4/5}}{8 \sqrt{2} \pi}\left( 1 + \dfrac{28 q}{27}+ \dfrac{4 q^2}{7}+ \dfrac{80 q^3}{57}+ \dfrac{5 q^4}{9}+ \dfrac{40 q^5}{29} + \cdots \right)\,,  \\
\nonumber Y_{\mathbf{3}',1}^{(0)} &=& y - \dfrac{5 }{\pi}\left( \dfrac{e^{-4 \pi y}}{ q}+ \dfrac{3\, e^{-8 \pi y}}{4 q^2}+ \dfrac{e^{-12 \pi y}}{ q^3}+ \dfrac{3\, e^{-16 \pi y}}{4 q^4} + \cdots  \right)   \\
\nonumber &&-\dfrac{5\sqrt{5}\log \phi}{2\pi}- \dfrac{5 }{\pi}\left( q + \dfrac{3 q^2}{4}+ \dfrac{q^3}{1}+ \dfrac{3 q^4}{4}+ \dfrac{6 q^5}{5} + \cdots \right)  \,, \\
\nonumber Y_{\mathbf{3}',2}^{(0)} &=& \dfrac{25 q^{2/5}}{3 \sqrt{2} \pi}\left( \dfrac{e^{-12 \pi y/5}}{ q}+ \dfrac{15\, e^{-32 \pi y/5}}{16 q^2}+ \dfrac{18\, e^{-52 \pi y/5}}{13 q^3}+ \dfrac{7\, e^{-72 \pi y/5}}{12 q^4} + \cdots  \right)   \\
\nonumber && + \dfrac{25 q^{2/5}}{4 \sqrt{2} \pi}\left( 1 + \dfrac{12 q}{7}+ \dfrac{ q^2}{1}+ \dfrac{32 q^3}{17}+ \dfrac{12 q^4}{11}+ \dfrac{40 q^5}{27} + \cdots \right) \,, \\
\nonumber Y_{\mathbf{3}',3}^{(0)} &=& \dfrac{25 q^{3/5}}{4 \sqrt{2} \pi}\left( \dfrac{e^{-8 \pi y/5}}{ q}+ \dfrac{12\, e^{-28 \pi y/5}}{7 q^2}+ \dfrac{e^{-48 \pi y/5}}{ q^3}+ \dfrac{32\, e^{-68 \pi y/5}}{17 q^4} + \cdots  \right)  \\
\nonumber && + \dfrac{25 q^{3/5}}{3 \sqrt{2} \pi}\left( 1 + \dfrac{15 q}{16}+ \dfrac{18 q^2}{13}+ \dfrac{7 q^3}{12}+ \dfrac{33 q^4}{23}+ \dfrac{27 q^5}{28} + \cdots \right)\,,  \\
\nonumber Y_{\mathbf{5},1}^{(0)} &=& y  - \dfrac{3 }{2 \pi}\left( \dfrac{e^{-4 \pi y}}{ q}+ \dfrac{3\, e^{-8 \pi y}}{2 q^2}+ \dfrac{4\, e^{-12 \pi y}}{3 q^3}+ \dfrac{7\, e^{-16 \pi y}}{4 q^4} + \cdots  \right) \\
\nonumber &&-\dfrac{5\log 5}{4\pi} - \dfrac{3 }{2 \pi}\left( q + \dfrac{3 q^2}{2}+ \dfrac{4 q^3}{3}+ \dfrac{7 q^4}{4}+ \dfrac{ q^5}{5} + \cdots \right)\,,   \\
\nonumber Y_{\mathbf{5},2}^{(0)} &=& \dfrac{35 \sqrt{3} q^{1/5}}{8 \sqrt{2} \pi}\left( \dfrac{e^{-16 \pi y/5}}{ q}+ \dfrac{52\, e^{-36 \pi y/5}}{63 q^2}+ \dfrac{48\, e^{-56 \pi y/5}}{49 q^3}+ \dfrac{80\, e^{-76 \pi y/5}}{133 q^4} + \cdots  \right)   \\
\nonumber && + \dfrac{5 \sqrt{3} q^{1/5}}{2 \sqrt{2} \pi}\left( 1 + \dfrac{2 q}{1}+ \dfrac{12 q^2}{11}+ \dfrac{31 q^3}{16}+ \dfrac{32 q^4}{21}+ \dfrac{21 q^5}{13} + \cdots \right) \,, \\
\nonumber Y_{\mathbf{5},3}^{(0)} &=& \dfrac{5 \sqrt{2} q^{2/5}}{\sqrt{3} \pi}\left( \dfrac{e^{-12 \pi y/5}}{ q}+ \dfrac{45\, e^{-32 \pi y/5}}{32 q^2}+ \dfrac{21\, e^{-52 \pi y/5}}{26 q^3}+ \dfrac{13\, e^{-72 \pi y/5}}{8 q^4} + \cdots  \right)  \\
\nonumber && + \dfrac{15 \sqrt{3} q^{2/5}}{4 \sqrt{2} \pi}\left( 1 + \dfrac{16 q}{21}+ \dfrac{14 q^2}{9}+ \dfrac{12 q^3}{17}+ \dfrac{12 q^4}{11}+ \dfrac{80 q^5}{81} + \cdots \right)\,,  \\
\nonumber Y_{\mathbf{5},4}^{(0)} &=& \dfrac{15 \sqrt{3} q^{3/5}}{4 \sqrt{2} \pi}\left( \dfrac{e^{-8 \pi y/5}}{ q}+ \dfrac{16\, e^{-28 \pi y/5}}{21 q^2}+ \dfrac{14\, e^{-48 \pi y/5}}{9 q^3}+ \dfrac{12\, e^{-68 \pi y/5}}{17 q^4} + \cdots  \right)  \\
\nonumber&& + \dfrac{5 \sqrt{2} q^{3/5}}{\sqrt{3} \pi}\left( 1 + \dfrac{45 q}{32}+ \dfrac{21 q^2}{26}+ \dfrac{13 q^3}{8}+ \dfrac{18 q^4}{23}+ \dfrac{3 q^5}{2} + \cdots \right) \,, \\
\nonumber Y_{\mathbf{5},5}^{(0)} &=& \dfrac{5 \sqrt{3} q^{4/5}}{2 \sqrt{2} \pi}\left( \dfrac{e^{-4 \pi y/5}}{ q}+ \dfrac{2\, e^{-24 \pi y/5}}{ q^2}+ \dfrac{12\, e^{-44 \pi y/5}}{11 q^3}+ \dfrac{31\, e^{-64 \pi y/5}}{16 q^4} + \cdots  \right)  \\
&& + \dfrac{35 \sqrt{3} q^{4/5}}{8 \sqrt{2} \pi}\left( 1 + \dfrac{52 q}{63}+ \dfrac{48 q^2}{49}+ \dfrac{80 q^3}{133}+ \dfrac{10 q^4}{7}+ \dfrac{120 q^5}{203} + \cdots \right)\,,
\end{eqnarray}
The tensor products of weight 2 modular multiplets $Y^{(2)}_{\mathbf{3}}$, $Y^{(2)}_{\mathbf{3}'}$, $Y^{(2)}_{\mathbf{5}}$ give rise to the weight $4$ modular forms of level $5$,
\begin{eqnarray}
\nonumber Y^{(4)}_{\mathbf{1}}&=& (Y^{(2)}_{\mathbf{3}}Y^{(2)}_{\mathbf{3}})_{\mathbf{1}} = Y_1^2 + 2 Y_2 Y_3  \,, \\
\nonumber Y^{(4)}_{\mathbf{3}}&=& -\dfrac{1}{2} (Y^{(2)}_{\mathbf{3}}Y^{(2)}_{\mathbf{5}})_{\mathbf{3}} = \dfrac{1}{2}
\begin{pmatrix}
2 Y_1 \widetilde{Y}_1 - \sqrt{3} (Y_2 \widetilde{Y}_5 + Y_3 \widetilde{Y}_2 ) \\
\sqrt{6} Y_3 \widetilde{Y}_3 - \sqrt{3} Y_1 \widetilde{Y}_2 -  Y_2 \widetilde{Y}_1  \\
\sqrt{6} Y_2 \widetilde{Y}_4 - \sqrt{3} Y_1 \widetilde{Y}_5 -  Y_3 \widetilde{Y}_1
\end{pmatrix} \,, \\
\nonumber Y^{(4)}_{\mathbf{3}'}&=&\dfrac{1}{\sqrt{3}} (Y^{(2)}_{\mathbf{3}}Y^{(2)}_{\mathbf{5}})_{\mathbf{3}'} = \dfrac{1}{3}
\begin{pmatrix}
3 Y_1 \widetilde{Y}_1 + \sqrt{3} ( Y_2 \widetilde{Y}_5 + Y_3 \widetilde{Y}_2 ) \\
\sqrt{3} Y_1 \widetilde{Y}_3 - \sqrt{6} (Y_2 \widetilde{Y}_2 + Y_3 \widetilde{Y}_4) \\
\sqrt{3} Y_1 \widetilde{Y}_4 - \sqrt{6} (Y_2 \widetilde{Y}_3 + Y_3 \widetilde{Y}_5)
\end{pmatrix} \,, \\
\nonumber Y^{(4)}_{\mathbf{4}}&=& (Y^{(2)}_{\mathbf{3}}Y^{(2)}_{\mathbf{3}'})_{\mathbf{4}} =
\begin{pmatrix}
\sqrt{2} Y_2 Y'_1 + Y_3 Y'_2 \\
-\sqrt{2} Y_1 Y'_2 - Y_3 Y'_3 \\
-\sqrt{2} Y_1 Y'_3 - Y_2 Y'_2 \\
\sqrt{2} Y_3 Y'_1 + Y_2 Y'_3
\end{pmatrix}\,, \\
\nonumber Y^{(4)}_{\mathbf{5}I}&=& \dfrac{1}{2} (Y^{(2)}_{\mathbf{3}}Y^{(2)}_{\mathbf{3}})_{\mathbf{5}} = \dfrac{1}{2}
\begin{pmatrix}
2 (Y_1^2 - Y_2 Y_3)  \\
-2\sqrt{3} Y_1 Y_2  \\
\sqrt{6} Y_2^2 \\
\sqrt{6} Y_3^2 \\
-2\sqrt{3} Y_1 Y_3
\end{pmatrix}\,,  \\
 Y^{(4)}_{\mathbf{5}II} &=& \dfrac{1}{2} (Y^{(2)}_{\mathbf{3}'}Y^{(2)}_{\mathbf{3}'})_{\mathbf{5}} = \dfrac{1}{2}
\begin{pmatrix}
2 (Y'_1{}^2 - Y'_2 Y'_3)  \\
\sqrt{6} Y'_3{}^2 \\
-2\sqrt{3} Y'_1 Y'_2  \\
-2\sqrt{3} Y'_1 Y'_3  \\
\sqrt{6} Y'_2{}^2
\end{pmatrix} \,.
\end{eqnarray}
Notice that $Y^{(4)}_{\mathbf{4}}$ and $Y^{(4)}_{\mathbf{5}I}-Y^{(4)}_{\mathbf{5}II}$ are cusp forms which can not be lifted to polyharmonic Maa{\ss} forms. The other modular multipelts $Y^{(4)}_{\mathbf{1}}$, $Y^{(4)}_{\mathbf{3}}$, $Y^{(4)}_{\mathbf{3}'}$ and $Y^{(4)}_{\mathbf{5}} = -\dfrac{1}{13} Y^{(4)}_{\mathbf{5}I} + \dfrac{14}{13} Y^{(4)}_{\mathbf{5}II}$ can be lifted to weight $-2$ polyharmonic Maa{\ss} forms through  Eq.~\eqref{eq:lift-cond} as follow,
\begin{eqnarray}
\nonumber \xi_{-2}(Y_{\mathbf{1}}^{(-2)}) &=& Y^{(4)}_{\mathbf{1}}\,,~~~~D^3(Y_{\mathbf{1}}^{(-2)})=-\dfrac{2}{(4\pi)^3} Y^{(4)}_{\mathbf{1}}\,, \\
\nonumber \xi_{-2}(Y_{\mathbf{3}}^{(-2)}) &=& \Omega_{\mathbf{3}} Y^{(4)}_{\mathbf{3}}\,,~~~~D^3(Y_{\mathbf{2}}^{(-2)})=-\dfrac{2}{(4\pi)^3} Y^{(4)}_{\mathbf{3}} \,, \\
\nonumber \xi_{-2}(Y_{\mathbf{3}'}^{(-2)}) &=& \Omega_{\mathbf{3}'} Y^{(4)}_{\mathbf{3}'}\,,~~~~D^3(Y_{\mathbf{3}'}^{(-2)})=-\dfrac{2}{(4\pi)^3} Y^{(4)}_{\mathbf{3}'} \,, \\
\xi_{-2}(Y_{\mathbf{5}}^{(-2)}) &=& \Omega_{\mathbf{5}} Y^{(4)}_{\mathbf{5}}\,,~~~~D^3(Y_{\mathbf{5}}^{(-2)})=-\dfrac{2}{(4\pi)^3} Y^{(4)}_{\mathbf{5}}\,.
\end{eqnarray}
From Eq.~\eqref{eq:polyHarmonic-expr-main}, we find the weight $k=-2$ polyharmonic Maa{\ss} forms of level 5 are
\begin{small}
\begin{eqnarray}
\nonumber Y_{\mathbf{1}}^{(-2)}&=& \dfrac{y^3}{3} - \dfrac{15\Gamma(3,4\pi y)}{4\pi^3 q} - \dfrac{135\Gamma(3,8\pi y)}{32\pi^3 q^2} - \dfrac{35\Gamma(3,12\pi y)}{9\pi^3 q^3} + \cdots \\
\nonumber&&-\dfrac{\pi}{12}\dfrac{\zeta(3)}{\zeta(4)} - \dfrac{15 q}{2\pi^3} - \dfrac{135 q^2}{16\pi^3} - \dfrac{70 q^3}{q\pi^3} - \dfrac{1095 q^4}{128\pi^3} - \dfrac{189 q^5}{25\pi^3} - \dfrac{35 q^6}{4\pi^3} + \cdots \,, \\
\nonumber  Y_{\mathbf{3},1}^{(-2)} &=& \dfrac{y^3}{3} - \dfrac{63 }{32 \pi^3}\left( \dfrac{\Gamma(3, 4 \pi y)}{ q}+ \dfrac{31 \Gamma(3, 8 \pi y)}{36 q^2}+ \dfrac{1612 \Gamma(3, 12 \pi y)}{1701 q^3}+ \dfrac{57 \Gamma(3, 16 \pi y)}{64 q^4} + \cdots  \right) \\
\nonumber &&-\dfrac{\pi L(3,\chi_3)}{12\sqrt{5}L(4,\chi_3)} - \dfrac{63 }{16 \pi^3}\left( q + \dfrac{31 q^2}{36}+ \dfrac{1612 q^3}{1701}+ \dfrac{57 q^4}{64}+ \dfrac{7813 q^5}{7875} + \cdots \right) \,, \\
\nonumber  Y_{\mathbf{3},2}^{(-2)} &=& \dfrac{7125 q^{1/5}}{2048 \sqrt{2} \pi^3}\left( \dfrac{\Gamma(3, 16 \pi y/5)}{ q}+ \dfrac{2368 \Gamma(3, 36 \pi y/5)}{2187 q^2}+ \dfrac{48 \Gamma(3, 56 \pi y/5)}{49 q^3}+ \dfrac{439040 \Gamma(3, 76 \pi y/5)}{390963 q^4} + \cdots  \right) \\
\nonumber&& + \dfrac{125 q^{1/5}}{16 \sqrt{2} \pi^3}\left( 1 + \dfrac{91 q}{108}+ \dfrac{1332 q^2}{1331}+ \dfrac{3641 q^3}{4096}+ \dfrac{988 q^4}{1029}+ \dfrac{3843 q^5}{4394} + \cdots \right) \,, \\
\nonumber  Y_{\mathbf{3},3}^{(-2)} &=& \dfrac{125 q^{4/5}}{32 \sqrt{2} \pi^3}\left( \dfrac{\Gamma(3, 4 \pi y/5)}{ q}+ \dfrac{91 \Gamma(3, 24 \pi y/5)}{108 q^2}+ \dfrac{1332 \Gamma(3, 44 \pi y/5)}{1331 q^3}+ \dfrac{3641 \Gamma(3, 64 \pi y/5)}{4096 q^4} + \cdots  \right) \\
\nonumber&& + \dfrac{7125 q^{4/5}}{1024 \sqrt{2} \pi^3}\left( 1 + \dfrac{2368 q}{2187}+ \dfrac{48 q^2}{49}+ \dfrac{439040 q^3}{390963}+ \dfrac{5915 q^4}{6156}+ \dfrac{520320 q^5}{463391} + \cdots \right) \,, \\
\nonumber  Y_{\mathbf{3}',1}^{(-2)} &=& \dfrac{y^3}{3} + \dfrac{31 }{16 \pi^3}\left( \dfrac{\Gamma(3, 4 \pi y)}{ q}+ \dfrac{441 \Gamma(3, 8 \pi y)}{496 q^2}+ \dfrac{91 \Gamma(3, 12 \pi y)}{93 q^3}+ \dfrac{57 \Gamma(3, 16 \pi y)}{64 q^4} + \cdots  \right) \\
\nonumber &&+\dfrac{\pi L(3,\chi_3)}{12\sqrt{5}L(4,\chi_3)} + \dfrac{31 }{8 \pi^3}\left( q + \dfrac{441 q^2}{496}+ \dfrac{91 q^3}{93}+ \dfrac{57 q^4}{64}+ \dfrac{126 q^5}{125} + \cdots \right)  \,, \\
\nonumber  Y_{\mathbf{3}',2}^{(-2)} &=& - \dfrac{1625 q^{2/5}}{432 \sqrt{2} \pi^3}\left( \dfrac{\Gamma(3, 12 \pi y/5)}{ q}+ \dfrac{945 \Gamma(3, 32 \pi y/5)}{1024 q^2}+ \dfrac{29646 \Gamma(3, 52 \pi y/5)}{28561 q^3}+ \dfrac{4921 \Gamma(3, 72 \pi y/5)}{5616 q^4} + \cdots  \right) \\
\nonumber && - \dfrac{875 q^{2/5}}{128 \sqrt{2} \pi^3}\left( 1 + \dfrac{2736 q}{2401}+ \dfrac{247 q^2}{252}+ \dfrac{39296 q^3}{34391}+ \dfrac{1332 q^4}{1331}+ \dfrac{151840 q^5}{137781} + \cdots \right) \,, \\
\nonumber  Y_{\mathbf{3}',3}^{(-2)} &=& - \dfrac{875 q^{3/5}}{256 \sqrt{2} \pi^3}\left( \dfrac{\Gamma(3, 8 \pi y/5)}{ q}+ \dfrac{2736 \Gamma(3, 28 \pi y/5)}{2401 q^2}+ \dfrac{247 \Gamma(3, 48 \pi y/5)}{252 q^3}+ \dfrac{39296 \Gamma(3, 68 \pi y/5)}{34391 q^4} + \cdots  \right) \\
\nonumber && - \dfrac{1625 q^{3/5}}{216 \sqrt{2} \pi^3}\left( 1 + \dfrac{945 q}{1024}+ \dfrac{29646 q^2}{28561}+ \dfrac{4921 q^3}{5616}+ \dfrac{164241 q^4}{158171}+ \dfrac{263169 q^5}{285376} + \cdots \right) \,, \\
\nonumber  Y_{\mathbf{5},1}^{(-2)} &=& \dfrac{y^3}{3} + \dfrac{315 }{416 \pi^3}\left( \dfrac{\Gamma(3, 4 \pi y)}{ q}+ \dfrac{9 \Gamma(3, 8 \pi y)}{8 q^2}+ \dfrac{28 \Gamma(3, 12 \pi y)}{27 q^3}+ \dfrac{73 \Gamma(3, 16 \pi y)}{64 q^4} + \cdots  \right) \\
\nonumber&&+\dfrac{5\pi}{312} \dfrac{\zeta(3)}{\zeta(4)} + \dfrac{315 }{208 \pi^3}\left( q + \dfrac{9 q^2}{8}+ \dfrac{28 q^3}{27}+ \dfrac{73 q^4}{64}+ \dfrac{2521 q^5}{2625} + \cdots \right) \,, \\
\nonumber  Y_{\mathbf{5},2}^{(-2)} &=& - \dfrac{45625 \sqrt{3} q^{1/5}}{26624 \sqrt{2} \pi^3}\left(\dfrac{\Gamma(3, 16 \pi y/5)}{ q}+ \dfrac{48448 \Gamma(3, 36 \pi y/5)}{53217 q^2}+ \dfrac{24768 \Gamma(3, 56 \pi y/5)}{25039 q^3}+ \dfrac{439040 \Gamma(3, 76 \pi y/5)}{500707 q^4} + \cdots  \right) \\
\nonumber && - \dfrac{625 \sqrt{3} q^{1/5}}{208 \sqrt{2} \pi^3}\left( 1 + \dfrac{7 q}{6}+ \dfrac{1332 q^2}{1331}+ \dfrac{4681 q^3}{4096}+ \dfrac{1376 q^4}{1323}+ \dfrac{9891 q^5}{8788} + \cdots \right) \,,\\
\nonumber  Y_{\mathbf{5},3}^{(-2)} &=& - \dfrac{4375 q^{2/5}}{936 \sqrt{6} \pi^3}\left( \dfrac{\Gamma(3, 12 \pi y/5)}{ q}+ \dfrac{15795 \Gamma(3, 32 \pi y/5)}{14336 q^2}+ \dfrac{4239 \Gamma(3, 52 \pi y/5)}{4394 q^3}+ \dfrac{757 \Gamma(3, 72 \pi y/5)}{672 q^4} + \cdots  \right) \\
\nonumber&& - \dfrac{5625 \sqrt{3} q^{2/5}}{1664 \sqrt{2} \pi^3}\left( 1 + \dfrac{2752 q}{3087}+ \dfrac{511 q^2}{486}+ \dfrac{4368 q^3}{4913}+ \dfrac{1332 q^4}{1331}+ \dfrac{163520 q^5}{177147} + \cdots \right) \,, \\
\nonumber  Y_{\mathbf{5},4}^{(-2)} &=& - \dfrac{5625 \sqrt{3} q^{3/5}}{3328 \sqrt{2} \pi^3}\left( \dfrac{\Gamma(3, 8 \pi y/5)}{ q}+ \dfrac{2752 \Gamma(3, 28 \pi y/5)}{3087 q^2}+ \dfrac{511 \Gamma(3, 48 \pi y/5)}{486 q^3}+ \dfrac{4368 \Gamma(3, 68 \pi y/5)}{4913 q^4} + \cdots  \right) \\
\nonumber && - \dfrac{4375 q^{3/5}}{468 \sqrt{6} \pi^3}\left( 1 + \dfrac{15795 q}{14336}+ \dfrac{4239 q^2}{4394}+ \dfrac{757 q^3}{672}+ \dfrac{82134 q^4}{85169}+ \dfrac{84753 q^5}{76832} + \cdots \right) \,, \\
\nonumber  Y_{\mathbf{5},5}^{(-2)} &=& - \dfrac{625 \sqrt{3} q^{4/5}}{416 \sqrt{2} \pi^3}\left( \dfrac{\Gamma(3, 4 \pi y/5)}{ q}+ \dfrac{7 \Gamma(3, 24 \pi y/5)}{6 q^2}+ \dfrac{1332 \Gamma(3, 44 \pi y/5)}{1331 q^3}+ \dfrac{4681 \Gamma(3, 64 \pi y/5)}{4096 q^4} + \cdots  \right) \\
&&  - \dfrac{45625 \sqrt{3} q^{4/5}}{13312 \sqrt{2} \pi^3}\left( 1 + \dfrac{48448 q}{53217}+ \dfrac{24768 q^2}{25039}+ \dfrac{439040 q^3}{500707}+ \dfrac{455 q^4}{438}+ \dfrac{1560960 q^5}{1780397} + \cdots \right)\,,
\end{eqnarray}
\end{small}
where $L(k,\chi_3) = \sum_{n=1}^{\infty} n^{-k} \chi_3(n)$ is the $L$-function with a Dirichlet character $\chi_3$ modulo $5$, and the value of the character $\chi_3(n)$ is given by
\begin{eqnarray}
\chi_3(n) =  \begin{cases}
1 \,,~~&n\equiv \pm 1 \,~({\rm mod}\, 5)  \\
-1 \,,~~&n\equiv \pm 2 \,~({\rm mod}\, 5)  \\
0 \,,~~&n\equiv 0 \,~({\rm mod}\, 5)
\end{cases} \,.
\end{eqnarray}
One can easily calculate the $L$-function with the software \texttt{Wolfram Mathematica}.

The weight $6$ modular forms of level $5$ can be arranged into nine multiplets of $A_5$: $Y^{(6)}_{\mathbf{1}}$, $Y^{(6)}_{\mathbf{3}I}$, $Y^{(6)}_{\mathbf{3}II}$, $Y^{(6)}_{\mathbf{3}'I}$, $Y^{(6)}_{\mathbf{3}'II}$, $Y^{(6)}_{\mathbf{4}I}$, $Y^{(6)}_{\mathbf{4}II}$, $Y^{(6)}_{\mathbf{5}I}$ and $Y^{(6)}_{\mathbf{5}II}$ which are given by
\begin{eqnarray}
\nonumber Y^{(6)}_{\mathbf{1}} &=& (Y^{(2)}_{\mathbf{3}}Y^{(4)}_{\mathbf{3}})_{\mathbf{1}}
=Y_1^2 \widetilde{Y}_1 + \dfrac{\sqrt{6}}{2} Y_2^2 \widetilde{Y}_4 + \dfrac{\sqrt{6}}{2} Y_3^2 \widetilde{Y}_3-\sqrt{3}Y_1Y_2\widetilde{Y}_5 - \sqrt{3} Y_1 Y_3 \widetilde{Y}_2 - Y_2 Y_3 \widetilde{Y}_1 \,,  \\
\nonumber Y^{(6)}_{\mathbf{3}I} &=& (Y^{(2)}_{\mathbf{3}}Y^{(4)}_{\mathbf{1}})_{\mathbf{3}}
=(Y_1^2 + 2 Y_2 Y_3)\begin{pmatrix}
Y_1 \\ Y_2 \\ Y_3
\end{pmatrix}\,, \\
\nonumber Y^{(6)}_{\mathbf{3}II} &=& \dfrac{1}{\sqrt{3}} (Y^{(2)}_{\mathbf{3}'}Y^{(4)}_{\mathbf{5}II})_{\mathbf{3}}
= \dfrac{1}{2}
\begin{pmatrix}
2 (Y'_1{}^3 - 3 Y'_1 Y'_2 Y'_3) \\
3\sqrt{2}\, Y'_1 Y'_3{}^2 - 2 Y'_2{}^3  \\
3\sqrt{2}\, Y'_1 Y'_2{}^2 - 2 Y'_3{}^3
\end{pmatrix} \,, \\
\nonumber Y^{(6)}_{\mathbf{3}'I} &=& (Y^{(2)}_{\mathbf{3}'}Y^{(4)}_{\mathbf{1}})_{\mathbf{3}'}
=(Y_1^2 + 2 Y_2 Y_3) \begin{pmatrix}
Y'_1 \\ Y'_2 \\ Y'_3
\end{pmatrix} \,, \\
\nonumber Y^{(6)}_{\mathbf{3}'II} &=& \dfrac{1}{\sqrt{3}} (Y^{(2)}_{\mathbf{3}}Y^{(4)}_{\mathbf{5}I})_{\mathbf{3}'}
= \dfrac{1}{2}
\begin{pmatrix}
2 (Y_1^3 - 3 Y_1 Y_2 Y_3)  \\
3\sqrt{2} Y_1 Y_2^2 - 2 Y_3^3 \\
3\sqrt{2} Y_1 Y_3^2 - 2 Y_2^3
\end{pmatrix} \,, \\
\nonumber Y^{(6)}_{\mathbf{4}I} &=& \dfrac{2}{3\sqrt{6}} (Y^{(2)}_{\mathbf{3}}Y^{(5)}_{\mathbf{5}I})_{\mathbf{4}}
=\begin{pmatrix}
-2 Y_1^2 Y_2 + Y_2^2 Y_3 \\
-Y_3^3 - \sqrt{2} Y_1 Y_2^2 \\
Y_2^3 + \sqrt{2} Y_1 Y_3^2 \\
2 Y_1^2 Y_3 - Y_2 Y_3^2
\end{pmatrix} \,, \\
\nonumber Y^{(6)}_{\mathbf{4}II} &=& \dfrac{2}{3\sqrt{6}} (Y^{(2)}_{\mathbf{3}'}Y^{(5)}_{\mathbf{5}II})_{\mathbf{4}}
=\begin{pmatrix}
Y'_2{}^3 + \sqrt{2} Y'_1 Y'_3{}^2 \\
-2 Y'_1{}^2 Y'_2 + Y'_2{}^2 Y'_3 \\
2Y'_1{}^2 Y'_3 - Y'_2 Y'_3{}^2  \\
- Y'_3{}^3 - \sqrt{2} Y'_1 Y'_2{}^2
\end{pmatrix} \,,  \\
\nonumber Y^{(6)}_{\mathbf{5}I} &=& (Y^{(2)}_{\mathbf{5}}Y^{(4)}_{\mathbf{1}})_{\mathbf{5}}
=(Y_1^2 + 2 Y_2 Y_3) \begin{pmatrix}
\widetilde{Y}_1 \\ \widetilde{Y}_2 \\ \widetilde{Y}_3 \\ \widetilde{Y}_4 \\ \widetilde{Y}_5
\end{pmatrix} \,, \\
Y^{(6)}_{\mathbf{5}II} &=& (Y^{(2)}_{\mathbf{3}}Y^{(4)}_{\mathbf{4}})_{\mathbf{5}}
=\begin{pmatrix}
\sqrt{6} Y_2^2 Y'_3 - \sqrt{6} Y_3^2 Y'_2 \\
4 Y_1 Y_2 Y'_1 - 2 Y_3^2 Y'_3 \\
2 Y_1^2 Y'_2 + \sqrt{2} Y_2^2 Y'_1 - 2\sqrt{2} Y_1 Y_3 Y'_3 - 2 Y_2 Y_3 Y'_2 \\
-2 Y_1^2 Y'_3 + 2\sqrt{2} Y_1 Y_2 Y'_2 + 2 Y_2 Y_3 Y'_3 - \sqrt{2} Y_3^2 Y'_1 \\
2 Y_2^2 Y'_2 - 4 Y_1 Y_3 Y'_1
\end{pmatrix}\,.
\end{eqnarray}
We can find that $Y^{(6)}_{\mathbf{3}I}-Y^{(6)}_{\mathbf{3}II}$, $Y^{(6)}_{\mathbf{3}'I}-Y^{(6)}_{\mathbf{3}'II}$, $Y^{(6)}_{\mathbf{4}I}$, $Y^{(6)}_{\mathbf{4}II}$ and $Y^{(6)}_{\mathbf{5}II}$ are cusp forms of level 5. Only the modular multiplets $Y^{(6)}_{\mathbf{1}}$, $Y^{(6)}_{\mathbf{3}}=\dfrac{1}{67}Y^{(6)}_{\mathbf{3}I} + \dfrac{66}{67}Y^{(6)}_{\mathbf{3}II}$, $Y^{(6)}_{\mathbf{3}'}=\dfrac{259}{268} Y^{(6)}_{\mathbf{3}'I} + \dfrac{9}{268} Y^{(6)}_{\mathbf{3}'II}$ and $Y^{(6)}_{\mathbf{5}}=Y^{(6)}_{\mathbf{5}I}+\dfrac{3\sqrt{3}}{62} Y^{(6)}_{\mathbf{5}II}$ can be lifted to weight $-4$ polyharmonic Maa{\ss} forms through  Eq.~\eqref{eq:lift-cond} as follow
\begin{eqnarray}
\nonumber \xi_{-4}(Y_{\mathbf{1}}^{(-4)}) &=& Y^{(6)}_{\mathbf{1}}\,,~~~~~D^5(Y_{\mathbf{1}}^{(-4)})= -\dfrac{24}{(4\pi)^5}} Y^{(6)}_{\mathbf{1} \,, \\
\nonumber \xi_{-4}(Y_{\mathbf{3}}^{(-4)}) &=& \Omega_{\mathbf{3}} Y^{(6)}_{\mathbf{3}}\,,~~~~D^5(Y_{\mathbf{3}}^{(-4)})=-\dfrac{24}{(4\pi)^5} Y^{(6)}_{\mathbf{3}} \,, \\
\nonumber \xi_{-4}(Y_{\mathbf{3}'}^{(-4)}) &=& \Omega_{\mathbf{3}'} Y^{(6)}_{\mathbf{3}'}\,,~~~~D^5(Y_{\mathbf{3}'}^{(-4)})=-\dfrac{24}{(4\pi)^5} Y^{(6)}_{\mathbf{3}'} \,, \\
\xi_{-4}(Y_{\mathbf{5}}^{(-4)}) &=& \Omega_{\mathbf{5}} Y^{(6)}_{\mathbf{5}}\,,~~~~D^5(Y_{\mathbf{5}}^{(-4)})=-\dfrac{24}{(4\pi)^5} Y^{(6)}_{\mathbf{5}}\,.
\end{eqnarray}
Using the general results of Eq.~\eqref{eq:polyHarmonic-expr-main}, we find the weight $k=-4$ polyharmonic Maa{\ss} forms of level $5$ are of the following form:
\begin{eqnarray}
\nonumber Y_{\mathbf{1}}^{(-4)}&=& \dfrac{y^5}{5} + \dfrac{63\Gamma(5,4\pi y)}{128\pi^5 q} + \dfrac{2079\Gamma(5,8\pi y)}{4096\pi^5 q^2} + \dfrac{427\Gamma(5,12\pi y)}{864\pi^5 q^3} + \dfrac{66591\Gamma(5,16\pi y)}{131072 \pi^5 q^4} + \cdots   \\
\nonumber &&+\dfrac{\pi}{80}\dfrac{\zeta(5)}{\zeta(6)} + \dfrac{189 q}{16\pi^5} + \dfrac{6237 q^2}{512 \pi^5} + \dfrac{427 q^3}{36 \pi^5} + \dfrac{199773 q^4}{16384 \pi^5} + \cdots \,, \\
\nonumber  Y_{\mathbf{3},1}^{(-4)} &=& \dfrac{y^5}{5} + \dfrac{7815 }{34304 \pi^5}\left( \dfrac{\Gamma(5, 4 \pi y)}{ q}+ \dfrac{24211 \Gamma(5, 8 \pi y)}{25008 q^2}+ \dfrac{378004 \Gamma(5, 12 \pi y)}{379809 q^3}+ \dfrac{993 \Gamma(5, 16 \pi y)}{1024 q^4} + \cdots  \right) \\
\nonumber &&+\dfrac{\pi L(5,\chi_3)}{80\sqrt{5}L(6,\chi_3)} + \dfrac{23445 }{4288 \pi^5}\left( q + \dfrac{24211 q^2}{25008}+ \dfrac{378004 q^3}{379809}+ \dfrac{993 q^4}{1024}+ \dfrac{4882813 q^5}{4884375} + \cdots \right) \,,  \\
\nonumber  Y_{\mathbf{3},2}^{(-4)} &=& - \dfrac{15515625 q^{1/5}}{35127296 \sqrt{2} \pi^5}\left( \dfrac{\Gamma(5, 16 \pi y/5)}{ q}+ \dfrac{60218368 \Gamma(5, 36 \pi y/5)}{58635657 q^2}+ \dfrac{5557184 \Gamma(5, 56 \pi y/5)}{5563117 q^3} + \cdots  \right) \\
\nonumber && - \dfrac{46875 q^{1/5}}{4288 \sqrt{2} \pi^5}\left( 1 + \dfrac{3751 q}{3888}+ \dfrac{161052 q^2}{161051}+ \dfrac{1016801 q^3}{1048576}+ \dfrac{1355684 q^4}{1361367} + \cdots \right) \,, \\
\nonumber  Y_{\mathbf{3},3}^{(-4)} &=& - \dfrac{15625 q^{4/5}}{34304 \sqrt{2} \pi^5}\left( \dfrac{\Gamma(5, 4 \pi y/5)}{ q}+ \dfrac{3751 \Gamma(5, 24 \pi y/5)}{3888 q^2}+ \dfrac{161052 \Gamma(5, 44 \pi y/5)}{161051 q^3} + \cdots  \right) \\
\nonumber && - \dfrac{46546875 q^{4/5}}{4390912 \sqrt{2} \pi^5}\left( 1 + \dfrac{60218368 q}{58635657}+ \dfrac{5557184 q^2}{5563117}+ \dfrac{2535526400 q^3}{2458766307}+ \dfrac{3844775 q^4}{3860784} + \cdots \right) \,, \\
\nonumber  Y_{\mathbf{3}',1}^{(-4)} &=& \dfrac{y^5}{5} - \dfrac{3905 }{17152 \pi^5}\left( \dfrac{\Gamma(5, 4 \pi y)}{ q}+ \dfrac{48453 \Gamma(5, 8 \pi y)}{49984 q^2}+ \dfrac{5731 \Gamma(5, 12 \pi y)}{5751 q^3}+ \dfrac{993 \Gamma(5, 16 \pi y)}{1024 q^4} + \cdots  \right) \\
\nonumber &&-\dfrac{\pi L(5,\chi_3)}{80\sqrt{5}L(6,\chi_3)} - \dfrac{11715 }{2144 \pi^5}\left( q + \dfrac{48453 q^2}{49984}+ \dfrac{5731 q^3}{5751}+ \dfrac{993 q^4}{1024}+ \dfrac{3126 q^5}{3125} + \cdots \right) \,, \\
\nonumber  Y_{\mathbf{3}',2}^{(-4)} &=& \dfrac{1890625 q^{2/5}}{4167936 \sqrt{2} \pi^5}\left( \dfrac{\Gamma(5, 12 \pi y/5)}{ q}+ \dfrac{7721325 \Gamma(5, 32 \pi y/5)}{7929856 q^2}+ \dfrac{45111978 \Gamma(5, 52 \pi y/5)}{44926453 q^3} + \cdots  \right) \\
\nonumber && + \dfrac{1453125 q^{2/5}}{137216 \sqrt{2} \pi^5}\left( 1 + \dfrac{537792 q}{521017}+ \dfrac{40051 q^2}{40176}+ \dfrac{45435392 q^3}{44015567}+ \dfrac{161052 q^4}{161051} + \cdots \right)\,, \\
\nonumber  Y_{\mathbf{3}',3}^{(-4)} &=& \dfrac{484375 q^{3/5}}{1097728 \sqrt{2} \pi^5}\left( \dfrac{\Gamma(5, 8 \pi y/5)}{ q}+ \dfrac{537792 \Gamma(5, 28 \pi y/5)}{521017 q^2}+ \dfrac{40051 \Gamma(5, 48 \pi y/5)}{40176 q^3} + \cdots  \right) \\
\nonumber && + \dfrac{1890625 q^{3/5}}{173664 \sqrt{2} \pi^5}\left( 1 + \dfrac{7721325 q}{7929856}+ \dfrac{45111978 q^2}{44926453}+ \dfrac{1823017 q^3}{1881792}+ \dfrac{71092323 q^4}{70799773} + \cdots \right) \,, \\
\nonumber  Y_{\mathbf{5},1}^{(-4)} &=& \dfrac{y^5}{5} - \dfrac{1563 }{15872 \pi^5}\left( \dfrac{\Gamma(5, 4 \pi y)}{ q}+ \dfrac{33 \Gamma(5, 8 \pi y)}{32 q^2}+ \dfrac{244 \Gamma(5, 12 \pi y)}{243 q^3}+ \dfrac{1057 \Gamma(5, 16 \pi y)}{1024 q^4} + \cdots  \right) \\
\nonumber &&-\dfrac{13\pi}{5208} \dfrac{\zeta(5)}{\zeta(6)} - \dfrac{4689 }{1984 \pi^5}\left( q + \dfrac{33 q^2}{32}+ \dfrac{244 q^3}{243}+ \dfrac{1057 q^4}{1024}+ \dfrac{1625521 q^5}{1628125} + \cdots \right)\,, \\
\nonumber  Y_{\mathbf{5},2}^{(-4)} &=& \dfrac{3303125 \sqrt{3} q^{1/5}}{16252928 \sqrt{2} \pi^5}\left( \dfrac{\Gamma(5, 16 \pi y/5)}{ q}+ \dfrac{60716032 \Gamma(5, 36 \pi y/5)}{62414793 q^2}+ \dfrac{17749248 \Gamma(5, 56 \pi y/5)}{17764999 q^3} + \cdots  \right) \\
\nonumber && + \dfrac{9375 \sqrt{3} q^{1/5}}{1984 \sqrt{2} \pi^5}\left( 1 + \dfrac{671 q}{648}+ \dfrac{161052 q^2}{161051}+ \dfrac{1082401 q^3}{1048576}+ \dfrac{4101152 q^4}{4084101} + \cdots \right) \,, \\
\nonumber  Y_{\mathbf{5},3}^{(-4)} &=& \dfrac{190625 q^{2/5}}{321408 \sqrt{6} \pi^5}\left( \dfrac{\Gamma(5, 12 \pi y/5)}{ q}+ \dfrac{8219475 \Gamma(5, 32 \pi y/5)}{7995392 q^2}+ \dfrac{45112221 \Gamma(5, 52 \pi y/5)}{45297746 q^3} + \cdots  \right) \\
\nonumber && + \dfrac{309375 \sqrt{3} q^{2/5}}{63488 \sqrt{2} \pi^5}\left( 1 + \dfrac{48896 q}{50421}+ \dfrac{64477 q^2}{64152}+ \dfrac{1376832 q^3}{1419857}+ \dfrac{161052 q^4}{161051} + \cdots \right)\,, \\
\nonumber  Y_{\mathbf{5},4}^{(-4)} &=& \dfrac{103125 \sqrt{3} q^{3/5}}{507904 \sqrt{2} \pi^5}\left( \dfrac{\Gamma(5, 8 \pi y/5)}{ q}+ \dfrac{48896 \Gamma(5, 28 \pi y/5)}{50421 q^2}+ \dfrac{64477 \Gamma(5, 48 \pi y/5)}{64152 q^3} + \cdots  \right) \\
\nonumber && + \dfrac{190625 q^{3/5}}{13392 \sqrt{6} \pi^5}\left( 1 + \dfrac{8219475 q}{7995392}+ \dfrac{45112221 q^2}{45297746}+ \dfrac{652223 q^3}{632448}+ \dfrac{391007898 q^4}{392616923} + \cdots \right) \,, \\
\nonumber  Y_{\mathbf{5},5}^{(-4)} &=& \dfrac{3125 \sqrt{3} q^{4/5}}{15872 \sqrt{2} \pi^5}\left( \dfrac{\Gamma(5, 4 \pi y/5)}{ q}+ \dfrac{671 \Gamma(5, 24 \pi y/5)}{648 q^2}+ \dfrac{161052 \Gamma(5, 44 \pi y/5)}{161051 q^3} + \cdots  \right) \\
&&\hskip-0.2in + \dfrac{9909375 \sqrt{3} q^{4/5}}{2031616 \sqrt{2} \pi^5}\left( 1 + \dfrac{60716032 q}{62414793}+ \dfrac{17749248 q^2}{17764999}+ \dfrac{2535526400 q^3}{2617236643}+ \dfrac{687775 q^4}{684936} + \cdots \right)\,.
\end{eqnarray}

\end{appendix}

%\bibliographystyle{utphys}
%\bibliography{references}

\providecommand{\href}[2]{#2}\begingroup\raggedright\endgroup

\end{document}